\documentclass{aa}  

\usepackage{graphicx}
\usepackage{txfonts}
\usepackage{multirow}

\newcommand{\kms}{km\,s$^{-1}$}
\newcommand{\vsini}{$v\sin i$}
\newcommand{\vrad}{$v_{\rm rad}$}
\newcommand{\msun}{M$_{\odot}$}
\newcommand{\rsun}{R$_{\odot}$}
\newcommand{\bl}{$B_{\ell}$}
\newcommand{\ddeg}{$^{\circ}$}

\begin{document}

\title{Discovery of new magnetic early-B stars\\ within the MiMeS HARPSpol survey
\thanks{Based on observations collected at the European Southern Observatory, Chile (Program ID 187.D-0917), and on observations obtained at the Canada-France-Hawaii Telescope (CFHT), which is operated by the National Research Council of Canada, the Institut National des Sciences de l'Univers (INSU) of the Centre National de la Recherche Scientifique (CNRS) of France, and the University of Hawaii}}
\titlerunning{Magnetic massive B stars discoveries in the MiMeS HARPSpol survey}

\author{E. Alecian
          \inst{1,2}
          \and
          O. Kochukhov \inst{3}
          \and
          V. Petit \inst{4}
          \and
          J. Grunhut \inst{5}
          \and
          J. Landstreet \inst{6,7}
          \and
          M.E. Oksala \inst{8}
          \and
          G.A. Wade \inst{9}
          \and\\
          G. Hussain \inst{5}
          \and
          C. Neiner \inst{2}
          \and
          D.~Bohlender \inst{10}
          \and
          the MiMeS Collaboration
          }

\institute{UJF-Grenoble 1 / CNRS-INSU, Institut de Plan\'etologie et d'Astrophysique de Grenoble (IPAG) UMR 5274, Grenoble, F-38041, France,
              \email{evelyne.alecian@obs.ujf-grenoble.fr}
              \and
              LESIA-Observatoire de Paris, CNRS, UPMC, Univ. Paris-Diderot, 5 place Jules Janssen, F-92195 Meudon Cedex, France
              \and
              Department of Physics and Astronomy, Uppsala University, Box 516, SE-751 20 Uppsala, Sweden
              \and
              Bartol Research Institute, Department of Physics \& Astronomy, University of Delaware, Newark, DE 19716, USA
              \and
              ESO, Karl-Schwarzschild-Strasse 2, 85748 Garching, Germany
              \and
              Armagh Observatory, College Hill, Armagh, BT61 9DG, Northern Ireland, UK
              \and
              Department of Physics and Astronomy, The University of Western Ontario, London, Ontario, N6A 3K7, Canada
              \and
              Astronomick\'y \'ustav, Akademie v\v{e}d \v{C}esk\'e republiky, Fri\v{c}ova 298, 251 65 Ond\v{r}ejov, Czech Republic
              \and
              Dept. of Physics, Royal Military College of Canada, PO Box 17000, Stn Forces, Kingston Ontario, K7K 7B4, Canada
              \and
              Dominion Astrophysical Observatory, Herzberg Astronomy and Astrophysics Program, National Research Council of Canada, 5071 West Saanich Road, Victoria, BC V9E 2E7, Canada
             }

   \date{Received September 15, 1996; accepted March 16, 1997}

 
\abstract
{The Magnetism in Massive Stars (MiMeS) project aims at understanding the origin of the magnetic fields in massive stars as well as their impact on stellar internal structure, evolution, and circumstellar environment. 
}
{One of the objectives of the MiMeS project is to provide stringent observational constraints on the magnetic fields of massive stars, however, identification of magnetic massive stars is challenging, as only a few percent of high-mass stars host strong fields detectable with the current instrumentation. Hence, one of the first objectives of the MiMeS project was to search for magnetic objects among a large sample of massive stars, and to build a sub-sample for in-depth follow-up studies required to test the models and theories of fossil field origins, magnetic wind confinement and magnetospheric properties, and magnetic star evolution.}
{We obtained high-resolution spectropolarimetric observations of a large number of OB stars thanks to three large programs (LP) of observations that have been allocated on the high-resolution spectropolarimeters ESPaDOnS, Narval, and the polarimetric module HARPSpol of the HARPS spectrograph. We report here on the methods and first analysis of the HARPSpol magnetic detections. We identified the magnetic stars using a multi-line analysis technique. Then, when possible, we monitored the new discoveries to derive their rotation periods, which are critical for follow-up and magnetic mapping studies. We also performed a first-look analysis of their spectra and identified obvious spectral anomalies (e.g., surface abundance peculiarities, H$\alpha$ emission), which are also of interest for future studies. 
}
{In this paper, we focus on eight of the 11 stars (from the HARPSpol LP sample) in which we discovered or confirmed a magnetic field from the HARPSpol LP sample (the remaining three were published in a previous paper). Seven of the stars were detected in early-type Bp stars, while the last star was detected in the Ap companion of a normal early B-type star. We report obvious spectral and multiplicity properties, as well as our measurements of their longitudinal field strengths, and their rotation periods when we are able to derive them. We also discuss the presence or absence of H$\alpha$ emission with respect to the theory of centrifugally-supported magnetospheres.}
{}


   \maketitle
%

\section{Introduction}

The properties of the magnetic fields of the {intermediate mass stars (1.5 to 8~\msun) of spectral type A and late-B (B4 and later) on the main sequence (A/B stars hereafter) are now well established \citep[see the reviews of ][as well as references hereinafter]{landstreet92,donati09}}. They are found in { a small fraction} of these stars, exclusively among the chemically peculiar Ap/Bp stars {\citep{wolff68,shorlin02,bagnulo06,auriere10,kochukhov13}}. They are mainly dipolar or low-order multi-polar fields, with polar strengths ranging from 300 G to 30 kG, most having fields of order 1~kG {\citep[e.g.][]{landstreet78,landstreet89,bohlender87,bohlender93,mathys97,wade00a,wade06,kochukhov06b,elkin10,bailey12}}. These fields are stable over many years, and even decades {\citep[for stars with sufficient observations, e.g.][]{landstreet89,khokhlova97,silvester14}}. While the field strengths seem to show a statistical decrease with stellar age, the field incidence does not depend on the stellar age \citep{kochukhov06a,landstreet08}. Such stable and large scale fields are called fossil fields \citep[i.e. they are not continuously maintained from dissipation,][]{cowling45,cowling53,borra82} and are observed in intermediate-mass stars from the pre-main-sequence phase (in Herbig Ae/Be stars), throughout the main-sequence phase, and very likely even until the red giant phase of  stellar evolution {\citep{auriere08,auriere11,alecian13}}.

Fossil fields have very different properties compared to the magnetic fields of the Sun and other cool, low-mass stars, which indicate a different origin. While in low-mass stars, the presence of a deep sub-surface convective zone allows for a dynamo to occur and produce very complex and unstable magnetic fields as observed at the surface of the Sun, intermediate-mass stars lack such a convective zone to generate their  fields. It is believed instead that the fossil fields are remnants from fields enhanced or accumulated during star formation \citep[e.g.][]{moss01}. Studies performed the last years have brought a variety of new evidence in favour of this theory and in the generation and history of the such fossil fields \citep{wade05,braithwaite06,wade07,folsom08,alecian08a,alecian08b,alecian09,duez10a,duez10b,alecian13}.

{Massive stars (above 8 \msun) of spectral type O and early-B on the main sequence (OB stars hereafter), similarly to intermediate-mass stars, possess a large radiative envelope}. We therefore {assume} that the origin of the field is similar in both types of stars. The {OB} stars are however hotter than {A/B} stars, and can drive significant radiative winds \citep{castor75}. Magnetic interaction between the star and the environment could therefore be significant \citep{uddoula02}. It has also been proposed that magnetic fields can have a strong impact on the structure and evolution of massive stars \citep[e.g. via enhanced or suppressed mixing, surface velocity braking,][]{maeder00,uddoula09,briquet12}. Until recently, our knowledge of the magnetic properties of the OB stars was very poor. Fields were only detected in a few peculiar cases, such as in He-strong or He-weak stars, as well as in a few non-peculiar stars. Such a lack of observational constraints motivated a large consortium to start the Magnetism in Massive Stars (MiMeS) project, {to study the origin and physics of the magnetic fields in massive stars.

One of the main objectives of the MiMeS consortium was to obtain stringent observational constraints on the magnetic properties of  massive stars. With this aim, we have performed a high-resolution spectropolarimetric survey of about 360 OB stars selected in the field of the Galaxy and in young clusters or associations. A large sample was required to detect a number of magnetic stars large enough for compiling good statistics on the magnetic properties of the OB stars. To perform this survey}, three spectropolarimetric large programmes (LPs) were allocated between mid-2008 and early 2013, on ESPaDOnS installed on the Canada-France-Hawaii Telescope (CFHT, Hawaii), on Narval installed on the Telescope Bernard Lyot (TBL, Pic du Midi, France), and on HARPS accompanied with the polarimetric module HARPSpol installed on the ESO~3.6m telescope (La Silla, Chile). {The Narval and ESPaDOnS magnetic detections were published in separate papers \citep[e.g.][]{grunhut09,grunhut13,oksala10,briquet13}, and the whole ESPaDOnS, Narval and HARPSpol survey will be published in forthcoming papers (Wade et al., Grunhut et al., Petit et al., Neiner et al., Alecian et al., in prep.). The present paper focuses {on the magnetic detections} of the HARPSpol sample.}

Within the HARPSpol survey, we detected nine new magnetic stars and confirmed the presence of the magnetic fields at the surface of two others - HD~105382 and HD~109026 - that had been previously reported \citep{briquet07,borra83}. The two new magnetic detections (HD~122451 and HD~130807) and the field confirmation in HD~105382 obtained during the first run (May 2011) of the LP have already been published in a Letter \citep{alecian11}. In this paper, we report seven new detections and one field confirmation in HD~109026 that were obtained during the four remaining runs (Dec. 2011, July 2012, Feb. 2013, June 2013).

This paper is structured as follows. In Section 2, we describe the observations and reduction we performed. In Section 3, we describe the spectral properties of the new magnetic stars. In Section 4, we analyse the polarised spectra and interpret them to propose magnetic field geometries. In, Section 5 we discuss the magnetospheric signatures observed (or not) in H$\alpha$, and in Section 6, we present a summary of our results.


\section{Observations and Reduction}

We used the polarisation properties of the Zeeman effect inside spectral lines to measure the magnetic fields of our sample. Because stellar magnetic fields generally produce a stronger circular than linear polarisation signal \citep[e.g.][]{wade00b}, we carried out observations in circular polarisation mode only. We obtained observations with the polarimetric module \citep{piskunov11} of the HARPS spectrograph \citep{mayor03} installed on the ESO 3.6m telescope (La Silla Observatory, Chile). This instrument configuration provides us with high-resolution spectropolarimetric data, covering a wavelength range from 378 nm to 691 nm with a gap between 526 and 534 nm. Depending on the magnitude of the star, one \emph{observation} consists of one or several successive \emph{polarimetric measurements}. To obtain one circularly polarised measurement, we acquired four successive \emph{individual spectra} between which we rotated the quarter-wave plate by 90\ddeg\, starting at 45\ddeg. The calibration spectra (bias, flat field, wavelength calibration) were obtained before the start of each night.

To reduce the data we used a modified version of the REDUCE package \citep{piskunov02,makaganiuk11}, which performs an optimal extraction of the cross-dispersed \'echelle spectra after bias subtraction, flat fielding and cosmic ray removal. We performed the wavelength calibration using the spectrum of a ThAr calibration lamp. Following the wavelength calibration, the spectra were shifted in velocity to the heliocentric frame. We normalised the optimally extracted spectra to the continuum following three successive steps. First, we divided the stellar spectra by the one-dimensional optimally extracted spectrum of the flat field to remove the blaze shape and fringing. Then, the resulting stellar spectra were corrected by the response function derived from observations of the solar spectrum. Finally, we determined the continuum level by fitting a smooth, slowly varying function to the envelope of the entire spectrum. Before this final step, we carefully inspected the spectrum of each star and masked out the strongest and broadest lines (including all Balmer lines, and the strongest He lines) from the fitting procedure. The resolving power of our spectra ranges from 95000 to 113000, depending on wavelength and order, with a median of 106000. Their signal-to-noise ratio at 500 nm ranges between $\sim100$ to $\sim1000$.

\begin{table}
\caption{Log of the observations.}
\label{tab:log}      
\centering          
\begin{tabular}{@{}l@{\;\;} l@{\;}c@{\;\;\;}c@{\;\;\;}c@{\;\;}r@{}}
\hline\hline       
ID & Date (d/m/y) & HJD             & $t_{\rm exp}$ & \# & S/N \\
& UT                & (2450000+) & (s)             &     &       \\
\hline
\object{HD 66765} & 13-12-2011 7:24 & 5\,908.80955 & 7200 & 2 & 500 \\
                 & 14-12-2011 3:42 & 5\,909.65571 & 7200 & 2 & 450 \\
                 & 15-12-2011 7:17 & 5\,910.80453 & 7200 & 2 & 600 \\
                 & 16-12-2011 2:32 & 5\,911.60699 & 3600 & 1 & 330 \\
                 & 16-12-2011 8:23 & 5\,911.85086 & 3600 & 1 & 310 \\
                 & 17-12-2011 2:35 & 5\,912.60934 & 3600 & 1 & 320 \\
                 & 17-12-2011 8:12 & 5\,912.84324 & 3600 & 1 & 430 \\
\object{HD 67621} & 12-12-2011 7:06 & 5\,907.79702 & 7200 & 2 & 680 \\
                 & 13-12-2011 3:42 & 5\,908.65554 & 3600 & 1 & 430 \\
                 & 14-12-2011 2:22 & 5\,909.59972 & 1800 & 1 & 230 \\
                 & 14-12-2011 8:35 & 5\,909.85897 & 2200 & 1 & 350 \\
                 & 15-12-2011 8:40 & 5\,910.86215 & 2200 & 1 & 400 \\
                 & 16-12-2011 7:31 & 5\,911.81444 & 2200 & 1 & 330 \\
                 & 17-12-2011 5:39 & 5\,912.73657 & 2200 & 1 & 380 \\
\object{HD 109026} & 15-02-2013 6:57 & 6\,338.79036 & 3600 & 2 & 1150 \\
                   & 16-02-2013 3:03 & 6\,339.62764 & 1800 & 1 & 740  \\
                   & 17-02-2013 4:36 & 6\,340.69253 & 1800 & 1 & 780  \\
                   & 18-02-2013 6:11 & 6\,341.75842 & 1800 & 1 & 730  \\
                   & 19-02-2013 5:12 & 6\,342.71703 & 1800 & 1 & 690  \\
                   & 20-02-2013 7:41 & 6\,343.82113 & 1800 & 1 & 740  \\
                   & 21-02-2013 3:13 & 6\,344.63486 & 1800 & 1 & 620  \\
                   & 21-02-2013 9:58 & 6\,344.91629 & 1800 & 1 & 570  \\
\object{HD 121743} & 21-07-2012 1:12   & 6\,129.55193 & 3600 & 1 & 710 \\
                   & 21-07-2012 23:29 & 6\,130.47991 & 3600 & 1 & 790 \\
\object{HD 133518} & 15-02-2013 9:20 & 6\,338.88872 & 3600 & 1 & 440 \\
                   & 16-02-2013 9:57 & 6\,339.91444 & 1200 & 1 & 260 \\
                   & 21-02-2013 7:19 & 6\,344.80544 & 1200 & 1 & 160 \\
\object{HD 147932} & 20-02-2013 9:02 & 6\,343.87595 & 7200 & 2 & 430 \\
                   & 21-02-2013 8:35 & 6\,344.85718 & 7200 & 2 & 250 \\
\object{HD 156324} & 19-07-2012   2:58 & 6\,127.62863 & 7200 & 2 & 150 \\
                   & 20-07-2012   6:38 & 6\,128.78131 & 3600 & 1 & 110 \\
                   & 19-06-2013 10:12$^{\ddag}$ & 6\,462.92963 & 4800 & 1 & 390 \\
\object{HD 156424} & 18-07-2012 4:20 & 6\,126.68582 & 7200 & 2 & 150 \\
                   & 19-07-2012 6:32 & 6\,127.77721 & 3600 & 1 & 100 \\
\hline                  
\end{tabular}
\tablefoot{
\tablefoottext{$\ddag$}{Observation obtained with ESPaDOnS.}
}
\end{table}

For one star, HD~156324, we recently acquired one additional observation using ESPaDOnS at CFHT, within another LP\footnote{The BinaMIcS LP aims at studying the magnetic fields in spectroscopic binary systems (see http://lesia.obspm.fr/binamics/)}. These data were obtained in a similar way to the HARPSpol data, and werereduced with the dedicated tool Libre-Esprit only available at CFHT. The resulting spectrum covers a wavelength range from 370 to 1048 nm, with a resolution of 65000 \citep[for additional information on ESPaDOnS spectropolarimetric data, see e.g.][]{silvester12}.

We obtained the polarised spectra by combining the four individual spectra using the ratio method \citep{bagnulo09}. A diagnostic null spectrum was also obtained by combining the spectra in such a way as to cancel the polarisation from the object, which reveals possible spurious polarisation contributions \citep{donati97}.

If successive polarimetric measurements of the same object were obtained, we combined them, pixel per pixel, using a weighted mean, where each weight is the square of the inverse of the error on each pixel. Finally, we performed a continuum re-normalisation order by order to all observations, using the Image Reduction and Analysis Facility (IRAF) routine \emph{continuum}, by fitting cubic spline functions to the line-free portions of the spectra. Table \ref{tab:log} summarises the log of the observations of the stars examined in this paper. Columns 1 to 6 give the stellar identification, date and UT time, heliocentric Julian date (HJD), total exposure time, number of polarimetric sequences, and signal-to-noise (SNR) of the observations.

\section{Spectral properties of the photospheres}

\subsection{Effective temperatures and surface gravities}

To interpret our data, it is necessary to have reliable estimates of effective temperature $T_{\rm eff}$ and surface gravity $\log g$. We obtained estimates of these quantities from available Johnson UBV, Str\"{o}mgren uvby$\beta$, and Geneva six-colour photometry obtained from the General Catalogue of Photometric Data \citep{mermilliod97}\footnote{http://obswww.unige.ch/gcpd/gcpd.html}. The stellar basic parameters have been estimated using transformations from photometry that have been calibrated with normal stars. The available UBV measurements have been used to derive $T_{\rm eff}$ estimates by computing the approximately reddening-independant parameter $Q = (U-B) - 0.72*(B-V)$. The calibration for this quantity was that for stars of $\log g = 4.0$ tabulated by \citet{worthey11}. UBV data cannot be used to obtain $\log g$ unless an independent estimate of the reddening parameter E(B-V) is available, and even using the reddening found from other photometry systems does not give us a very precise value for $\log g$. We transformed the Str\"{o}mgren photometry to estimates of $T_{\rm eff}$ and $\log g$ using the FORTRAN calibration programme of \citet{napiwotzki93}. This calibration is valid for main-sequence stars up to $T_{\rm eff} \sim 30000$~K. Geneva six-colour photometry was used to derive estimates of $T_{\rm eff}$ and $\log g$ using the FORTRAN calibration programme of \citet{kunzli97}, which is also valid up to about 30000~K. The derived basic parameters are listed in Table~\ref{tab:phot}. The inter-agreement of parameters obtained using the three photometric systems are in good overall agreement. We checked that the derived values are consistent with our spectra. To this aim, we calculated synthetic spectra in the local thermodynamic equilibrium (LTE) approximation using the code SYNTH of \citet{piskunov92}. SYNTH requires, as input, atmosphere models, obtained using the ATLAS 9 program \citep{kurucz93}, and a list of spectral line data obtained from the Vienna Atomic Line Database\footnote{http://ams.astro.univie.ac.at/$\sim$vald/} \citep[VALD;][]{piskunov95,kupka99}. All synthetic spectra have been computed using the solar metallicity of \citet{grevesse93} and have been convolved to a rotation function dependent on the projected rotational velocity \vsini\ \citep{gray92}. We compared, by eye, the wings of the Balmer lines and the metallic lines of our data to synthetic spectra and adjusted the temperature, gravity, \vsini, and \vrad\ until we obtained a reasonable match. For few stars, multiple pairs of $[T_{\rm eff},\log g]$ are equally able to fit our data. {When variability or distortion is observed in the spectral lines \vsini\ and \vrad\ have been obtained by fitting the blue and red edges of the metallic lines only.} Our estimated and adopted values of $T_{\rm eff}$, $\log g$, \vsini, and \vrad\ are presented in Table \ref{tab:phot}. No significant difference is found between the photospheric and spectroscopic determinations of $T_{\rm eff}$ and $\log g$. 

\begin{table*}
\caption{Effective temperatures, surface gravities, \vsini, and \vrad\ of our sample.}             
\label{tab:phot}      
\centering          
\begin{tabular}{@{}l@{\;\;}l@{\;\;}l@{\;\;}ll@{\;\;}l|l@{\;\;}l@{\;\;}l@{\;\;}l@{\;\;}l@{\;}|l@{\;\;}l@{}}
\hline\hline
ID & Johnson        & \multicolumn{2}{c}{Str\"omgren} & \multicolumn{2}{c|}{Geneva} & \multicolumn{5}{c|}{Spectroscopy} & \multicolumn{2}{c}{Adopted}\\
    & $T_{\rm eff}$ & $T_{\rm eff}$ & $\log g$             & $T_{\rm eff}$ & $\log g$       & St. & $T_{\rm eff}$ & $\log g$ & \vsini & $v_{\rm rad}$ & $T_{\rm eff}$ & $\log g$ \\
    & (K)                 & (K)                  & (cgs)                 & (K)                  & (cgs)            & comp. & (K)                 & (cgs)      & (\kms) & (\kms) & (K)                 & (cgs)      \\
(1) & (2)                & (3)                   & (4)                    & (5)                  & (6)                 & (7)    & (8)                 & (9)          & (10)      & (11)     & (12)         & (13)       \\    
\hline
HD~66765 & 21000 & (...) & (...) & 19820 & 4.32 & & 20000 & 4.0 & 100 & 21 & 20000 & 4.0\\
HD~67621 & 21500 & (...) & (...) & 20580 & 4.22 & & 21000 & 4.0 & 23 & 18 & 21000 & 4.0 \\
\multirow{2}{*}{HD~109026\tablefootmark{$\dag$}} & \multirow{2}{*}{16000} & \multirow{2}{*}{16310} & \multirow{2}{*}{4.12} & \multirow{2}{*}{16080} & \multirow{2}{*}{4.00} & A & [15000,17000] & 4.0 & 180\tablefootmark{$*$} & 0\tablefootmark{$*$} & \multirow{2}{*}{16000} & \multirow{2}{*}{4.0} \\
 & & & & & & B & (...) & (...) & 20 & 0 & & \\
HD~121743 & 22000 & 20810 & 3.79 & 21180 & 4.07 & & 21000 & 4.0 & 60 & 10 & 21000 & 4.0 \\
HD~133518 & 22000 & 20670 & 3.98 & 19400 & 4.29 & & 19000 & [3.5,4.0] & 0 & -5 & 19000 & 4.0 \\
HD~147932 & 18000 & 18260 & 4.74 & 18060 & 4.04 & & [16000,18000] & [3.75,4.0] & 140\tablefootmark{$*$} & -10\tablefootmark{$*$} & 17000 & 4.0 \\
\multirow{3}{*}{HD~156324\tablefootmark{$\dag$}} & \multirow{3}{*}{21000} & \multirow{3}{*}{(...)} & \multirow{3}{*}{(...)} & \multirow{3}{*}{19580} & \multirow{3}{*}{4.38} & A & 22000 & 4.0 & 60\tablefootmark{$*$} & [10,-40,30]\tablefootmark{$**$} & \multirow{3}{*}{22000} & \multirow{3}{*}{4.0} \\
 & & & & & & B & [14000-17000] & (...) & 30\tablefootmark{$*$} & [-30,140,-80]\tablefootmark{$**$} & & \\
 & & & & & & C & [14000-17000] & (...) &  5  & [10,8,0]\tablefootmark{$**$} & & \\
HD~156424 & 21000 & 21340 & 4.20 & (...) & (...) & & 20000 & 4.0 & 12 & 0 & 20000 & 4.0 \\
\hline                  
\end{tabular}
\tablefoot{Column (1): star identification. Column (2), (3), and (5) effective temperature of the stars determined using Johnson, Str\"omgren, and Geneva photometry, respectively. Columns (4) and (6): surface gravity of the stars obtained from Str\"omgren and Geneva photometry. Column (7): spectroscopic stellar component in the cases of a multiple spectroscopic system (P for primary, S for secondary, T for tertiary). Columns (8) to (11) effective temperature, $\log g$, \vsini\ and radial velocities estimated from our data. Columns (12) and (13): adopted effective temperatures and surface gravities for the least-squares deconvolution (LSD, Sec. 4.1). \\
'(...)': Not determined, because of unavailable photometric measurements. \tablefoottext{$\dag$}{Spectroscopic multiple systems, approximate.} \tablefoottext{$*$}{Approximate (large and variable spectral lines).} \tablefoottext{$**$}{Radial velocities ordered in increasing HJD. Approximate values for the component A and C due to broad, variable, or faint lines.}
}
\end{table*}

\subsection{Spectrum description and peculiarities}

In order to facilitate the interpretation of our data, we performed a literature search of each star, and listed important results that could affect spectroscopic observations of each object, including pulsations that can have an important impact in some cases \citep[e.g.][and references hereinafter]{neiner12}. Our findings are summarised for each star below. We then noted the peculiarities observed in our data when compared with synthetic spectra, and propose an interpretation for each one of them. Sometimes, at temperatures above 15000~K, non-LTE effects could significantly affect the strength of the synthetic lines \citep[e.g.][]{mihalas73}. In particular, in our sample, they could affect the lines of \ion{He}{i}, \ion{C}{iii}, and \ion{Si}{iii} \citep[e.g.][]{przybilla11}. We therefore also compared our data with non-LTE synthetic spectra computed using TLUSTY non-LTE atmosphere models and the SYNSPEC code \citep{hubeny88,hubeny92,hubeny95}. We checked that non-LTE effects cannot explain the peculiarities observed in the He, C, and Si lines. We summarise our findings in Table \ref{tab:sppec}, and we detail below our analysis for each star.

\begin{table*}
\caption{{Main spectral peculiarities, and spectroscopic classification of our sample stars.}}
\label{tab:sppec}      
\centering          
\begin{tabular}{llllll|l}
\hline\hline
ID & St. & He str. & Sp. var. & Sp. & Notes & Class \\
    & comp.        & & & distor. & & \\
(1) & (2) & (3) & (4) & (5) & (6) & (7) \\
\hline
HD~66765      & & strong & high & yes & & He-strong \\
HD~67621      & & weak   & faint & yes & & He-weak \\
\multirow{2}{*}{HD~109026} & P & solar ? & faint ? & no ? & & Normal \\
 & S & (...)      & high & no ? & Very strong metallic lines & Ap \\
HD~121743    & & strong  & high & yes & & He-strong \\
HD~133518    & & strong  & no    & no ? & & He-strong \\
HD 147932     & & faint     & high  & yes  & & He-weak  \\
\multirow{3}{*}{HD 156324} & P  & strong  & high  ? & yes ? & & He-strong \\
 & S  & (...)       & (...)     & (...)     & & (...)            \\
 & T  & weak ? & (...)   & (...)       & Numerous strong \ion{P}{ii/iii} lines & He-weak PGa \\
HD 156424      & & strong  & faint & no         & & He-strong       \\
\hline                  
\end{tabular}
\tablefoot{ Column (1): star identification. Column (2): spectroscopic stellar component in the cases of a multiple spectroscopic system (P for primary, S for secondary, T for tertiary). Column (3): apparent strength of the He lines relative to a solar abundance. Column (4): variability of the spectral lines. Column (5): distortion of the spectral lines. Column (6): additional notes on the peculiarities of the spectra. Column (7): classification of the star in terms of spectral peculiarity. \\
'(...)': cannot be determined using our data. '?': not accurate due to the blending of the spectral lines by the companion (for multiple spectroscopic systems), or because of the absence of rotational Doppler resolution (for HD 133518).
}
\end{table*}

\subsubsection{HD~66765}

HD~66765 is a B1/B2 V star member of the Vel OB2 association \citep{houk78,dezeeuw99}. Vel OB2 is an OB association found in the vicinity of the Wolf-Rayet WC8+O8-8.5III $\gamma^2$~Vel binary system, at a distance of about 410 pc, and with an age of $\sim20$ Myr \citep{dezeeuw99}. Strong and variable He lines of HD~66765 have been reported by various authors \citep{balona75,garrison77,wiegert98}. \citet{garrison77} propose HD~66765 to be a double-lined spectroscopic binary (SB2) system, but neither \citeauthor{wiegert98}'s data nor ours confirm this suggestion. 

Our spectra are consistent with a single star of a temperature of about 20000 K. The radial velocities of our two spectra do not vary, are found around 21 \kms, and differ by less than 2$\sigma$ from the mean value ($15.5\pm3.2$) of the members of Vel OB2 published by \citet{kharchenko05}. Our spectra display very strong and variable He lines with broad wings. Variable and distorted \ion{Si}{iii} and \ion{C}{ii} lines are also observed, revealing abundance spots at the surface of the star, and suggesting that HD~66765 belongs to the class of He-strong stars.

\subsubsection{HD~67621}

HD~67621 is a B2 IV star, a member of the Vel OB2 association \citep{dezeeuw99}, and has been poorly studied up to now. We detected a Zeeman signature in the $V$ spectrum of HD~67621 on the night of Dec. 12$^{\rm th}$ 2011, which allowed us to follow the star, hence the variation of the magnetic signature, during the following nights.

Our observations are consistent with a synthetic spectrum of $T_{\rm eff}=21000$~K, and a radial velocity around 18 \kms, consistent with the mean values of the members of Vel OB2 \citep[$15.5\pm3.2$,][]{kharchenko05}. \ion{He}{i}, \ion{Si}{ii}, and \ion{S}{ii} appear to be under-abundant compared to solar abundances, while the \ion{Si}{iii} triplet at 4552, 4567, and 4574 \AA\ is stronger than predicted. The shape and strengths of the line profiles of some elements (e.g. \ion{Fe}{iii}, \ion{Fe}{ii}, \ion{Si}{ii}, \ion{He}{i}, \ion{S}{ii}) present slight changes from one observation to another, suggesting the presence of abundance spots at the surface of the star. We therefore propose the HD~67621 belongs to the class of He-weak stars.

\subsubsection{HD~109026}

HD~109026 ($=\gamma$ Mus) is classified as a B5 V He-weak type star \citep{renson09}, is situated at a distance of $\sim$99~pc \citep{vanleeuwen07}, and is rotating with a projected velocity of $188\pm10$~\kms \citep{brown97}. \citet{waelkens98} classify the star as a slowly pulsating B star (SPB) based on Hipparcos and Geneva photometric observations. The photometric determinations of the effective temperature reported by different authors range from 15000 K to 17700 K \citep[e.g.][]{molenda04,kochukhov06a,zorec09}. \citet{niemczura03} found a value of 16000 K from ultra-violet (UV) spectra.  \citet{borra83} measured the magnetic field of the star in the wings of H$\beta$ and obtained three positive measurements at a level higher than 3$\sigma$, with values varying between 140 and 470 G, and error bars around 90~G. 

HD~109026 belongs to the sample of stars used by \citet{jilinski06} to analyse the radial velocities of B stars in the Sco-Cen association. The Scorpius-Centaurus OB association is the closest OB association, and consists of three subgroups, Upper Scorpius (US), Upper Centaurus-Lupus (UCL), and Lower Centaurus-Crux (LCC), which have ages of about 5, 17, and 16 Myr, and distances of 145, 140 and 118~pc, respectively. \citet{jilinski06} argue that HD~109026 could be a member of the association based on radial velocities. Based on the position and proper motions, \citet{bertiau58} placed this star inside the association. \citet{jones71}, to analyse the kinematics of the Sco-Cen association, excluded all stars close to the border of the association, including HD~109026, to avoid contamination by field stars. As a consequence \citet{dezeeuw99} did not include it in their analysis. \citet{thompson87}, based on the radial velocity and proper motions, find that HD~109026 has a low probability of membership. 

We searched the new Hipparcos catalogue \citep{perryman97,vanleeuwen07} for all Sco-Cen members (de Zeeuw et al. 1999) within roughly 11.5 degrees of HD~109026. Twelve HIP association members were found. The mean and standard deviation of the parallax and the two components of proper motion for this sample are $\pi = 9.85 \pm 0.70$~mas, $\mu_{\alpha} \cos \delta = -38.75 \pm 2.48$~mas, and $\mu_{\delta} = -9.86 \pm 2.54$~mas. These average values are very similar to those of the nearest HIP member, HIP 60561, roughly 2.5 degrees away on the sky. The corresponding quantities for HD~109026 are 10.04, $-51.33$, and $-5.40$~mas, all with uncertainties of about $\pm 0.12$~mas. Although the parallax of HD~109026 agrees well with that of the Sco-Cen stars, and the proper motion in declination is within $2\sigma$ of the mean of nearby Sco-Cen stars, the proper motion in right ascension differs from the mean of nearby association stars by five times the observed dispersion of member motions. It therefore appears unlikely that HD~109026 is a member of Sco-Cen.


Our spectroscopic data revealed, in addition to the rapidly rotating B star, a secondary component with a low \vsini\ ($\sim20$~\kms), highly variable metallic lines, and a lower temperature. The magnetic field is only detected in this second component (Sec. 4). The secondary contributes only weakly to the photospheric Balmer lines but displays very strong metallic lines, implying that the secondary is chemically peculiar, and belongs to the magnetic Ap class. As a consequence, the temperature of the secondary is difficult to determine. Depending on the temperature of the secondary our spectra are consistent with a primary of 15000 to 17000 K, with solar abundances (including He) and a rotation velocity of about 180 \kms. We therefore do not confirm the membership of the primary star in the He-weak class. 

\citet{waelkens98} reported the discovery of SPB-type pulsations in the Hipparcos photometric data of HD~109026, with a period of 2.73 days. In our spectra, the lines of the primary vary slightly but are contaminated with the strong and variable lines of the secondary, which does not allow us to disentangle between pulsations in the primary and chemical spots at the surface of the secondary. No variation of the radial velocities of the components could be measured in our data, which we obtained over seven nights. The orbital period of the system is therefore much longer than seven days, and the photometric variability observed by Waelkens et al. cannot be explained by a partial eclipse of the primary by the secondary. The shape of the magnetic signatures changes strongly from one night to the other (Sec. 4), as well as the individual intensity profile shapes, and their variations could be consistent with a period of 2.73 days. The photometric variations reported by \citet{waelkens98} could instead be due to the Ap nature of the secondary, and the measured periodicity could be its rotational period.

\subsubsection{HD~121743}

HD~121743 ($\phi$~Cen) is a B2 IV-V star member of the Upper Centaurus Lupus region of the Sco-Cen association \citep{dezeeuw99}. While no photometric variation has been detected \citep{jakate79,adelman01}, HD~121743 is known to show radial velocity variations with an amplitude of few tens of km/s \citep{buscombe60,vanhoof63,levato87}. No visual or spectroscopic companion could be found in the close vicinity of HD~121743 \citep{jones71,levato87,shatsky02}, however, pulsations of $\beta$~Cep type with high order modes have been detected in the spectrum of this star \citep{telting06}. The photometric measurements of the effective temperature range from 20200 to 23050 K \citep{nissen74,wolff85,morossi85,degeus89,wolff90,castelli91,sokolov95}, while the value of $\log g$ determined from fitting the Balmer lines with LTE atmosphere models varies from 3.9 to 4.2 \citep{wolff85,wolff90}.

Our data are consistent with a star of a temperature of 21000~K. We found no evidence of radial velocity variations in our spectra taken one day apart. The cores of the lines are clearly distorted, showing bumps that vary from one night to another. These bumps could be caused by either stellar pulsations, or abundance patches. No strong abundance anomalies are observed in the spectrum except a slight enhancement in He.

\subsubsection{HD~133518}

HD~133518 is a B2-B3 He-strong star \citep{zboril00,groote80,garrison77,macconnell70} with narrow lines ($v\sin i<30$ \kms, e.g. \citealt{walborn83}). \citet{smith01} mention the presence of redshifted emission in the \ion{C}{IV} resonance lines observed in ultra-violet spectra, which is proposed to be the result of downflows arising in magnetically confined wind shocks (e.g. \citealt{babel97}). They therefore suggest HD~133518 as a candidate magnetic B star, however, the magnetic measurements performed by \citet{borra79} and \citet{bohlender87} were all below detection with uncertainties on the longitudinal fields ranging from 230~G to 280~G.

Our spectra are consistent with an effective temperature around 19000 K. They show very strong abundance peculiarities in the \ion{He}{i}, \ion{Si}{ii}, \ion{S}{ii}, \ion{Fe}{ii}, and \ion{Fe}{iii} lines, consistent with He-strong stars. No variability is observed in our three observations obtained over six nights. We also note that no rotational broadening is required to fit the spectral lines, the $v\sin i$ measured from our high-resolution spectra is consistent with zero.


\begin{figure*}
\centering
\includegraphics[width=9cm,angle=90]{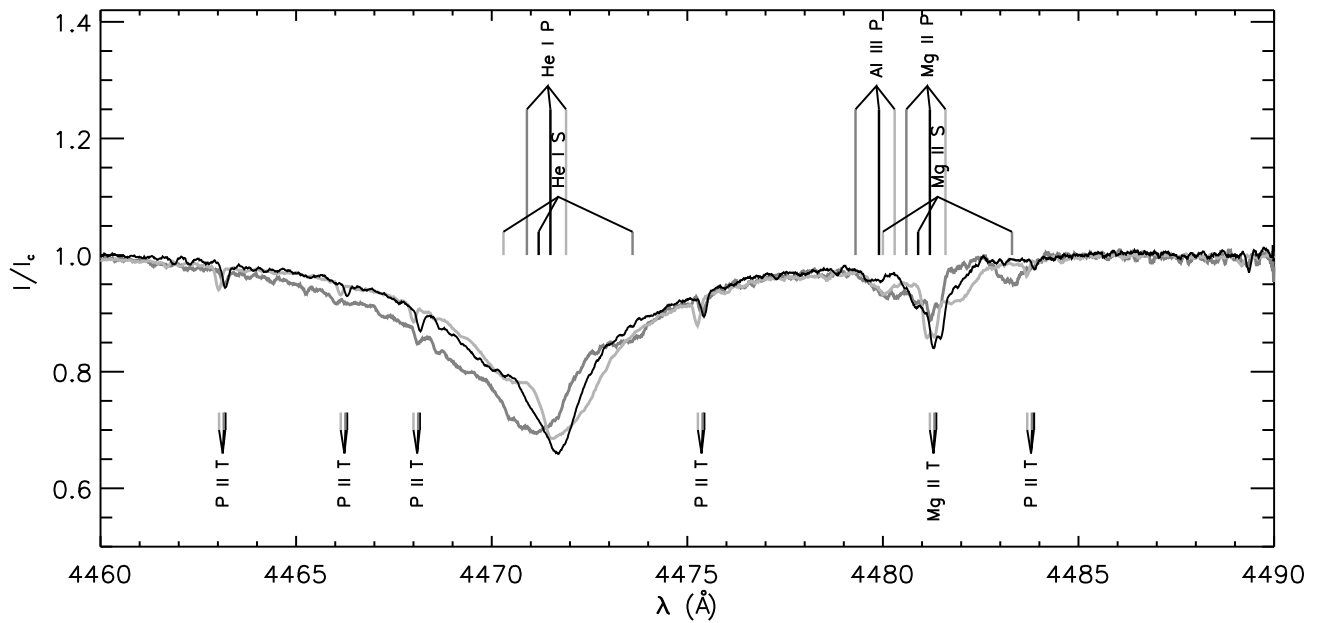}
\caption{Intensity spectrum of HD~156324 around \ion{He}{i}~4471~\AA\ and \ion{Mg}{II}~4481~\AA\ of the first (black), second (dark grey) and third (light grey) observations. The dominant ions of the primary (P), secondary (S) and tertiary (T) components are labelled. The positions of the ions of the component T are almost identical for the two first observations.}
\label{fig:hd156324}
\end{figure*}

\subsubsection{HD~147932}

HD~147932 is an early-type star, and a member of the Upper Scorpius region of the Sco-Cen association \citep{dezeeuw99}. The spectral type and effective temperature determinations found in the literature are highly scattered (from 12000 K to 18000 K, e.g. \citealt{hernandez05,degeus89}), most certainly due to the peculiarity of the spectrum (see below). \citet{koen02} report a photometric variability with a frequency 1.15758~d$^{-1}$, (i.e. 20.7~h), but were not able to identify the origin of the variability.

The wings of the Balmer lines (except H$\alpha$) of our spectra are consistent with an effective temperature ranging from 16000 to 18000~K. Whatever the temperature, \ion{He}{I} lines are much fainter than expected, suggesting that the star belongs to the He-weak class. The He and metallic lines (mainly Si and Fe) are highly distorted and show strong variability between the two observations obtained one night apart. These variations could be due to either abundance patches at the surface of the star, pulsations, or a combination of the two. A careful examination of the spectra does not allow us to determine the origin, and additional observations at different rotation phases are required to be able to interpret these distortions.

\subsubsection{HD~156324}

HD~156324 is a B2 V star member of the Sco OB4 association \citep{kharchenko04}, situated at about 1~kpc, with an age of about 7~Myr \citep{melnik09,kharchenko05}. \citet{roslund68} measured relatively fast radial velocity variations (from -33 to 73 km/s within 8 days), indicating an SB1 nature. Later, speckle observations revealed the presence of a component at about 0.4\arcsec \citep{hartkopf93,hartkopf96,tokovinin10}. At a distance of 1100~pc \citep{kharchenko05} the speckle companion would be separated by about 400 AU from the companion star, and would orbit with a period of few thousands of years. Therefore, the speckle companion cannot be at the origin of the radial velocity variations observed in the spectrum of HD~156324. On the other hand, with a separation of only 0.4", the light from the speckle companion must have entered the 0.8" fibre during our observations. \citet{tokovinin10} measured a magnitude difference of about 1.4, which implies that if the light from the companion has also entered the fibre, it would have contaminated our spectra.

Our spectra of HD~156324 are rather complex, revealing three different stellar components (Fig. \ref{fig:hd156324}). The dominant one (the primary) is rotating at about 60 km/s, and the two others are rotating at about 30 km/s (the secondary) and 5 km/s (the tertiary). The radial velocities of the primary and secondary show variations on a timescale of one day, while the radial velocity variations of the tertiary are much smaller, but do show small changes between July 2012 and June 2013. The secondary component of our spectrum is therefore very likely at the origin of the radial velocity variations observed by \citet{roslund68}. On the other hand, because of its small radial velocity changes, it is not yet clear if the tertiary component corresponds to the companion detected with speckle imaging.

We attempted to fit our observations with the composite synthetic spectrum of a triple system (Appendix A), to estimate the temperature of the stars, but the line blending between all components and the line variations of the primary make the task difficult. We propose an effective temperature of about 22000 K for the primary, and a temperature range of 14000-17000 K for the two others. Additional observations at various orbital phases are required in order to disentangle the spectral lines of all the components and give a better estimate of the stellar parameters. 

The shape and strength of the lines of the primary vary from one observation to the other, suggesting spots and/or pulsations. The strength of the He lines of the primary are much greater than the prediction, indicating an over-abundance of He. The depths of these lines cannot be reproduced by changing the temperature, by taking NLTE effects into account, or by changing the luminosity ratio between the components. The primary component of HD~156324, therefore, belongs to the class of He-strong stars. The \ion{Si}{ii/iii} lines appear weak compared to the model, and the \ion{He}{i}, \ion{C}{ii}, and \ion{Si}{ii/iii} lines appear distorted with bumps. The spectral peculiarities of the third stellar component identified in the spectrum (see below), however, makes it ambiguous whether these bumps originate in the photosphere of the primary.

A large number of narrow lines with a depth much higher than predicted are observed all over the spectrum. These lines have the same \vsini\ and radial velocities of the tertiary and, therefore, are associated with the tertiary component. Most of these lines have been identified and associated with the elements \ion{P}{ii} and \ion{P}{iii} (Fig. \ref{fig:hd156324}). Changing the effective temperature of the tertiary or the luminosity ratio between the components cannot account for the presence of these lines. We note also the difficulty in detecting this component in the He lines, with respect to the ease with which it is detected in the \ion{Mg}{II}~4481 line (Fig. \ref{fig:hd156324}). The tertiary could therefore be a chemically peculiar star belonging to the class of He-weak PGa. But the He-strong nature of the primary, the current large uncertainty on the luminosity ratio between the components and the temperature of the tertiary do not allow us to analyse the He lines of the tertiary in detail. We were unable to identify any Ga lines in the spectrum. Our classification of the tertiary as a CP4 PGa star is therefore tentative at this stage, and requires additional observations accompanied by a more extensive analysis of the spectra to be confirmed.

\subsubsection{HD~156424}

HD~156424 is a B2 V star, member of the Sco OB4 association \citep{kharchenko04}. Speckle observations obtained in 1990-1992 revealed the presence of a companion at a position angle of about 20\ddeg\ with a separation of about 0.78" \citep{hartkopf93}. In 2008, new speckle observations confirmed the presence of a companion that moved to a position angle of about 50\ddeg\ and a separation of 0.35" \citep{tokovinin10}. \citeauthor{tokovinin10} estimate the magnitude difference to be about 2.3 mag, but flag the difference to be highly uncertain and probably overestimated.

Despite the small separation between both components, no obvious indication of the presence of a companion could be found in our spectra, which could be due to the large magnitude difference measured with speckle imaging. Our data are consistent with a single star with a temperature of about 20000~K, and a radial velocity of about 0~\kms\ consistent with the mean value ($2.3\pm 4.2$ \kms) of the members of Sco OB4 published by \citet{kharchenko04}. The \ion{He}{i} lines appear very strong, which suggests that the star belongs to the magnetic He-strong class. Only slight variability is observed in the cores of few \ion{He}{i} and \ion{Si}{ii} lines between both observations obtained one night apart. 


\begin{figure*}
\centering
\includegraphics[width=6cm]{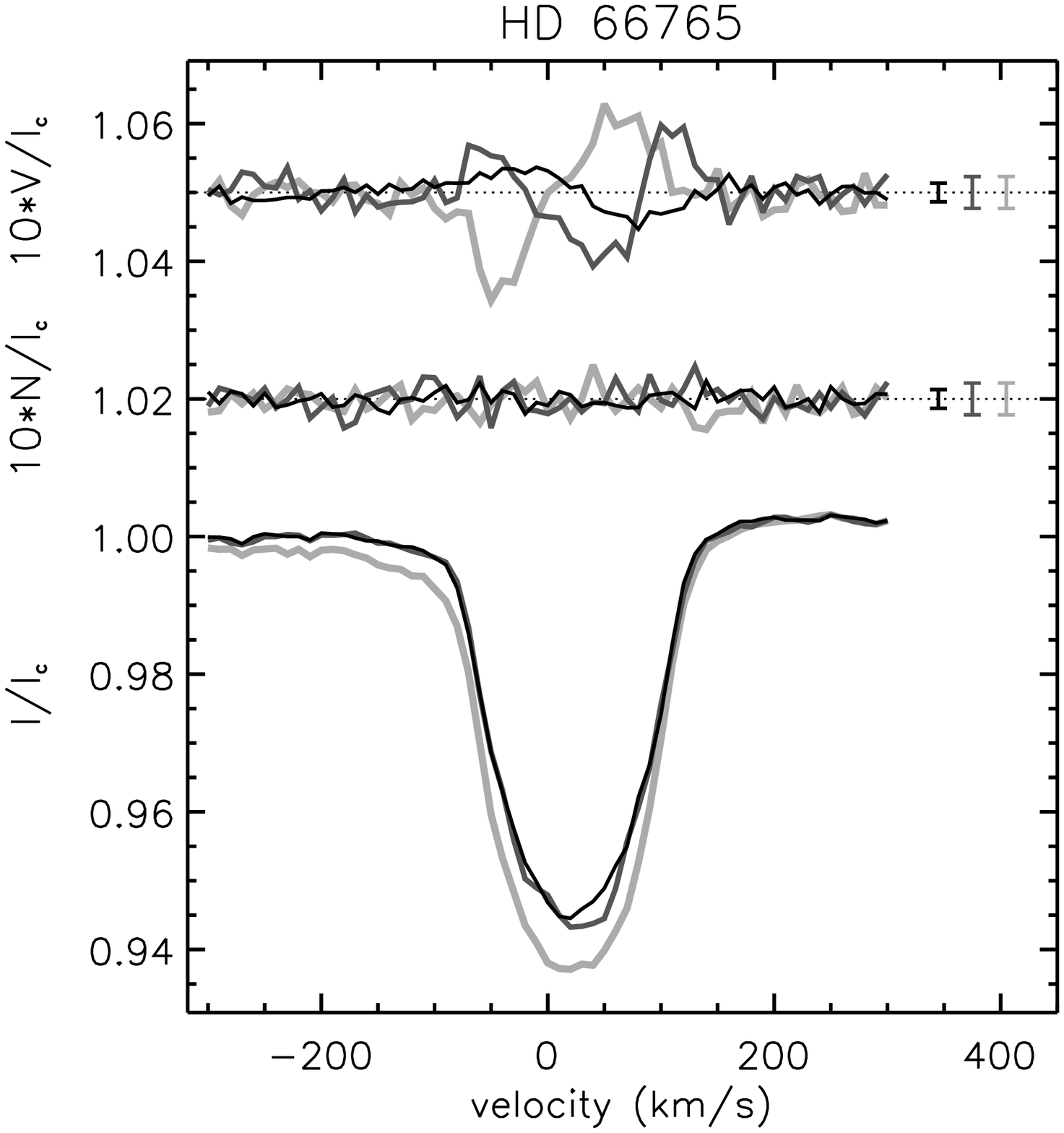}
\includegraphics[width=6cm]{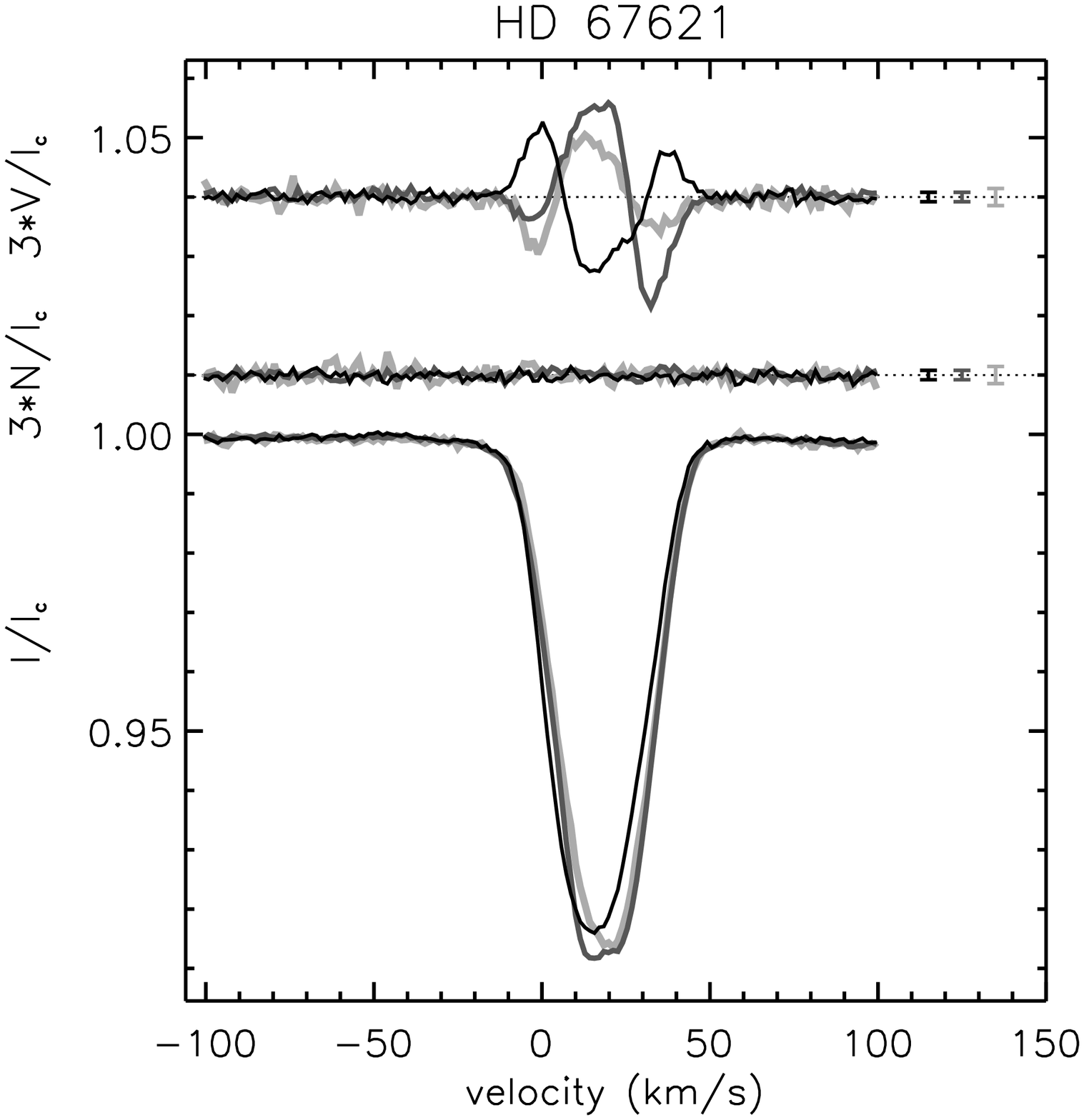}
\includegraphics[width=6cm]{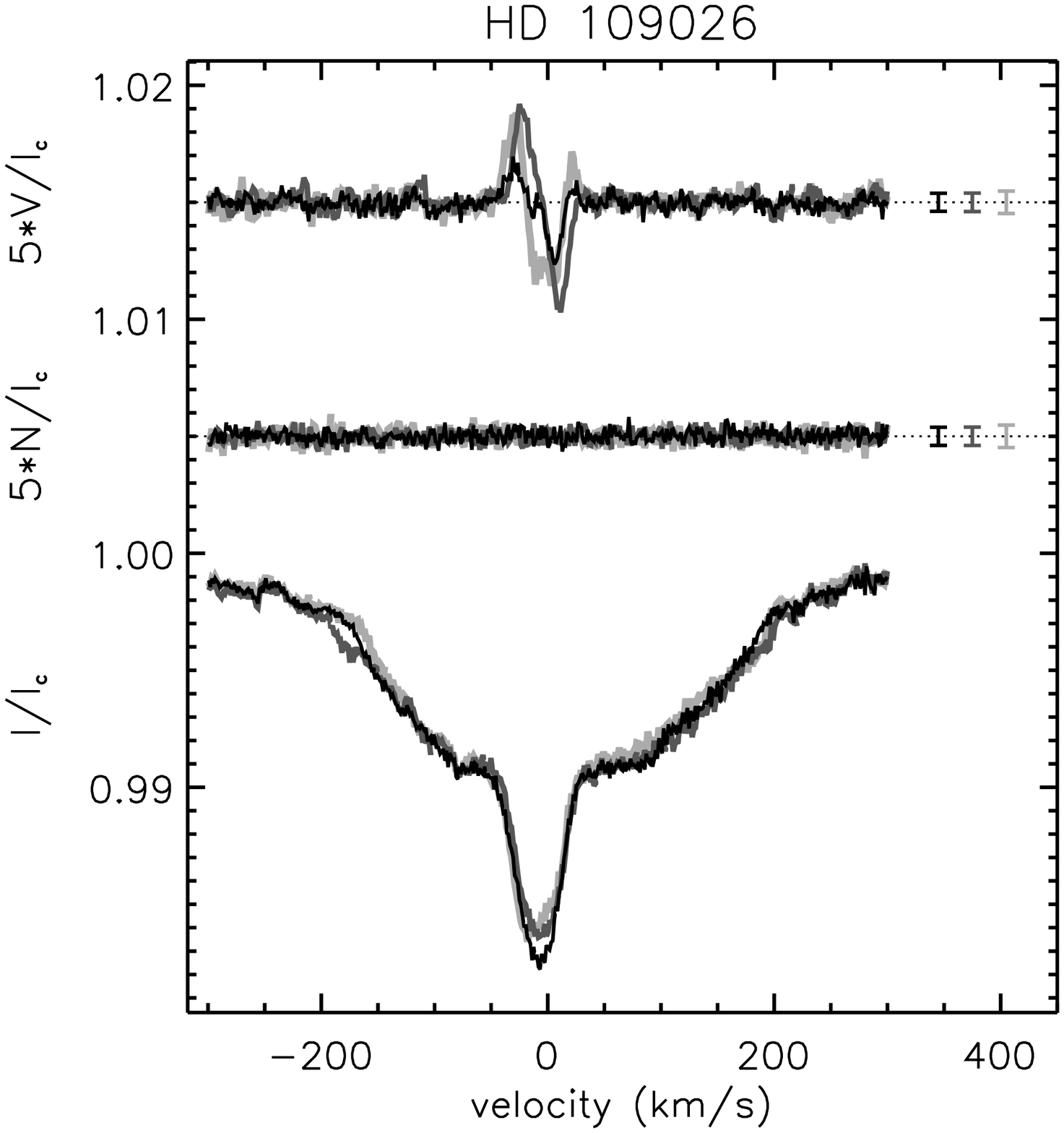}
\includegraphics[width=6cm]{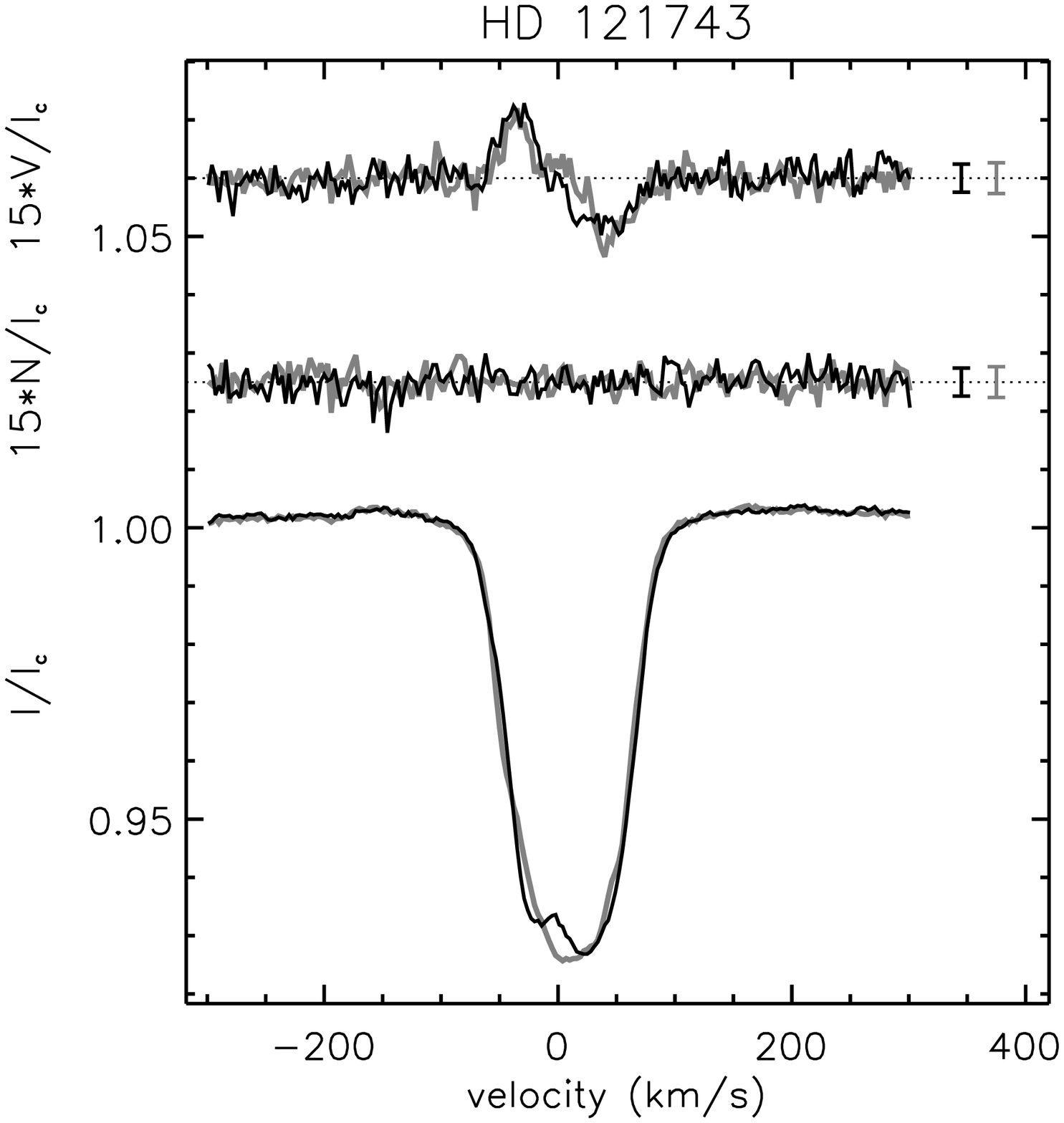}
\includegraphics[width=6cm]{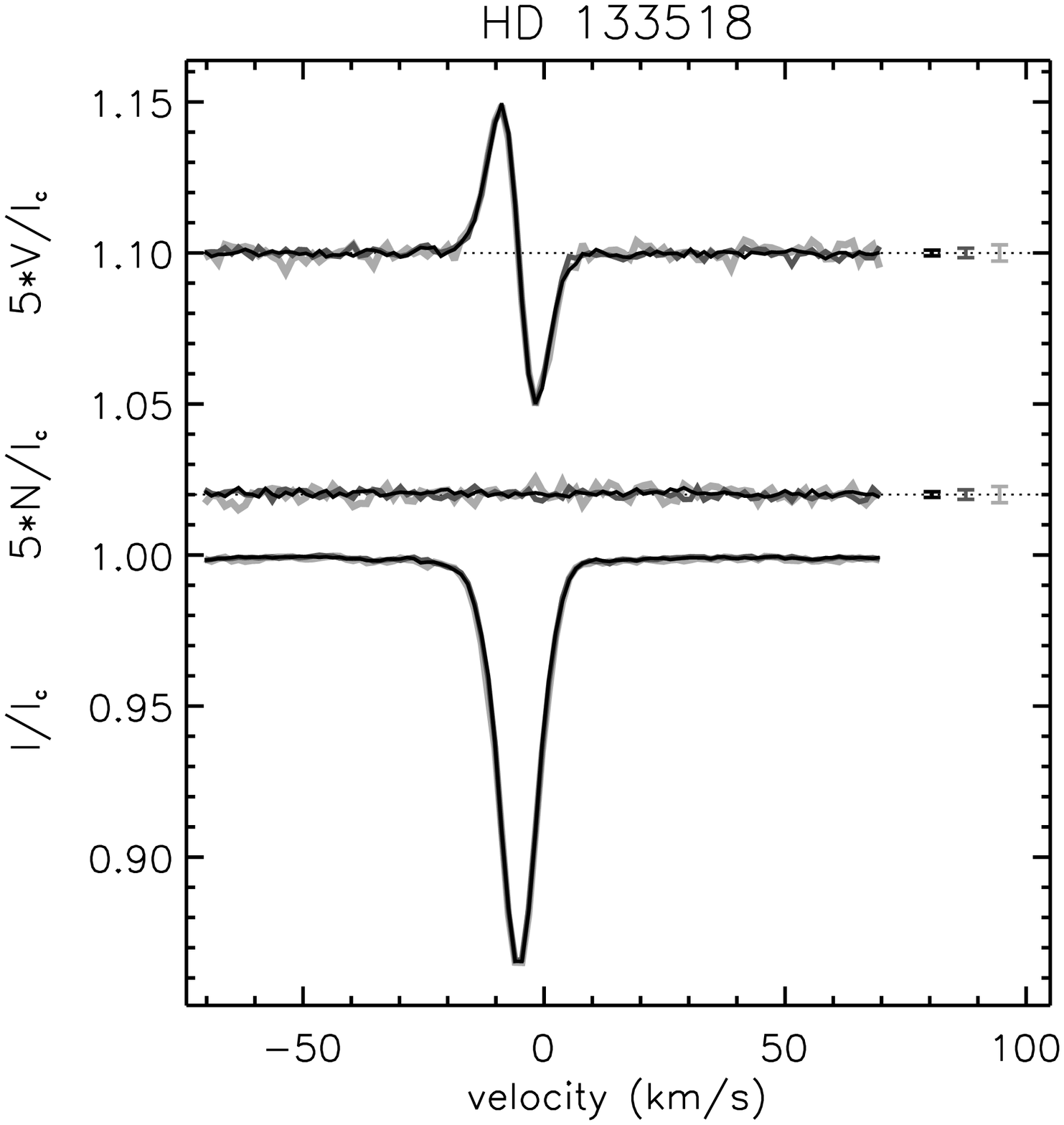}
\includegraphics[width=6cm]{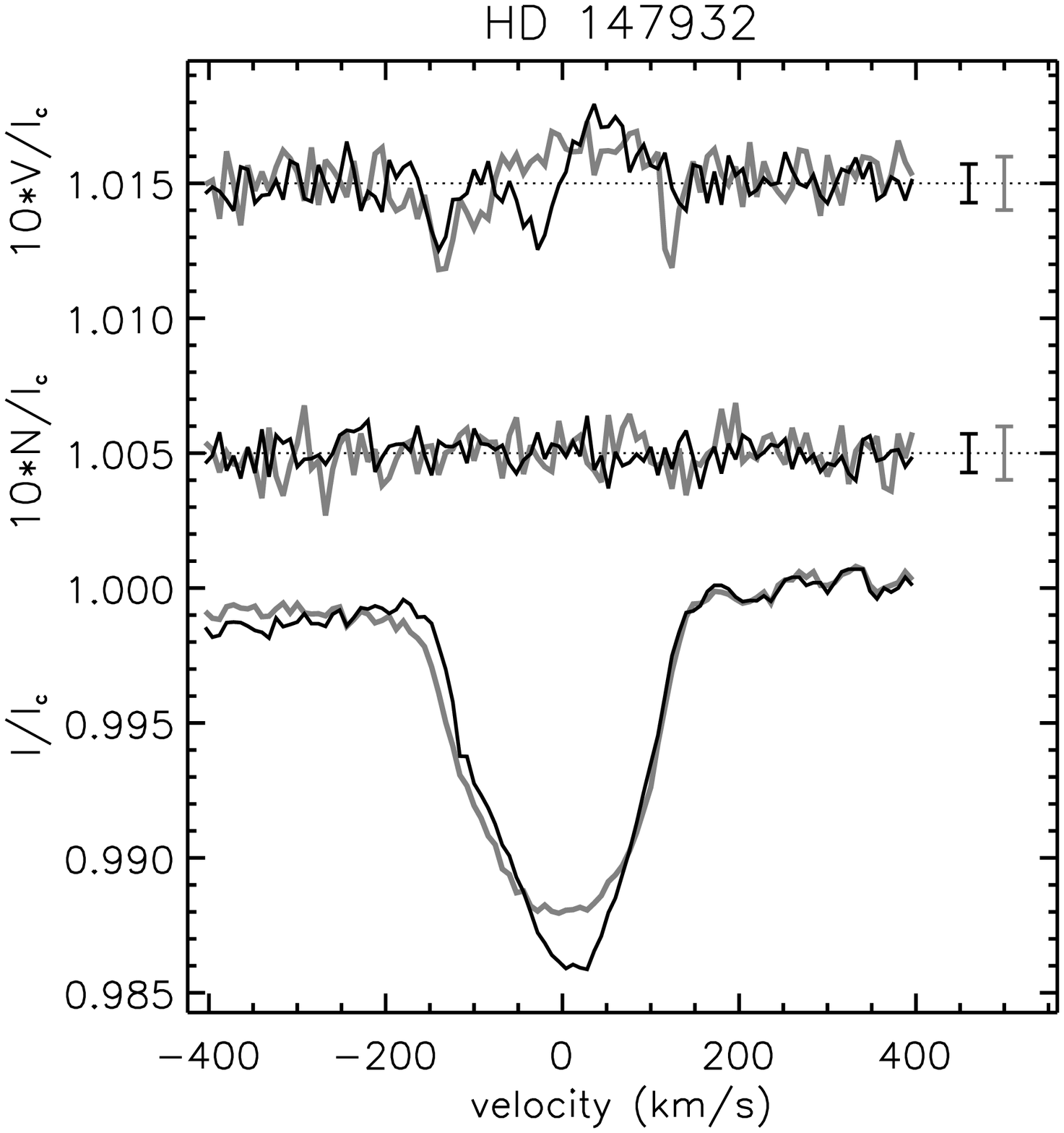}
\includegraphics[width=6cm]{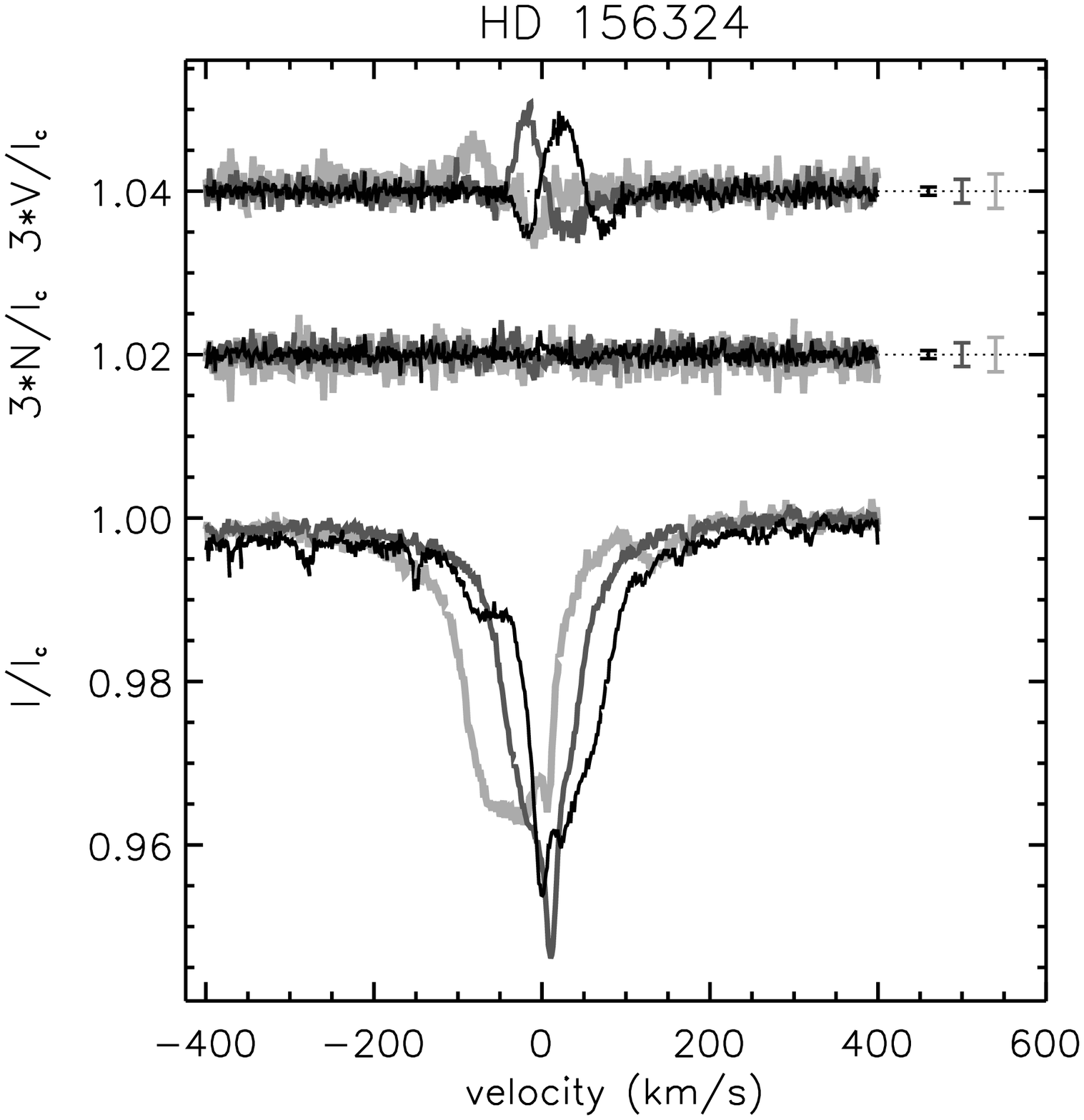}
\includegraphics[width=6cm]{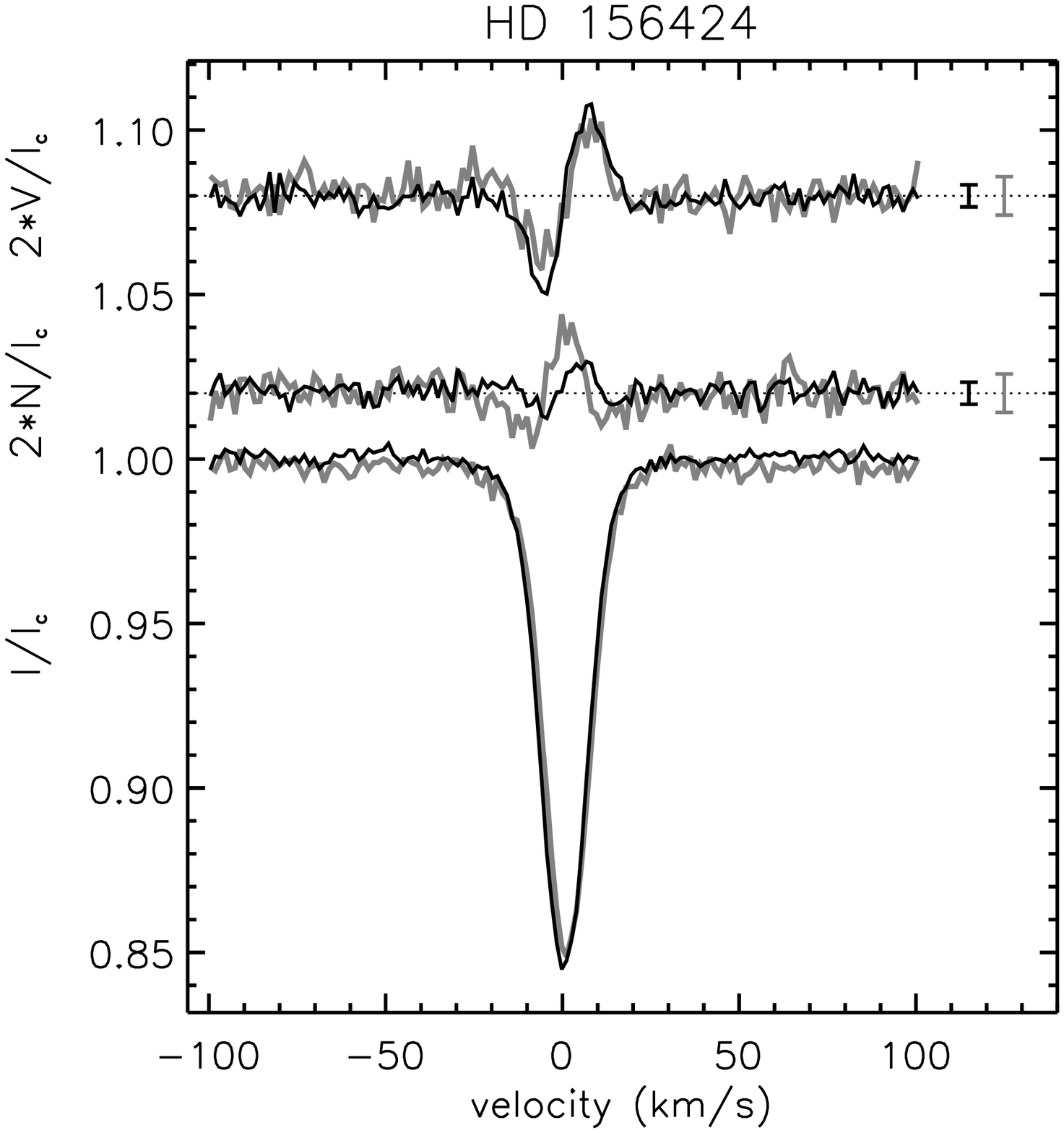}
\includegraphics[width=6cm]{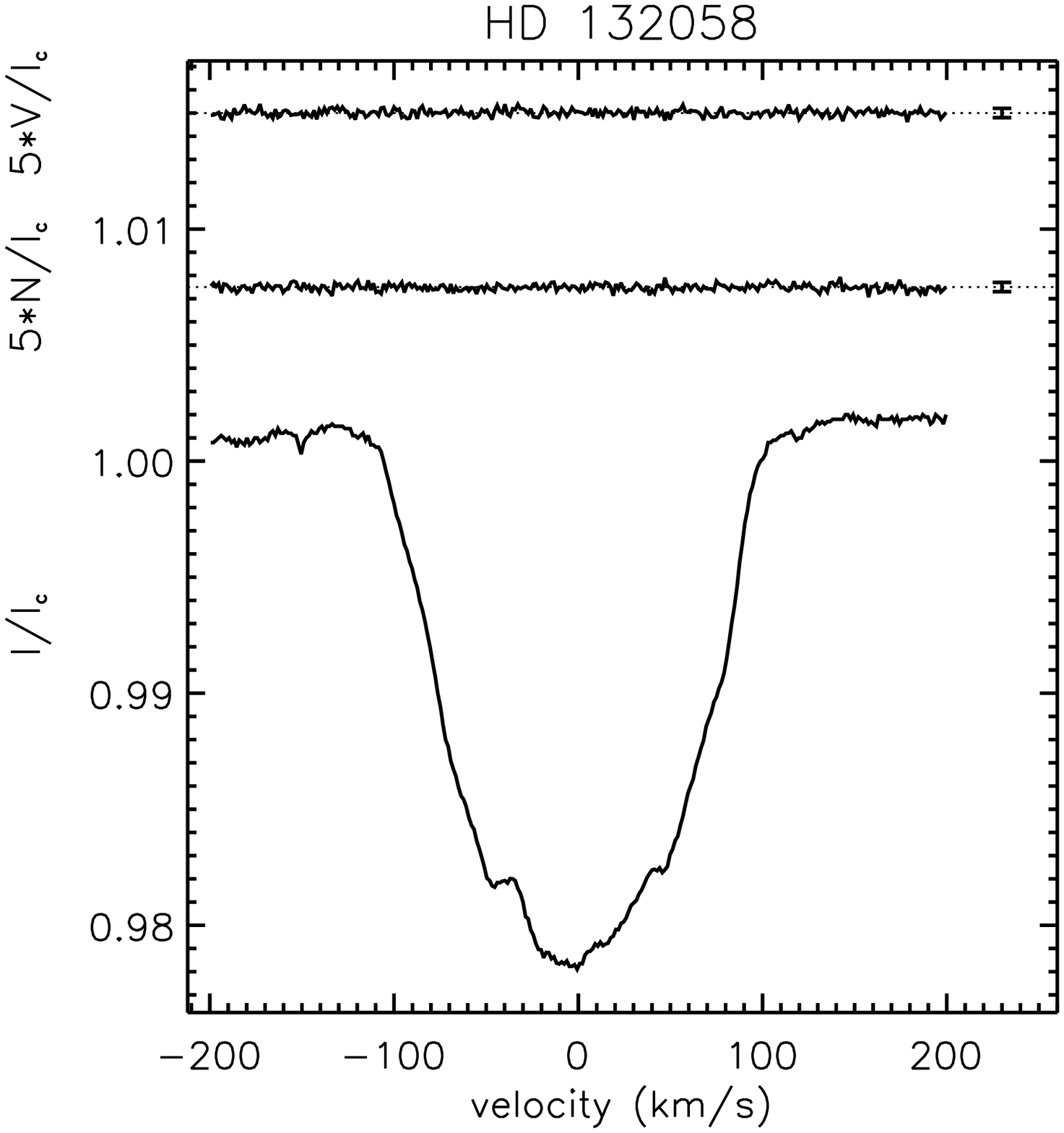}
\caption{LSD $I$ (bottom), $V$ (top) and $N$ (middle) profiles of the stars of our sample. For comparison, the profiles of a star with no detected field, {HD~132058}, are also plotted. The $V$ and $N$ profiles have been shifted and multiplied by a magnification factor for display purpose. The mean error bars in $V$ and $N$ are plotted on the side of each profile. HD~66765: the observations obtained at phases 0.06 (black), 0.43 (light grey) and 0.81 (dark grey) are plotted. HD~67621: the observations obtained at phases 0.11 (dark grey), 0.37 (light grey) and 0.73 (black) are plotted. HD~109026: the observations obtained at phases 0.42 (black), 0.71 (light grey) and 0.95 (dark grey) are plotted. For all other stars, the profiles have been plotted in order of decreasing darkness for decreasing $V$ SNR. The three observations of HD~133518 are almost perfectly superimposed.}
\label{fig:lsd}
\end{figure*}

\section{Magnetic field analysis}

\subsection{Calculation of LSD profiles}

\begin{table*}
\caption{LSD and magnetic data.}
\label{tab:lsd}      
\centering          
\begin{tabular}{l@{\;}c r r c r@{\,$\pm$\,}l c r}
\hline\hline       
ID & HJD             & \multicolumn{1}{c}{$v_{\rm bin}$}        & \multicolumn{1}{c}{S/N}     & $v_{\rm int}$               & \multicolumn{2}{c}{$B_{\ell}$ (G)} & $\Phi_{\rm rot}$ \\
     & (2450000+) & \multicolumn{1}{c}{${\rm km.s}^{-1}$} & (LSD)                                 &  ${\rm km.s}^{-1}$        & \multicolumn{2}{c}{}                      & \\
\hline
\object{HD 66765} & 5\,908.80955 & 10.0 & 7400 & [-120,150] &    529 &   83 & 0.06 \\
                 & 5\,909.65571 & 10.0 & 6200 & [-120,150] &   -910 &   85 & 0.60 \\
                 & 5\,910.80453 & 10.0 & 8900 & [-120,150] &   -591 &   61 & 0.31 \\
                 & 5\,911.60699 & 10.0 & 4300 & [-120,150] &   -122 & 140 & 0.81 \\
                 & 5\,911.85086 & 10.0 & 4400 & [-120,150] &    639 & 137 & 0.96 \\
                 & 5\,912.60934 & 10.0 & 4300 & [-120,150] & -1045 & 115 & 0.43 \\
                 & 5\,912.84324 & 10.0 & 6500 & [-120,150] & -1003 &   80 & 0.57 \\
\object{HD 67621} & 5\,907.79702 & 1.4 & 6400 & [-15,55] & 428 & 13 & 0.87 \\
                 & 5\,908.65554 & 1.4 & 3800 & [-15,55] & 439 & 21 & 0.11 \\
                 & 5\,909.59972 & 1.4 & 2100 & [-15,55] &   50 & 42 & 0.37 \\
                 & 5\,909.85897 & 1.4 & 3300 & [-15,55] &  -81 & 28 & 0.45 \\
                 & 5\,910.86215 & 1.4 & 3800 & [-15,55] & 148 & 23 & 0.73 \\
                 & 5\,911.81444 & 1.4 & 3100 & [-15,55] & 510 & 26 & 0.99 \\
                 & 5\,912.73657 & 1.4 & 3500 & [-15,55] & 201 & 24 & 0.25 \\
\object{HD 109026} & 6\,338.79036 & 1.4 & 20000 & [-54,40] & 194 & 15\tablefootmark{$\dag$} & 0.65 \\
                   & 6\,339.62764 & 1.4 & 13000 & [-54,40] & 484 & 24\tablefootmark{$\dag$} & 0.95 \\
                   & 6\,340.69253 & 1.4 & 13000 & [-54,40] & 231 & 22\tablefootmark{$\dag$} & 0.32 \\
                   & 6\,341.75842 & 1.4 & 12000 & [-54,40] & 277 & 24\tablefootmark{$\dag$} & 0.70 \\
                   & 6\,342.71703 & 1.4 & 12000 & [-54,40] & 465 & 26\tablefootmark{$\dag$} & 0.03 \\
                   & 6\,343.82113 & 1.4 & 13000 & [-54,40] & 179 & 22\tablefootmark{$\dag$} & 0.42 \\
                   & 6\,344.63486 & 1.4 & 10000 & [-54,40] & 245 & 29\tablefootmark{$\dag$} & 0.71 \\
                   & 6\,344.91629 & 1.4 & 10000 & [-54,40] & 358 & 29\tablefootmark{$\dag$} & 0.81 \\
\object{HD 121743} & 6\,129.55193 & 3.0 & 5700 & [-80,90] & 288 & 29 & (...) \\
                   & 6\,130.47991 & 3.0 & 6200 & [-80,90] & 302 & 26 & (...) \\
\object{HD 133518} & 6\,338.88872 & 1.4 & 6300 & [-18,7.2] & 490 &   7 & (...) \\
                   & 6\,339.91444 & 1.4 & 3200 & [-18,7.2] & 468 & 12 & (...) \\
                   & 6\,344.80544 & 1.4 & 1900 & [-18,7.2] & 470 & 20 & (...) \\
\object{HD 147932} & 6\,343.87595 & 8.0 & 14000 & [-172,196] & -860 & 180 & (...) \\
                   & 6\,344.85718 & 8.0 & 10000 & [-172,196] & -990 & 250 & (...) \\
\object{HD 156324} & 6\,127.62863 & 1.4 & 2100 & (...) & \multicolumn{2}{c}{(...)} & (...) \\
                   & 6\,128.78131 & 1.4 & 1400 & (...) & \multicolumn{2}{c}{(...)} & (...) \\
                   & 6\,462.92963 & 1.4 & 6000 & (...) & \multicolumn{2}{c}{(...)} & (...) \\
\object{HD 156424} & 6\,126.68582 & 1.4 & 600 & [-25.4,25] & -768 & 96 & (...) \\
                   & 6\,127.77721 & 1.4 & 340 & [-26.8,23.6] & -193 & 167 & (...) \\
\hline                  
\end{tabular}
\tablefoot{(...) Values not derivable from our data.
\tablefoottext{$\dag$}{Unnormalised values.}
}
\end{table*}

In order to increase the SNR of our line profiles, we applied the least-squares deconvolution (LSD) procedure to all spectra \citep{donati97}. This procedure combines the information contained in many metal lines of the spectrum, to extract the mean intensity (Stokes $I$) and polarised (Stokes $V$) line profiles. In Stokes $I$, each line is weighted according to its central depth, while in Stokes $V$ the profiles are weighted according to the product of the central depth, wavelength, and Land\'e factor. These parameters are contained in a 'line mask' derived from a synthetic spectrum corresponding to the effective temperature and gravity of the star. The construction of the line mask for each star involved several steps. First, we used Kurucz ATLAS~9 models \citep{kurucz93} to obtain \emph{generic} masks of solar abundances, and of T$_{\rm eff}$/$\log g$ following the Kurucz models grid. Our masks contain only lines with intrinsic depths larger or equal to 10\% of the continuum level. We do not need to use a lower cutoff for detecting magnetic fields, which was the primary purpose of the MiMeS survey. For {mapping the magnetic fields of the stars} we recommend using a cutoff of about 1\%, allowing for the inclusion of the faintest predicted lines. In a second step, we excluded hydrogen Balmer lines and strong resonance lines from the generic masks, providing us with a set of \emph{corrected generic} masks of various T$_{\rm eff}$/$\log g$.Then we chose the corrected generic masks of T$_{\rm eff}$/$\log g$ adapted to our stars, and built masks specific to each star by modifying the intrinsic line depths to take the relative depth of the lines of the observed spectra into account. For the SB2 HD~109026, we did not perform the last step, as both components make important contributions to the line profiles. For the two spectroscopic multiple systems, HD~109026 and HD~156324, we chose corrected generic masks of temperature and gravity adapted to the primary component. At this stage, the masks contain only predicted lines satisfying the initial assumption of the LSD procedure, i.e. a similar shape for all spectral lines considered in the procedure. Columns 3 and 4 of Table \ref{tab:lsd} summarise the velocity step used in the LSD computation and the resulting SNR in the LSD $V$ profiles. The LSD profiles are plotted in Fig. \ref{fig:lsd}.

Clear Zeeman signatures are detected in the LSD $V$ profiles of all our observations. To assess the significance of a signature compared to local noise, we compute the false alarm probability (FAP) of detection as described by \citet{donati97}. The FAP allows us to estimate the probability that a signal in the $V$ profile could occur by chance. We consider that if the FAP is lower than 0.00001 it is a definite detection (DD). If The FAP  is situated between 0.001 and 0.00001, it is a marginal detection (MD) and the star must be re-observed to confirm the detection. If the FAP is higher than 0.001, no detection is achieved. We consider that if we obtain at least one DD for an object, that object is magnetic. If no MD or DD has been obtained for the object, we consider that the star is not magnetic (or hosts a magnetic field too low to be detected with our data). All FAPs of the data presented here have a value lower than 0.00001. For comparison, we have also plotted the LSD profiles of the star {HD~132058 \citep[B2 III-IV][]{zorec09}}, for which, we detected no magnetic field (Fig. \ref{fig:lsd} lower right panel)\footnote{The reduction and analysis of these data have been performed in the same way as the data presented in this paper, and will be published in a future survey paper (Wade et al. in prep.)}. The FAP of this observation is {0.9998}, and we observe indeed a flat $V$ profile, contrasting with the profiles of the detected stars.

\subsection{Origin of the LSD $V$ signatures}

The Zeeman signatures are all found within the bounds of the photospheric $I$ profile (Fig. \ref{fig:lsd}). Except for two stars, the null $N$ profiles are flat, indicating that no significant spurious signals are affecting our data, and this allows us to be confident that stellar surface magnetic fields are at the origin of the $V$ signatures. For HD~156324 and HD~156424, features are observed in $N$, coinciding with the photospheric $I$ and $V$ profiles. These features are often detected in short-period binaries or pulsating stars. {\citet{neiner12} report the analysis of similar spurious signatures observed in the null profiles of the magnetic star HD~96446. They showed that the radial velocities of the spectral lines have changed between the four sub-exposures of one polarised observation, and that  such rapid changes (relative to the exposure time) in the radial velocities of the spectral lines, with an amplitude of about 7~\kms, create spurious signatures in the null profiles. The results of this work are consistent with a similar modelling performed by \citet{schnerr06}. In both studies we observe that the shape of the observed or modelled spurious null signatures are very similar to those of HD~156324 and HD~156424, while the amplitudes are of the same order or lower. This indicates that the signatures observed in the null profiles of HD 156324 and HD 156424 are most likely due to radial velocity changes during one polarised observation, and that the radial velocity variations are of the order of few \kms. The simulations performed in \citet{schnerr06} and \citet{neiner12} have shown that while the radial velocity changes can affect the shape of the $V$ profiles slightly, they do not significantly affect the longitudinal field measurements, and they do not put into question the magnetic field detection at the surface of the stars. We are therefore very confident in our detections of the magnetic fields at the surface of HD~156324 and HD~156424.}

For HD~109026, the broadening of the Zeeman signatures corresponds to the broadening of the narrow component of the $I$ profile. The magnetic field is therefore only detected in the secondary component. For HD~156324, the centroid of the signatures is changing from one observation to the other in the same direction as the primary. The broadening of the $V$ signatures is similar to the broadening of the intensity profile of the primary, indicating that a magnetic field is detected in the photosphere of the primary. As the lines of the tertiary component are always blended with the lines of the primary, however we are unable to exclude the possibility of contamination of the Zeeman signature by the tertiary component. Finally, no Zeeman signature is detected in the secondary component. For HD~156324, the LSD $I$ profiles, in addition to the deep primary component, reveal the secondary and the tertiary components. The secondary component is easily detected in the second and third observations at radial velocities of $\sim$140~\kms\ and $\sim-80$~\kms, respectively. It is blended with the primary component in the first observation. The tertiary component is also detected at radial velocities of $\sim$10~\kms\ in the first and second observations, and at $\sim$0~\kms\ in the third observation.

As already mentioned in Sec. 3, the individual lines of the spectrum of HD~121743 are distorted and show clear variations from one observation to the other (Sec. 3). Such variable distortions are also visible in the LSD $I$ profiles (Fig. \ref{fig:lsd}), indicating that they affect a large number of spectral lines. Abundance patches at the surface of magnetic early Bp stars, because of the rotation of the stars, lead usually to a correlation between the amplitude of the line distortion and the longitudinal field \citep[e.g.][]{landstreet78}, while pulsations, which induce surface deformations on time scales unrelated to the stellar rotation period, do not. We therefore suggest that the observed line profile variations of HD~121743 are not likely to be due to abundances patches, as the longitudinal field and the Zeeman signatures have varied insignificantly between the two observations. We propose instead that the line distortions are mainly caused by $\beta$-Cep pulsations as proposed by Telting et al. (2006). Pulsating stars have distorted surfaces, which affect the intensity spectra, but can also affect the shapes of the Stokes V spectra if the modes and periods of the pulsations affect the shape of the photospheric spectral lines significantly \citep{neiner13}. We indeed observe small features within the Zeeman signatures of HD~121743 (Fig. \ref{fig:lsd}), at the velocity bins where the LSD $I$ profile is at its most distorted (between -20 \kms and 30 \kms). While pulsations seem to perturb the Stokes $V$ spectra, it does not put into question a surface magnetic field as the dominant shaper of the Zeeman signatures observed in Fig. \ref{fig:lsd}. They should be taken into account in future spectropolarimetric observations and analyses of this star.

For HD~147932, we also observe distortions in $I$ affecting mainly the centre of the profiles. As suggested for HD~121743, these distortions might be produced either by pulsations, by abundance patches, {or by another unknown source}. They could have affected the $V$ profiles, and they could explain the variability observed in some pixels of the $V$ profile. Additional observations are required to reach a conclusion on this point.

\subsection{Longitudinal field measurements}

We measured the longitudinal magnetic fields (\bl, i.e the line-of-sight component of the magnetic field integrated over the stellar surface) of each observation of the stars with single-lined spectra, by computing the first order moment of $V$ and the equivalent width of $I$ using the method described by \citet{wade00c} and the integration limits given in Table \ref{tab:lsd}. 

For HD~109026 and HD~156324, the LSD $I$ and $V$ profiles need to be corrected from the secondary component. Such a correction requires a knowledge of the effective temperatures of the components as well as the luminosity ratio (e.g. \citealt{alecian08a}). Because both are either highly uncertain or unknown, we are not able to compute realistic values of longitudinal fields for the magnetic stars of both systems. 

Columns 5 and 6 of Table \ref{tab:lsd} summarise the integration ranges used in the \bl\ computations and the resulting \bl\ values with error bars for all of the other stars. The measured values range from $\sim1$~kG to $\sim600$~G, with errors being as low as 7~G and reaching up to 250~G. The errors depend strongly on the quality of the observations, but also on the temperature and \vsini\ of the stars and on the cleaning procedure applied to the mask when the LSD profiles were computed. We emphasise here that because of the Doppler resolution of high-resolution spectra, we are able to obtain a definite magnetic detection even if the measurement of \bl\ is consistent with 0. Indeed, \bl\ is an integrated value over the whole stellar surface, and therefore over the entire width of the line profile. As a result, \bl\ can give a null value, even if a non-zero polarisation signal is present in individual velocity {(or wavelength) bins of the mean LSD (or individual spectral line)} profiles. The combination of high-resolution spectropolarimetry and FAP analysis is therefore more efficient for detecting magnetic fields at the surface of stars than measuring the longitudinal field alone.

For HD~109026, we are not able to correct the LSD profiles from the non-magnetic component's continuum and measure the \emph{normalised} longitudinal values. The \emph{unnormalised} longitudinal field values can be measured by integrating the $I$ and $V$ profiles of the system over the line width of the magnetic star only, as if it was a single star. Such values does not reflect the real field strength at the surface of the star, but the relative \bl\ values can allow us to derive the rotation period of the star (Sec. 4.4.1).

\subsection{Magnetic properties}

\begin{table*}
\caption{\bl\ curves fitting parameters}             
\label{tab:blfit}      
\centering          
\begin{tabular}{lr@{\,$\pm$\,}lr@{\,$\pm$\,}lr@{\,$\pm$\,}lr@{\,$\pm$\,}ll}
\hline\hline
ID & \multicolumn{2}{c}{$P_{\rm rot}$} & \multicolumn{2}{c}{$T_{0}-2\,450\,000$} & \multicolumn{2}{c}{$A_{\rm B}$} & \multicolumn{2}{c}{$B_{0}$} & $\chi^2_{\rm min}$ \\
    & \multicolumn{2}{c}{(d)}                  & \multicolumn{2}{c}{(d)}                             & \multicolumn{2}{c}{(G)}               & \multicolumn{2}{c}{(G)}         & \\
\hline
HD~66765 & 1.62 & 0.15 & 5\,908.68 & 0.09 & 860 & 200 & -250 & 140 & 0.54\\[3pt]
HD~67621 & \multicolumn{2}{c}{$3.60^{+0.26}_{-0.20}$} & 5\,908.25 & 0.09 & 300 & 50 & 210 & 34 & 0.68 \\[6pt]
HD~109026 &  \multicolumn{2}{c}{$2.84^{+0.18}_{-0.22}$} & 6\,339.79 & 0.13 & 170 & 54\tablefootmark{$\dag$} & 308 & 35\tablefootmark{$\dag$} & 0.77 \\[3pt]
\hline                  
\end{tabular}
\tablefoot{\tablefoottext{$\dag$}{Unnormalised values}}
\end{table*}


\begin{figure*}
\centering
\includegraphics[width=3cm,angle=90]{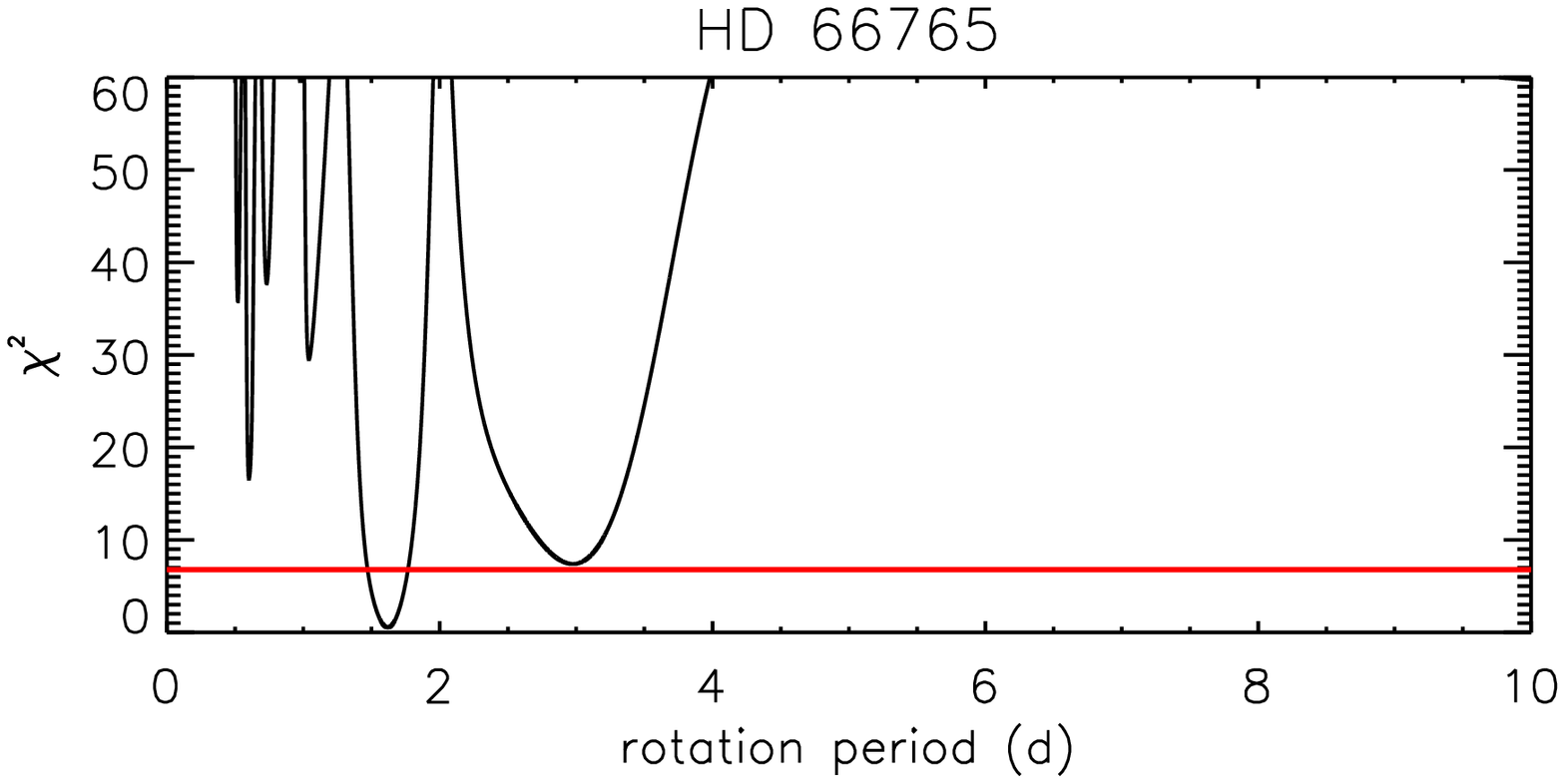}
\includegraphics[width=3cm,angle=90]{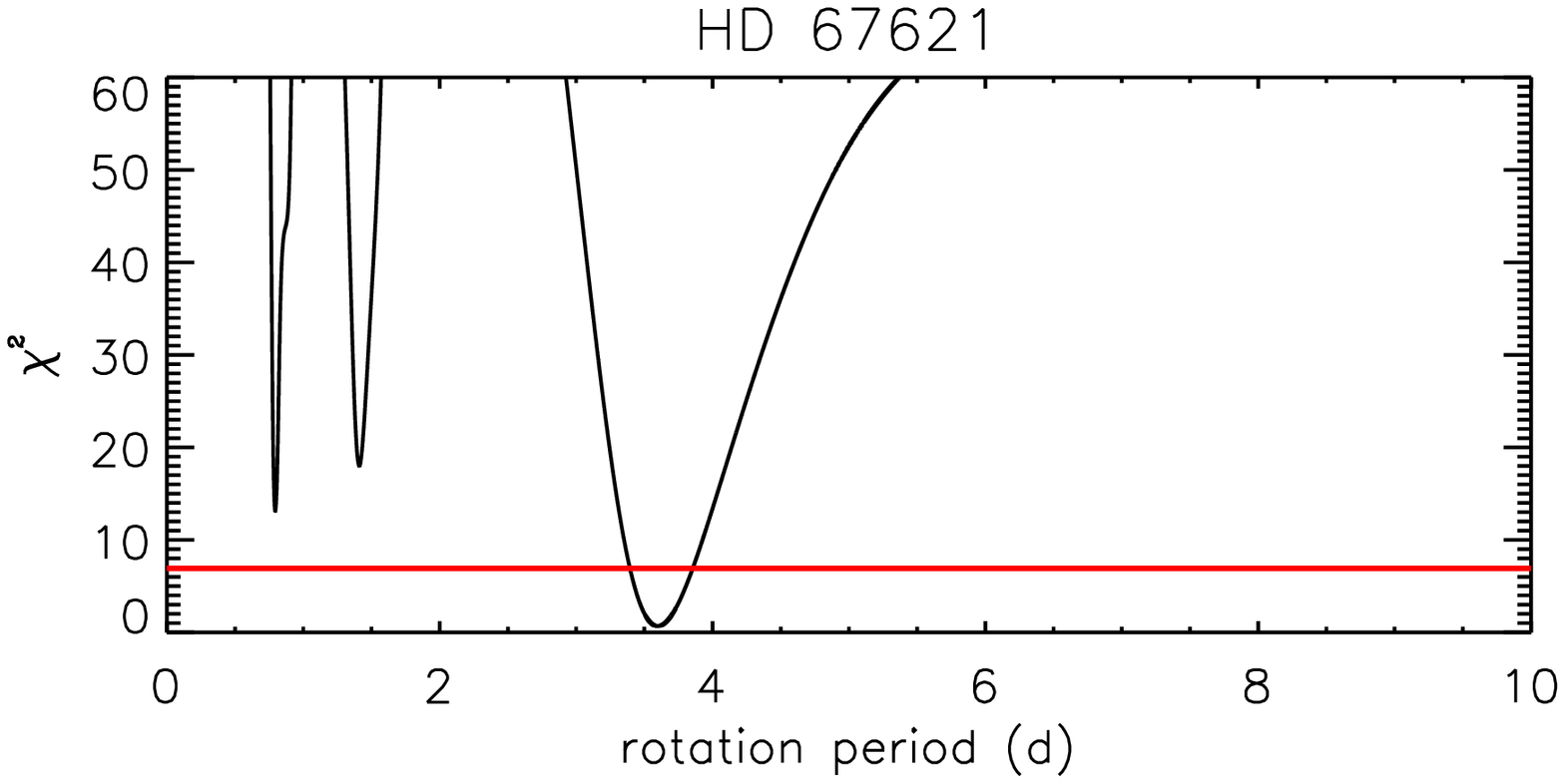}
\includegraphics[width=3cm,angle=90]{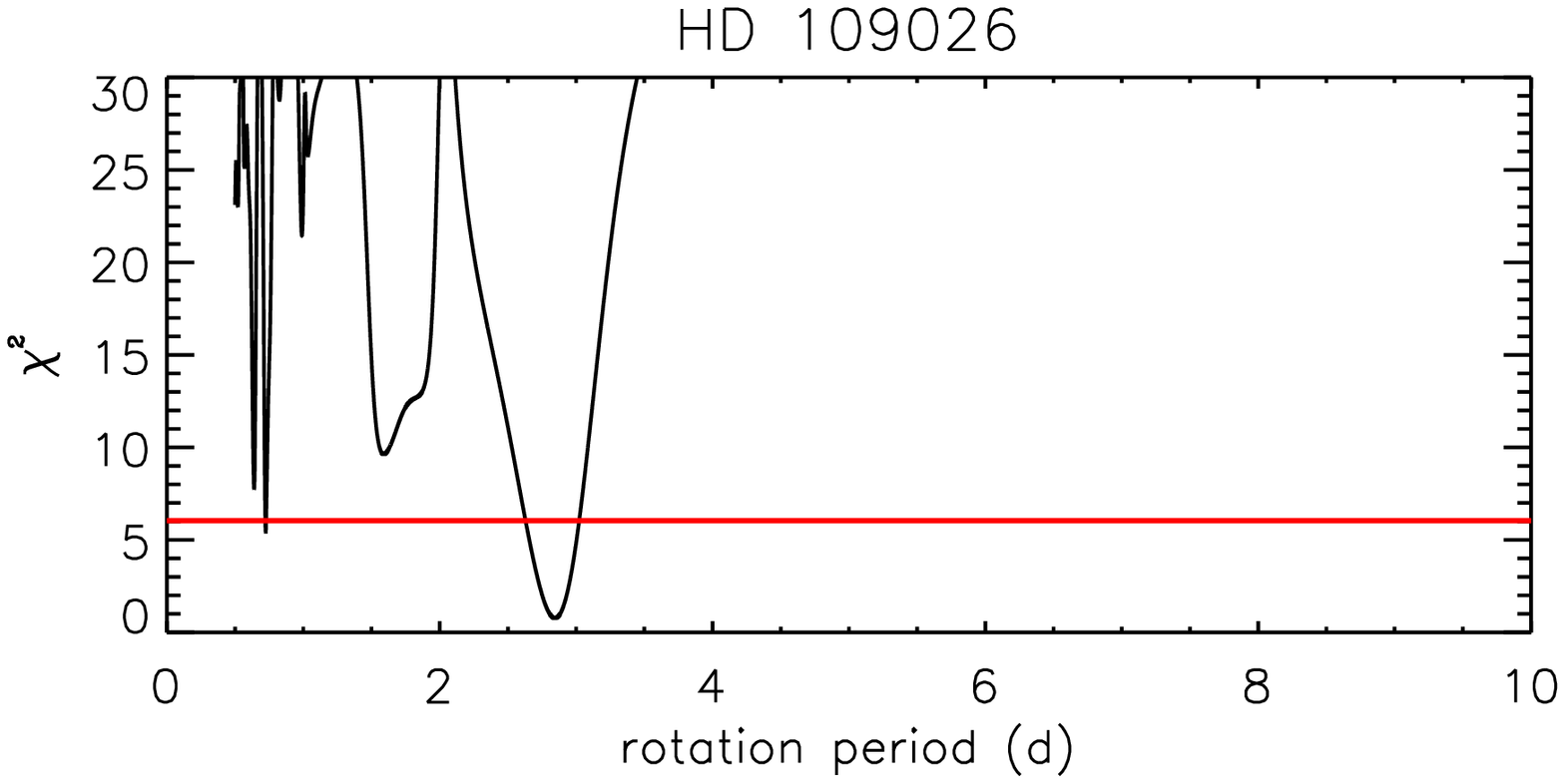}
\caption{$\chi^2$ of the \bl\ curve fits as a function of the rotation period for HD~66765, HD~67621, and HD~109026. The horizontal red line indicates the 99.97\% significance level.}
\label{fig:periodo}
\end{figure*}


\begin{figure*}
\centering
\includegraphics[width=3.5cm,angle=90]{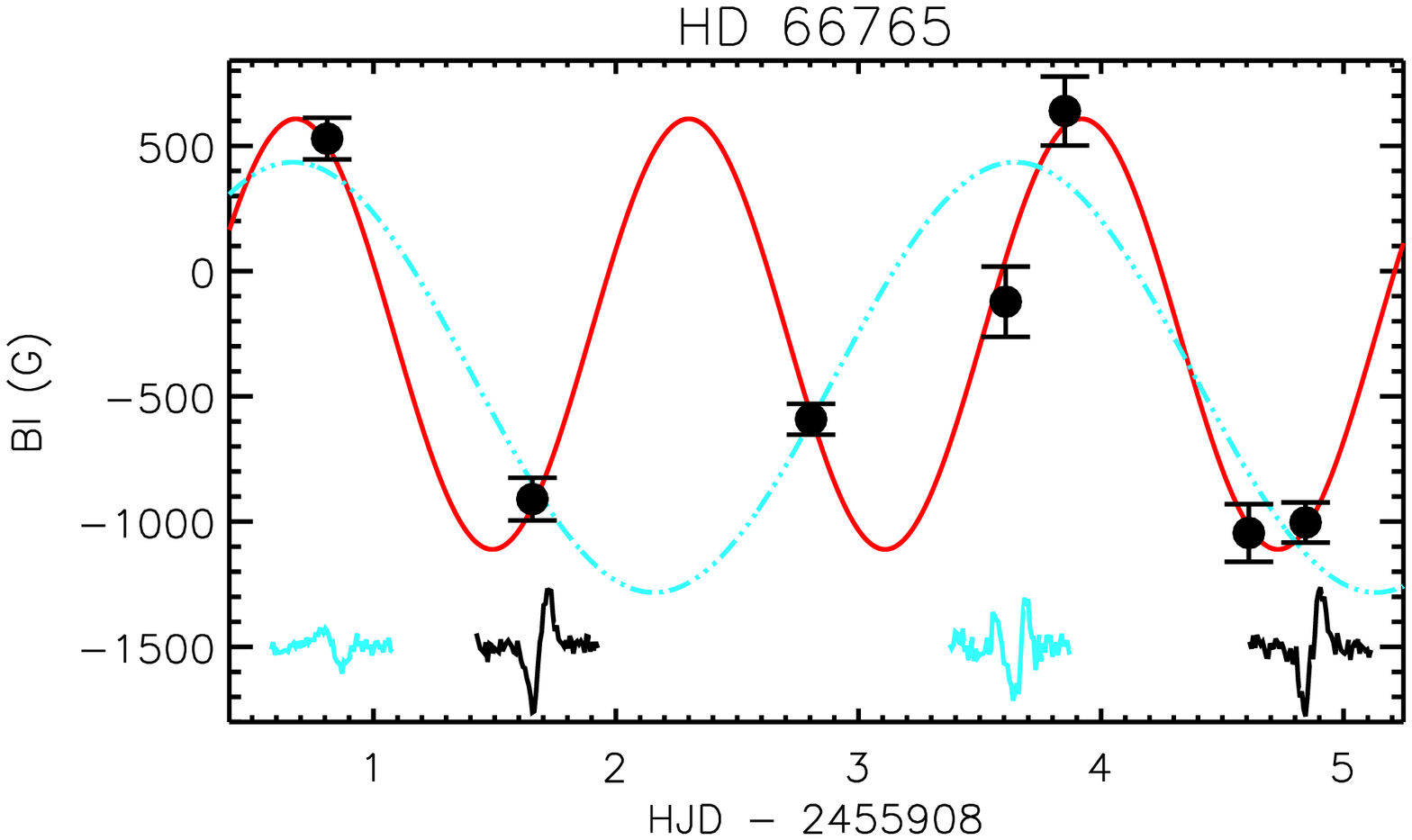}
\includegraphics[width=3.5cm,angle=90]{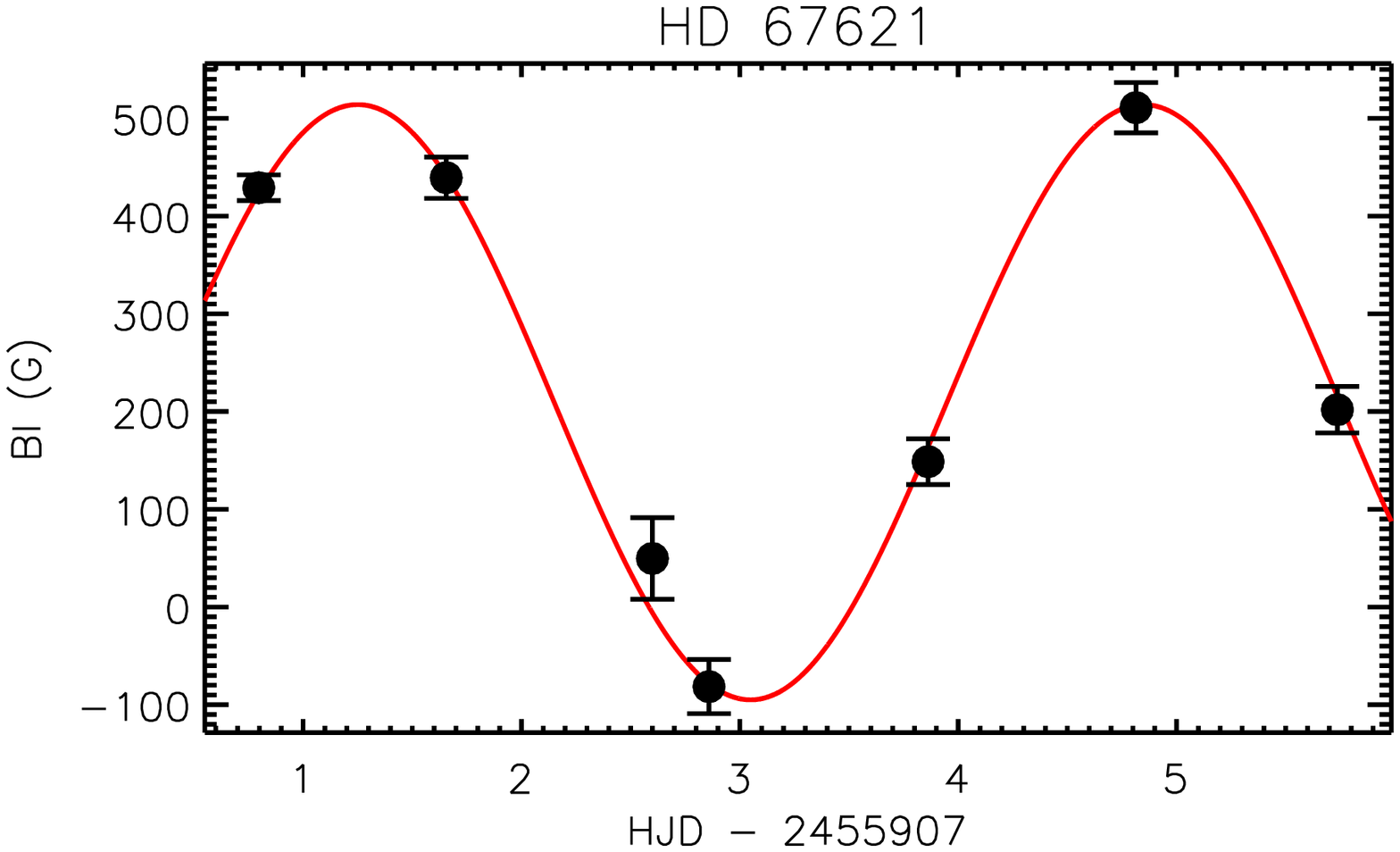}
\includegraphics[width=3.5cm,angle=90]{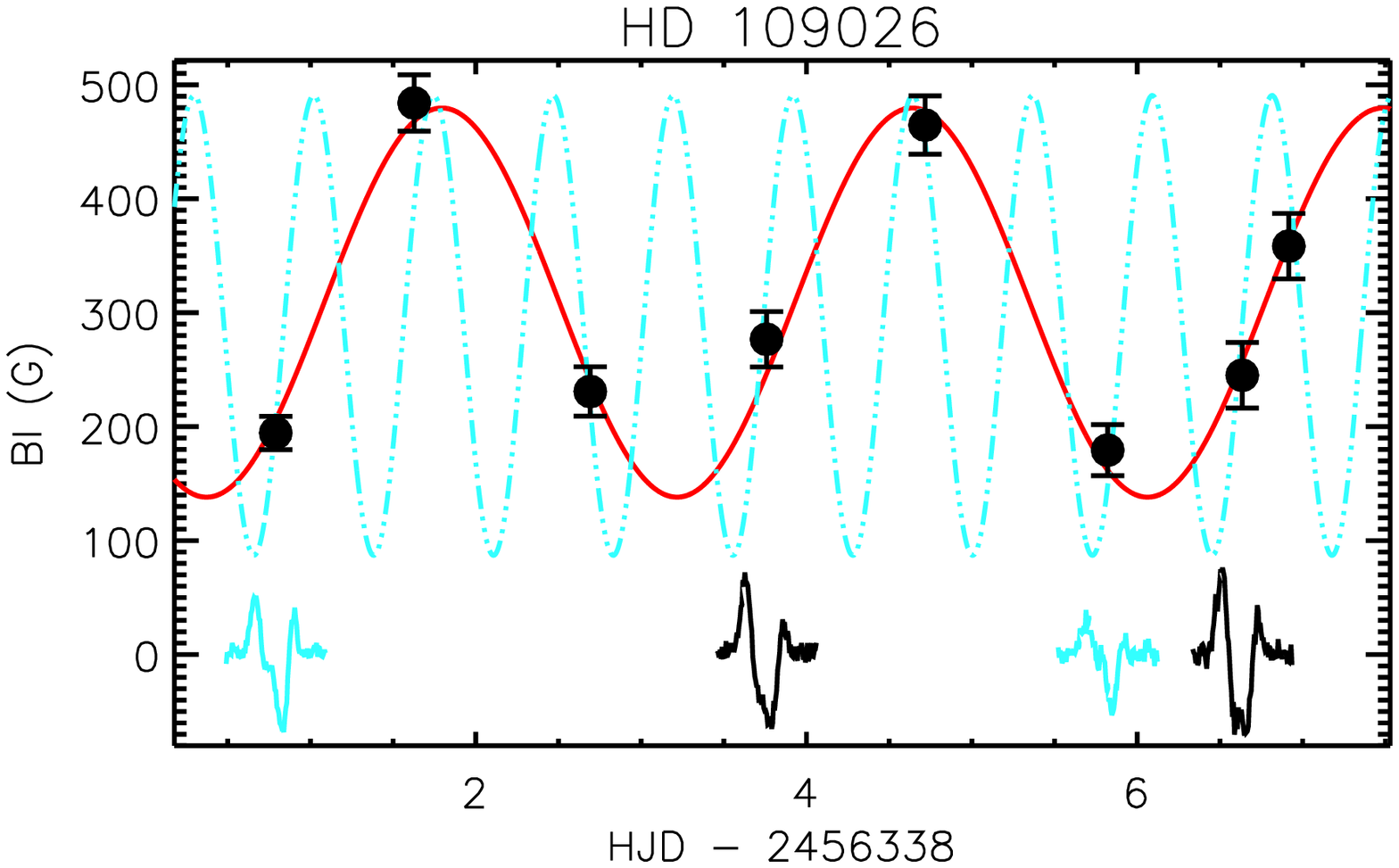}
\caption{Normalised or unnormalised longitudinal field values of HD~66765, HD~67621, and the magnetic component of HD~109026 as a function of julian date. The best-fit solution is plotted with a full red line. For HD 66765 and HD 109029, the solution close to the 3-$\sigma$ level is plotted with a dashed-doted light-blue line. LSD $V$ profiles are plotted for few observations centred on their julian date. The $V$ profiles have been shifted and/or multiply by a magnification factor for display purpose. The black profiles are associated with a similar phase according to our red fit, while the light-blue profiles are associated with a similar phase according to the blue fits.}
\label{fig:blfit}
\end{figure*}


\begin{figure*}
\centering
\includegraphics[width=6cm]{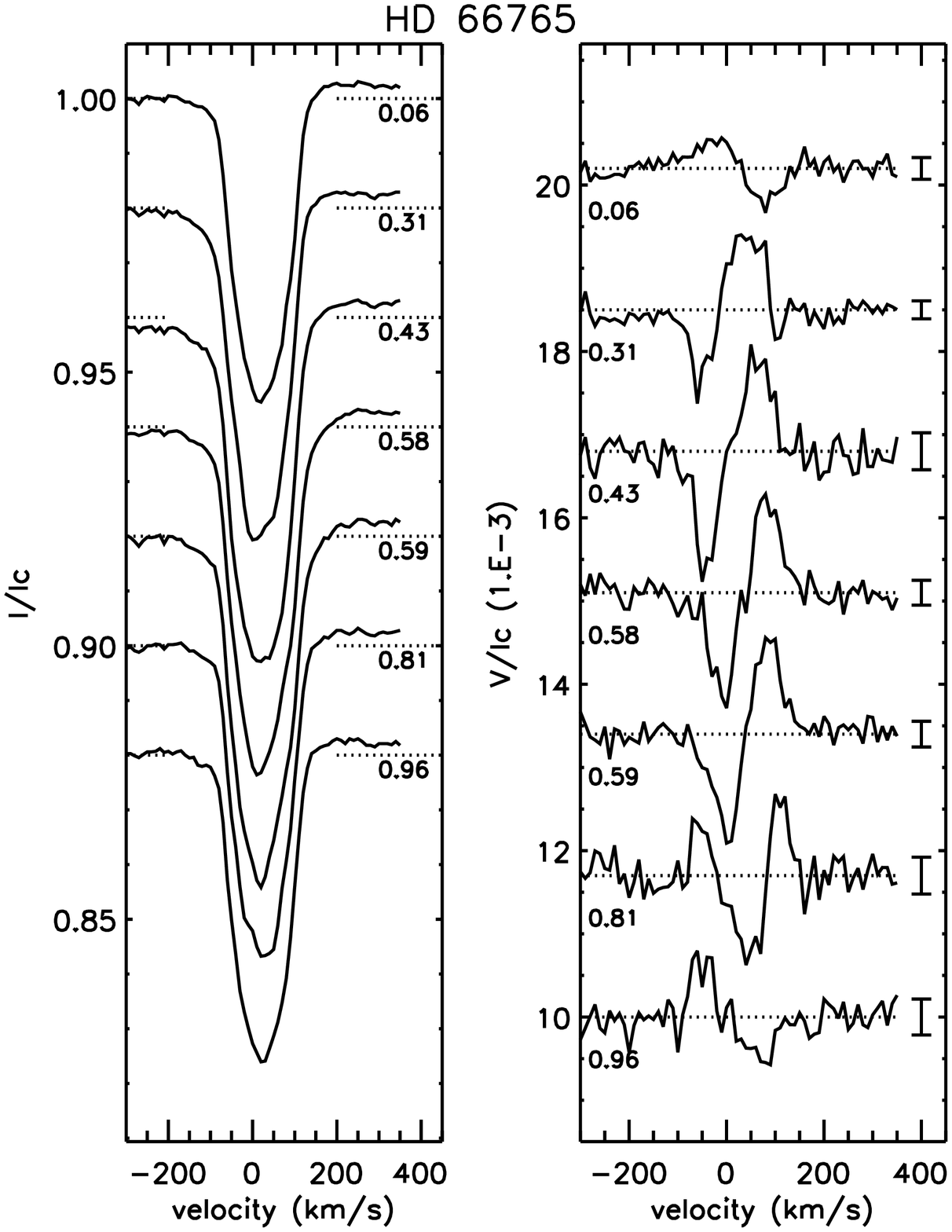}
\includegraphics[width=6cm]{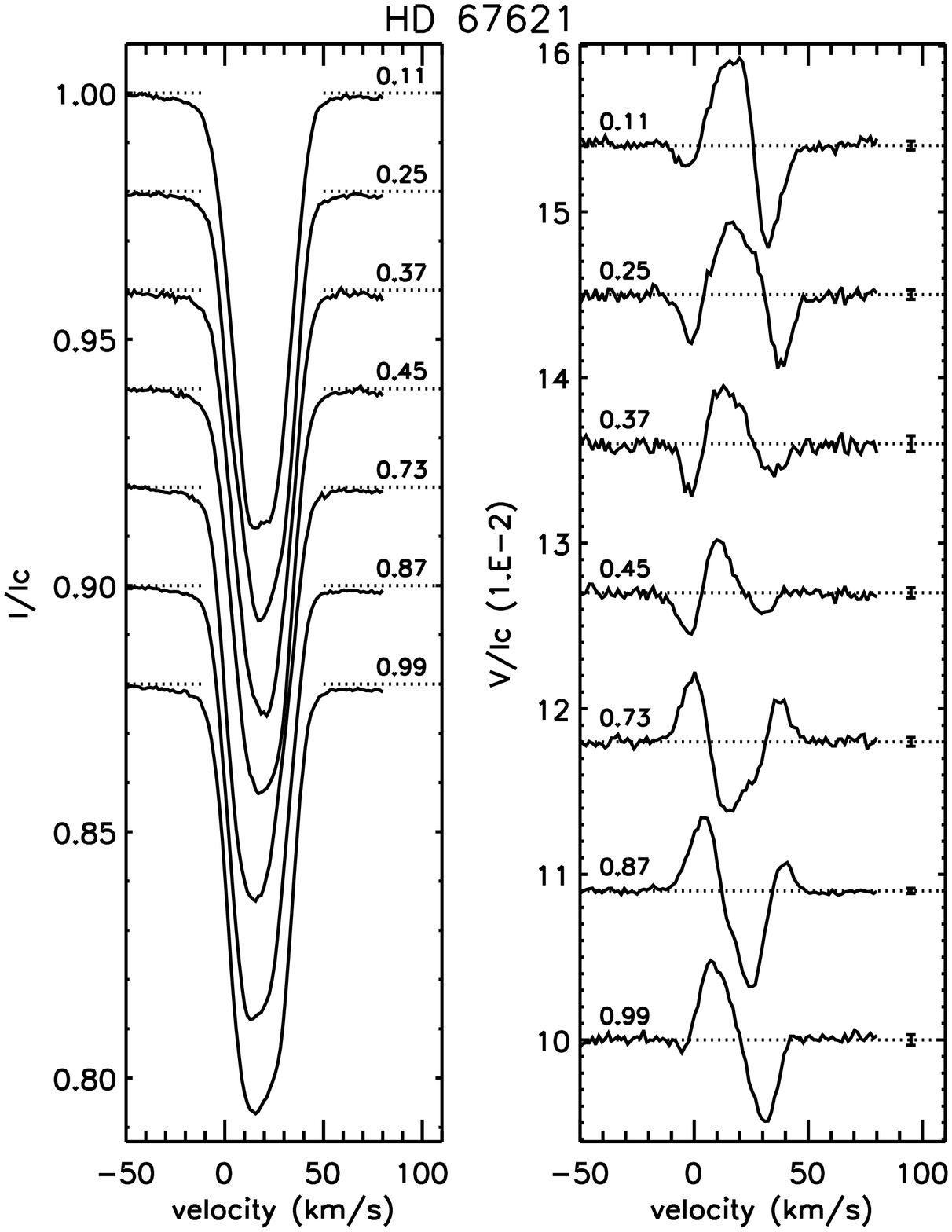}
\includegraphics[width=6cm]{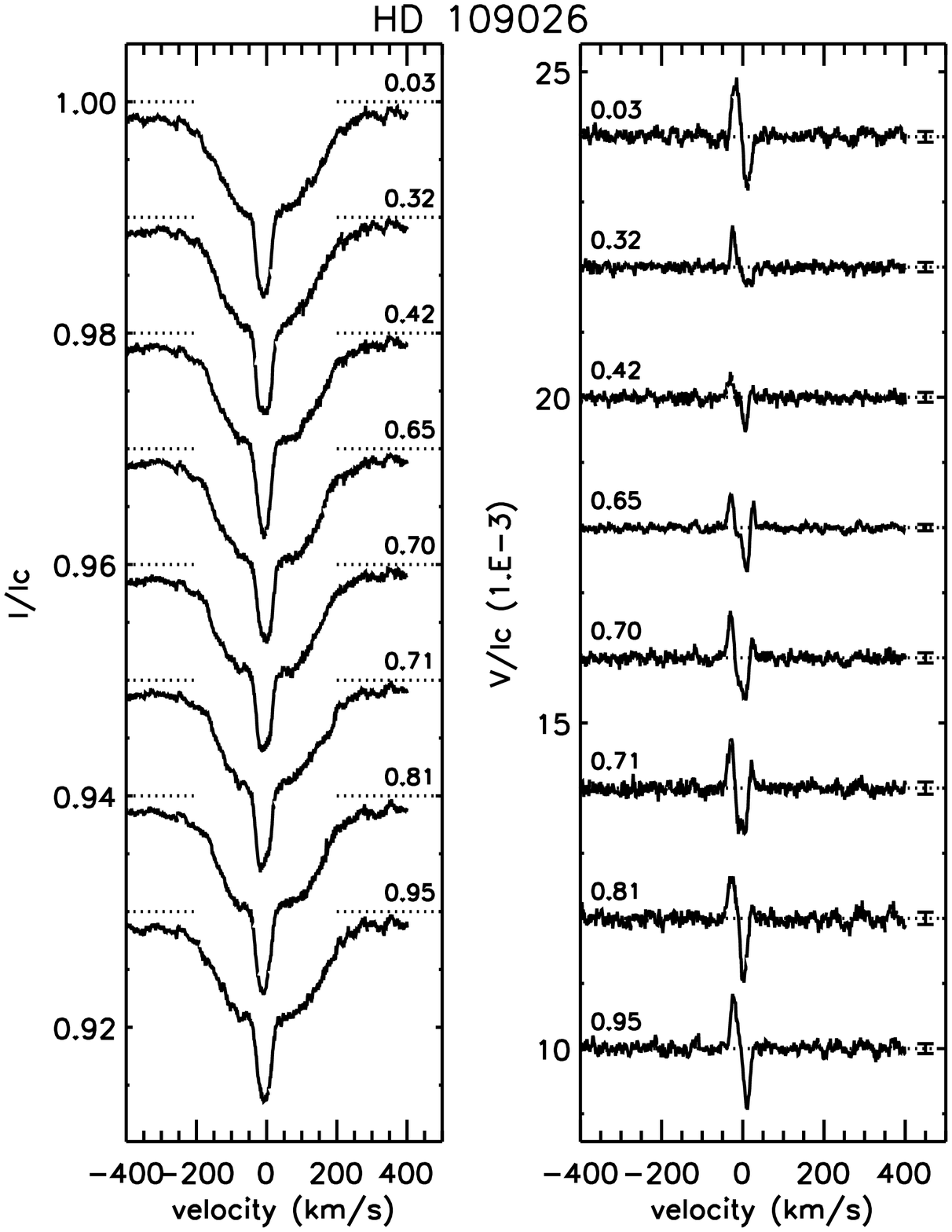}
\caption{LSD $I$ (left) and $V$ (right) profiles of HD~66765, HD~67621, and HD~109026 ordered by increasing phase (from top to bottom). The phase and the mean $V$ error bars are indicated on the side of each profile.}
\label{fig:lsdphase}
\end{figure*}

\subsubsection{Analysis of the Stokes $V$ variable stars}

For four stars (HD~66765, HD~67621, HD~109026, HD~156324) the Zeeman signatures vary on timescales of few hours to one day. These variations are interpreted in terms of the oblique rotator model (ORM, \citealt{stibbs50}), in which the magnetic axis is inclined with respect to the rotation axis. As the star rotates the magnetic configuration of the observed stellar surface changes with time, producing variable magnetic signatures.

For three stars (HD~66765, HD~67621, HD~109026) we attempted to derive the rotation period of the stars from the \bl\ variations. If their magnetic fields are dominantly dipolar (which is common in such {early-B} stars), the longitudinal magnetic field curve, as a function of time (or rotation phase), should be very close to a sinusoidal function. While unlikely, their magnetic fields could potentially be dominantly quadrupolar. In this case, the \bl\ curve would be represented with 2nd-order Fourier expansion. Because of the number of data points and the observing window, we are not able to constrain such a model, and we will only fit our \bl\ curves with a single-sinusoidal function.

To estimate the rotation period of the star we used the following method. We performed successive $\chi^2$ fits on the longitudinal field values using a sinusoidal function by fixing the period to a value between 0.5 and 10 d with a 3.e-4 d step. The derivatives of our \bl\ curves show at least two sign reversals over a maximum time-lapse of five days, and the minimum time lapse between two observations is few hours. The chosen period range is covering all possible periods we can derive with our data. The three other parameters of the sinusoid, namely the amplitude ($A_{\rm B}$), the vertical ($B_0$), and horizontal ($T_0$) shifts of the curve, are free parameters in each fitting iteration. The fitted parameters are summarised in Table \ref{tab:blfit}, while the curves of the $\chi^2$ as a function of phase and the best sinusoidal curves are shown in Fig.~\ref{fig:periodo} and \ref{fig:blfit}, respectively. In the case of HD~66765 and HD~67621, the result of the fitting procedure provides us with only one significant minimum at a level of 99.97\%, which corresponds to respective periods of $1.62\pm0.15$~d and $3.60^{+0.26}_{-0.20}$~d. For HD~109026, the $\chi^2$ curve provides us with a minimum at 2.84$^{+0.18}_{-0.22}$~d. A second minimum falls below the 99.97\% level, but this value gives a less good agreement between the model and our data (Fig. \ref{fig:blfit}). 

To further establish the robustness of these results, we also computed periodograms using the Lomb-Scargle technique, as implemented by \citet{townsend10}, and the CLEAN algorithm \citep[][]{hogbom74,roberts87,gutierrez09}. In each case, the resulting periodograms achieved maximum power at the periods identified with our $\chi^2$ analysis. The periodogram obtained with the Lomb-Scargle technique provided nearly identical results to our $\chi^2$ analysis, while the results of our CLEANed analysis were greatly improved; the maximum power achieved in all other peaks in the CLEANed periodogram were found to be $<10$\% of the peak power. Furthermore, the uncertainties estimated from our $\chi^2$ analysis were considerably more conservative than similar estimates using the FWHM of the Lomb-Scargle periodogram, which were much closer to the 68.3\% confidence limits that would be estimated from our $\chi^2$ analysis.

Even if the number of measurements is small (7 or 8 points), and the observing window is short (four to six days), we are very confident in the periods we derived for the following reasons. First, we have very good agreement between our data and our best solutions (Fig. \ref{fig:blfit}) for the three stars. Secondly, we obtained continuous (nightly sampling) measurements of \bl, and we even obtained some of the data during the same night, improving the sampling. In two cases (HD~66765 and HD~67621), we observe that \bl\ is changing sign, and in the third case (HD~109026), we see two apparent cycles. We therefore clearly follow the \bl\ evolution almost in real time, and it is really unlikely that periods larger than those we derive can fit our data. In theory, shorter periods could also potentially reproduce our \bl\ variations. For HD 67621 and HD 109026, two peaks in the $\chi^2$ curves are below or close to the 3-$\sigma$ level, corresponding to periods of 2.98~d and 0.72~d, respectively). These periods can be discarded however, as these solutions reproduce far less well our \bl\ measurements, as illustrated in Fig. \ref{fig:blfit}. Third, we observe in Fig. \ref{fig:lsdphase} that profiles obtained at similar rotation phases, but during different cycles, are very similar. To emphasise this point we have plotted in the same panel as our \bl\ curves, the Stokes $V$ profiles obtained at two almost identical phases as predicted by our best-fit solution for HD 66765 and HD 109026 (in black). We have also plotted in grey, two other profiles obtained at very similar phases as predicted by the solutions with respective periods of 2.98~d and 0.72~d. We observe that while the black profiles are very similar, the grey profiles do not look the same. This constitutes a nearly independent check of our solutions as for given magnetic configuration and rotation phase, only one Zeeman Stokes $V$ signature can be produced. Finally, the derived period for HD 109026 is consistent with the photometric period found by \citet[][2.73~d]{waelkens98}, suggesting that the photometric period is related to the rotation of the star, and that the magnetic star might be spotted enough to produce detectable photometric variations with Hipparcos.

Our dataset is however too small to say anything about the potential contribution from a higher order field. In the extreme case of a dominant quadrupolar the real period of the star would be twice our derived period. We note that for HD 109026, the agreement between the two independent \bl\ and photometric periods makes this possibility very unlikely.

\subsection{Possible origins for the lack of Stokes $V$ variability}

For four stars (HD~121743, HD~133518, HD~147932, HD~156424) the Zeeman signatures do not show significant variations. For HD~121743 and HD~147932, while variability is found in some pixels of the profile centre, the amplitudes and the global shape of the $V$ signatures are similar in both observations of each star. We therefore propose that the profile changes have not been produced mainly by the rotation of the star. Such a lack of variability can be interpreted with four different effects:
\begin{enumerate}[(i)]
\item The rotation period could be much (at least 10 times) longer than the time-lapse between the first and last observations, which would allow the rotation phases to not change significantly from one night to another. 
\item The star is seen close to pole-on. 
\item The magnetic and rotation axes are almost aligned, and the magnetic field is \emph{axisymmetric} (i.e. is symmetric about the magnetic axis). 
\item The rotation period is close to the time lapse between the two observations (for HD~121743 and HD~156424).
\end{enumerate}

For HD~121743, option (i) can be ruled out for the following reasons. The \vsini\ of the star is about 80~\kms, and the radius of the star is estimated to be 4.7~\rsun\ \citep{petit13}. The maximum rotation period allowed is therefore 71~h. Even if the error on the radius is large, it appears unlikely that the rotation period can be larger than $\sim$10~d. Option (ii) is also unlikely due to the large value of \vsini. For HD~147932, we can also rule out both hypotheses because of its high value of \vsini. For HD~133518 and HD~156424, however, the \vsini\ is too low to be able to reject the two first hypotheses.

Even if few exceptions exist, the magnetic fields in early-B stars are usually of simple axisymetric configuration. Furthermore, the shape of the Zeeman signatures of HD~121743, HD~133518, and HD~156424 tend to favour an axisymetric dipolar configuration , and there is no {\em a priori} physical reason preventing the spin and magnetic axes to be aligned. Therefore, option (iii) is plausible for HD~121743, HD~133518, and HD~156424. For HD~147932, the shape of the $V$ profiles are not as simple as those observed in stars with a centred dipole, which suggests that the field might be a little more complex (quadrupolar or a higher order), or that the $V$ signatures are strongly modified by an inhomogeneous surface abundance distribution{, or by pulsations}. In the latter cases, option (iii) would also be possible for HD~147932.

Finally, only two observations have been obtained for HD~121743, HD~147932, and HD~156424 with respective time lapses of 22.3~h, 23.5~h, and 26.2~h. If we assume respective stellar masses of 8.0~\msun\ and 8.5~\msun\ and respective radii of 4.7~\rsun\ and 4.8~\rsun\ \citep{petit13} for HD~121743 and HD~156424, the break-up velocities at the surfaces of the stars are 571~\kms and 583~\kms, equivalent to rotation periods of 8.9~h and 10.0~h. If the rotation periods are close to the time lapse between the two observations, i.e.  $\sim$22~h and $\sim$26~h respectively, it would allow the star to rotate at a velocity lower than the break-up velocity. Similarly, for HD~147932, we can estimate the luminosity of the star to be between $\log L/L_{\odot}=[2.3 , 2.8]$ using the Hipparcos parallax ($7.40\pm0.59$~mas, \citealt{vanleeuwen07}), the Hipparcos magnitude and colour converted into the Johnson system ($V=7.27$, $(B-V)=0.32$, \citealt{esa97}), and the colour-temperature calibration of \citet{worthey11} that predict an intrinsic colour $(B-V)_0$ between -0.239 and -0.157, and a bolometric correction between -2.04 and -1.36. According to main sequence CESAM evolutionary tracks for a solar metallicity \citep{morel97}, we estimate the mass between 4.5 and 6.0 \msun, and the radius between 2.0 and 3.4 \rsun. With a \vsini\ around 140 \kms and a period of 23.5 h, the minimum allowed rotation rate would be 35\% of the break-up velocity, which allows option (iv). The photometric variability of a period of 20.7~h, reported by \citet{koen02}, could therefore be related to the stellar rotation, and be due to abundance patches at the surface of the star.

While some options can be easily ruled out, we are not able to favour one of them with our current dataset. Additional observations at various times are required to determine the geometry of the fields at the surface of these stars.

\section{H$\alpha$ magnetospheric signatures}


\begin{figure*}
\centering
\includegraphics[width=4.9cm,angle=90]{IV_halpha_hd66765.ps}
\hfill
\includegraphics[width=4.9cm,angle=90]{IV_halpha_hd67621.ps}
\includegraphics[width=4.9cm,angle=90]{IV_halpha_hd109026.ps}
\hfill
\includegraphics[width=4.9cm,angle=90]{IV_halpha_hd121743.ps}
\includegraphics[width=4.9cm,angle=90]{IV_halpha_hd133518.ps}
\hfill
\includegraphics[width=4.9cm,angle=90]{IV_halpha_hd147932.ps}
\includegraphics[width=4.9cm,angle=90]{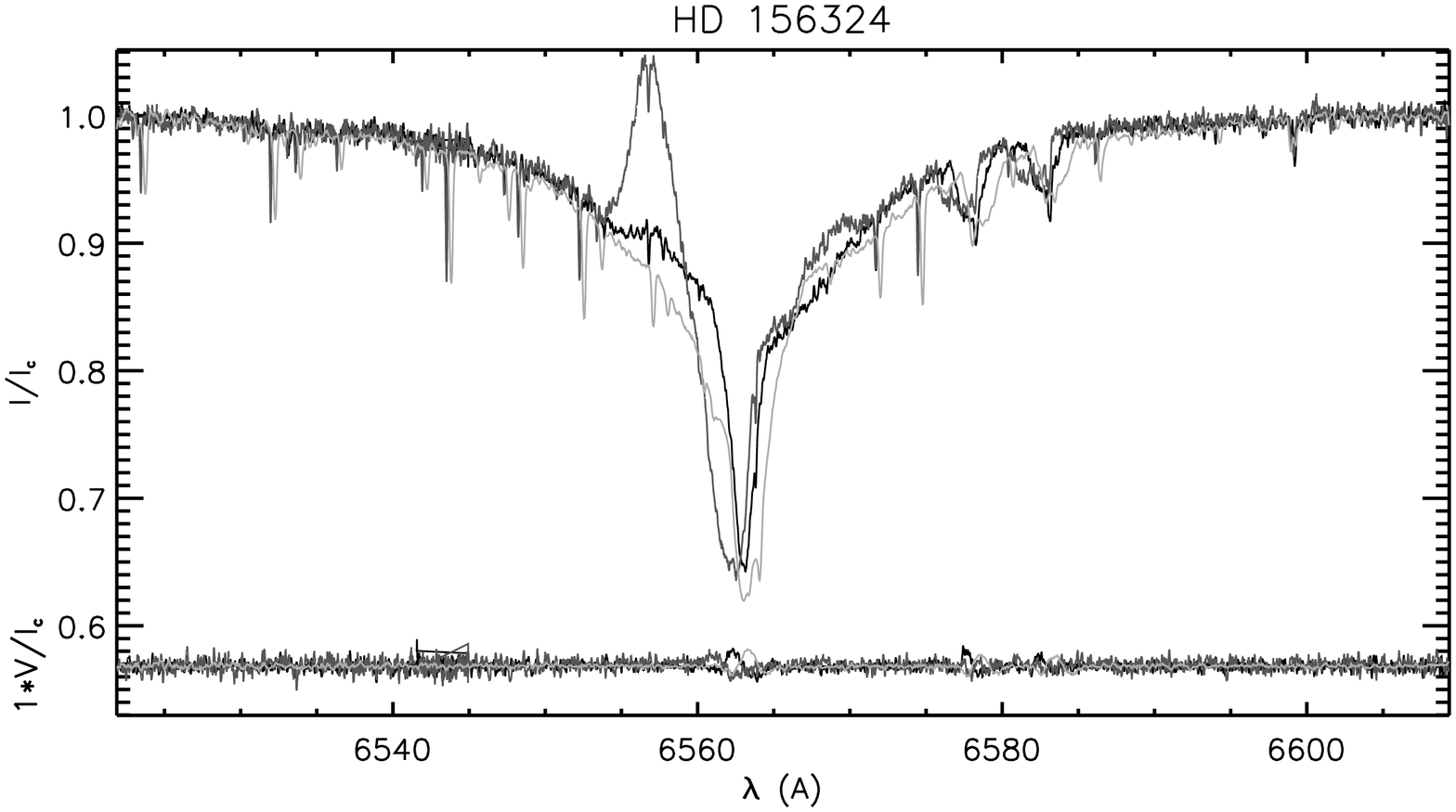}
\hfill
\includegraphics[width=4.9cm,angle=90]{IV_halpha_hd156424.ps}
\caption{Stokes $I$ (top) and $V$ (bottom) spectra of our sample stars around H$\alpha$. HD~66765: the 1st and 7th observations are plotted. HD~67621: the 5th and 7th observations are plotted. HD~109026: the 2nd and 6th observations are plotted. HD~133518: the 1st and 2nd observations are plotted. HD~156324: the 1st, 2nd and 3rd observations are plotted.}
\label{fig:halpha}
\end{figure*}

We searched for H$\alpha$ emission and variability in our data. For this purpose, we used the normalised reduced spectra described in Sec. 2, but not re-normalised by cubic spline functions, which could affect the shape of broad H$\alpha$ profiles. The almost perfect superimposed H$\alpha$ profiles observed in a few stars (e.g. in HD~133518) show that the reduction technique applied to our data is reliable enough for comparing spectra obtained at different dates. The quality of the continuum normalisation, and the reality of the inferred emission/emission variability, are supported by the excellent agreement of spectra acquired at different dates in regions outside the H$\alpha$ line wings, and the almost perfectly superimposed H$\alpha$ profiles of stars with no detected emission (e.g. in HD 133518). Indeed, we have confirmed that continuum normalisation does not contribute to the appearance of variable emission by examining the unnormalised H$\alpha$ profiles, which show the same general shape and variability as those illustrated in Fig, 6.

In Fig. \ref{fig:halpha}, we plot the spectra of our sample of stars around H$\alpha$ obtained at two different dates. In four stars (HD~66765, HD~147932, HD~156324, HD~156424) variable emission in the wings of the H$\alpha$ profiles is observed (Fig. \ref{fig:halpha}). These emissions look very similar to H$\alpha$ emissions observed in other magnetic OB stars. Such emissions are normally observed to be in phase with the rotation of the stars \citep[e.g.][]{oksala10,oksala12,grunhut12b}. In Fig. \ref{fig:halpha} we overplotted the H$\alpha$ profiles of HD~66765 obtained at the two rotation phases (around 0.0 and 0.5) where the observed variations are most extreme, and where the longitudinal magnetic field strength values are also extreme. Furthermore, we also find in our data that the H$\alpha$ profiles at similar rotation phases do not show significant differences, confirming that the H$\alpha$ emission in HD~66765 is correlated with the rotation of the star.

For HD~67621, we also overplotted the H$\alpha$ profiles in Fig. \ref{fig:halpha} at two opposite phases to give us the best chance to detect some variability. The two profiles are very similar and do not show significant differences. We also find that all profiles obtained at other phases are similar to those plotted in Fig. \ref{fig:halpha}.

For HD~66765 and HD~147932, the variations are small and might be due to the reduction method. We believe that these variations are real because (i) the data have been obtained using the same instrument and have been reduced in the same way ; (ii) the H$\alpha$ profiles are well superimposed at almost all pixel bins, except where emission or metallic lines are present; and (iii) in the case of HD~66765 no significant differences are observed in the H$\alpha$ profiles observed at different dates but at similar rotation phases. 

For HD~156324, while H$\alpha$ is affected by the spectroscopic companions, it is clear that emission is present in the first two observations, and that this emission has disappeared in the third observation. The two $V$ signatures plotted at the bottom show indeed that these observations have been obtained at three different rotation phases. This suggests that the H$\alpha$ emission might be connected to the primary component of the system, and coming from a centrifugal magnetosphere. Additional observations, providing a uniform sampling of the rotation of the primary and obtained at different rotation cycles, are required to confirm this conclusion.

Such H$\alpha$ emission is believed to result from magnetically confined winds, which collide in the magnetic equatorial plane in the close vicinity of the star \citep[e.g.][]{townsend05}. As a result, depending on the angle between the rotation and magnetic axis and on the geometry of the magnetic field, circumstellar clouds or discs are forming that rotate rigidly with the star. \citet{uddoula08} predict that such {\em centrifugal magnetospheres} can only exist if the stellar magnetic strength is high enough to confine the wind, and if the rotation rate of the star is high enough to centrifugally support the material confined into the clouds and prevent it from falling onto the star. The first condition (wind confinement) is reached if the magnetic confinement parameter, $\eta_*$, measuring the magnetic to kinetic energy ratio, is larger than unity \citep{uddoula02}, and the second condition (centrifugal support) is reached if the Alfv\'en radius ($R_A$) is larger than the Kepler co-rotation radius ($R_K$, \citealt{uddoula08}). \citet{petit13} discussed such theoretical predictions with respect to the observational constraints on the magnetospheres that were available before July 2012. In particular, they find that while all magnetic B stars with H$\alpha$ emission satisfy the condition $R_A>R_K$, not all magnetic B stars with $R_A>R_K$ show emission in $H\alpha$. They find that H$\alpha$ emission is predominantly observed in the most luminous B stars, and at a given luminosity it seems that H$\alpha$ emission is preferentially detected in the most extended magnetospheres.

In this paper, we report the discoveries of new magnetic stars whose fields have not been discussed by \citet{petit13}, as well as new determinations of polar fields and rotation periods. We therefore propose to discuss the rotational, magnetic, and magnetospheric properties that the data presented here allowed us to derive with respect to the theoretical predictions of centrifugal magnetospheres. In Table \ref{tab:eta}, we summarise the four important parameters for the following discussion: the polar field strength ($B_{\rm P}$), $\eta_*$, $R_A$, and $R_K$. We only considered single-lined spectroscopic stars here. A lower limit only on $B_{\rm P}$ could be derived by assuming that the fields are predominantly dipolar. Hence, the polar field is higher than three times the largest value of the absolute value of the measured \bl\ (Table \ref{tab:lsd}). The four objects HD~66765, HD~67621, HD~121743, and HD~156424 were part of the sample of \citet{petit13}. We therefore took the published values of $\eta_*$, $R_A$ and $R_K$ and corrected them to take our new values of $B_{\rm P}$, \vsini\ or $P_{\rm rot}$ into account. For HD~133518 and HD~147932, which have been recently discovered, we applied the same method as \citet{petit13} to derive those values. For HD~121743, HD~133518, HD~147932, and HD~156424, the rotation period being unknown, we derived lower limit on $R_K$ using their \vsini.

\begin{table}
\caption{Magnetospheric parameters of the non-spectroscopic binaries}             
\label{tab:eta}      
\centering          
\begin{tabular}{llllllllll}
\hline\hline
ID & $B_{\rm p}$ & $\eta_*$ & $R_{\rm A}$ & $R_{\rm K}$ \\
     & (kG)               &               & ($R_*$)      & ($R_*$)        \\
\hline
\underline{HD~66765}   & $> 3$ & $> 2.4$\,e4 & $>12.7$ & 2.2 \\
HD~67621   & $> 1.5$ & $> 1.4$\,e4 & $>11.2$ & 4.4 \\
HD~121743 & $> 0.9$ & $> 1.1$\,e3 & $>6.1$ & $< 3.7$ \\
HD~133518 & $> 1.5$ & $> 1.3$\,e4 & $>11$ & $<17$ \\
\underline{HD~147932} & $> 3$ & $> 2.8$\,e4 & $>13$ & $<1.9$ \\
\underline{HD~156424} & $>2.3$ & $> 7.1$\,e3 & $>9.5$ & $< 13$ \\
\hline                  
\end{tabular}
\tablefoot{Underlined IDs refer to stars with H$\alpha$ emission.}
\end{table}

Among the three single stars with H$\alpha$ emission, for two of them (HD~66765 and HD~147932) we find lower limits on $\eta_*$ and $R_A$ and upper limits on $R_K$ that satisfy the conditions for centrifugal magnetospheres. The third one, HD~156424, has a predicted Alfv\'en radius larger than 9.5~$R_*$, but a Kepler radius lower than 13~$R_*$. Although these values are not very constraining, in any case they allow for a centrifugal magnetosphere to be present. The reason why $R_K$ is not well constrained is because the \vsini\ is low, but, such a low \vsini\ does not necessarily imply that the star is rotating slowly. In fact, this star could rotate much faster than the minimum speed derived from the inclination angle. Indeed, we argue in Sec. 4 that the surface of the star would rotate at a velocity lower than the break-up velocity for rotation periods larger than 10~h. As we do observe variable emission in H$\alpha$ that is likely coming from a centrifugal magnetosphere, we propose to use the condition $R_A>R_K$ to estimate the limit below which the rotation period should be to allow for a centrifugal magnetosphere. According to  equations (2) and (4) of \citet{petit13}, and the criteria $R_A>R_K$, we find the following relation:
\begin{equation}
P < R_{\rm A}^{3/2}\left(\frac{4\pi^2}{GM_{\star}}\right)^{1/2}
\end{equation}
where $M_{\star}$ is the mass of the star. Using the values of the mass and radius of \citet{petit13}, we find that the Alfv\'en radius can be higher than the Kepler radius only if the rotation period of the star is lower than about 12 days. If the polar field strength of HD~156424 is much higher than 3~kG, this upper limit would be even larger. In any case, it appears that even with a moderate rotation, a centrifugally supported magnetosphere can exist around HD~156424 and can therefore be at the origin of the emission and variability of the H$\alpha$ profile.

For HD~67621, HD~121743, and HD~133518, while they show no evidence of circumstellar emission in the H$\alpha$ profiles, their Alfv\'en radii are larger than their Kepler co-rotating radii (Table \ref{tab:eta}). It therefore appears that these stars could have centrifugal magnetospheres. But, as discussed by \citet{petit13}, the stellar luminosity needs to be sufficient, and the magnetosphere large enough to be able to detect such magnetospheres in H$\alpha$. The lack of emission in our data might indicate that if centrifugal magnetospheres are present in the vicinity of the stars, they are too faint to be detected in our observations. \citet{groote81} detected some IR excess in the direction of HD~133518, which could indeed indicate the presence of circumstellar material.

\section{Summary}

We report the discovery of magnetic fields at the surface of seven early-B stars, as well as confirming the magnetic field of HD~109026, from observations obtained within the MiMeS project with HARPSpol. We do not detect the magnetic field in the early-B primary component of HD~109026, we reject the primary as being a He-weak star, and we are able to associate the previously reported magnetic field with the secondary Ap component only. Taking into account the discoveries obtained during the first run \citep{alecian11}, the HARPSpol LP allowed us to discover a total of nine new magnetic B stars, and to confirm the magnetic fields in one (HD 105382) additional early-B star. This brings us to a total of $\sim40$ MiMeS magnetic discoveries, illustrating the important contribution of the HARPSpol LP to the MiMeS project.

Our data also allowed us to analyse the spectral properties of the newly discovered magnetic early-B stars, which led to the discovery of one double-lined spectroscopic binary (HD~109026) and one triple-lined spectroscopic binary (HD~156324). For three of the new magnetic stars (HD~66765, HD~67621, HD~109026) we acquired enough data to derive their rotation period, which will facilitate future spectropolarimetric follow-up for magnetic mapping.

All magnetic stars discussed in this paper show abundance peculiarities: six display He-strong peculiarities, two He-weak peculiarities, and it is not yet clear to what class of CP stars the magnetic secondary of HD~109026 belongs (He-weak, Ap Si or Ap SrCrEu). It appears that, as in intermediate-mass stars, magnetic fields in B-type stars in the temperature range $18000-22000$~K are preferentially detected in chemically peculiar stars. Variable H$\alpha$ emission is detected in four out of the eight stars discussed in this paper. We argue that their emissions are very likely coming from centrifugal magnetospheres as is the case for many other magnetic OB stars \citep{petit13}. We also note the absence of H$\alpha$ emission in the other stars, while the simplest form of the theory predicts that their magnetic fields are strong enough and their rotation fast enough to host centrifugal magnetospheres. The absence of emission in H$\alpha$ is not in contradiction with the theoretical predictions of centrifugal magnetospheres, but confirms the global complexity of the formation and dynamics of such magnetospheres, as discussed by \citet{petit13}.

A full statistical analysis of the whole HARPSpol, ESPaDOnS, and Narval MiMeS sample is in progress and will treat in particular the relation between the magnetic fields properties in massive stars and the stellar parameters, chemical peculiarities, pulsations, rotation, and age (Wade et al., Grunhut et al., Petit et al., Neiner et al., Landstreet et al., Alecian et al., in prep.). Those studies will improve our knowledge of the impact of the magnetic fields on massive star formation, structure and evolution.

\begin{acknowledgements}
We thank the referee for helping us to improve this paper. EA, CN, and the MiMeS collaboration acknowledge financial support from the Programme National de Physique Stellaire (PNPS) of INSU/CNRS. MEO acknowledges financial support from GA\v{C}R under grant number P209/11/1198 and the postdoctoral program of the Czech Academy of Sciences.  The Astronomical Institute Ond\v{r}ejov is supported by the project RVO:67985815. JDL and GAW acknowledge financial support from the Natural Science and Engineering Research Council of Canada (NSERC). We thank B. Leroy for providing us with the CLEAN-NG algorithm.
\end{acknowledgements}

\bibliographystyle{aa}
\bibliography{harps_detection}

\begin{thebibliography}{136}
\expandafter\ifx\csname natexlab\endcsname\relax\def\natexlab#1{#1}\fi

\bibitem[{{Adelman}(2001)}]{adelman01}
{Adelman}, S.~J. 2001, \aap, 367, 297

\bibitem[{{Alecian} {et~al.}(2008a){Alecian}, {Catala}, {Wade}, {Donati},
  {Petit}, {Landstreet}, {B{\"o}hm}, {Bouret}, {Bagnulo}, {Folsom}, {Grunhut},
  \& {Silvester}}]{alecian08a}
{Alecian}, E., {Catala}, C., {Wade}, G.~A., {et~al.} 2008a, \mnras, 385, 391

\bibitem[{{Alecian} {et~al.}(2011){Alecian}, {Kochukhov}, {Neiner}, {Wade}, {de
  Batz}, {Henrichs}, {Grunhut}, {Bouret}, {Briquet}, {Gagne}, {Naze}, {Oksala},
  {Rivinius}, {Townsend}, {Walborn}, {Weiss}, \& {Mimes
  Collaboration}}]{alecian11}
{Alecian}, E., {Kochukhov}, O., {Neiner}, C., {et~al.} 2011, \aap, 536, L6

\bibitem[{{Alecian} {et~al.}(2008b){Alecian}, {Wade}, {Catala}, {Bagnulo},
  {Boehm}, {Bohlender}, {Bouret}, {Donati}, {Folsom}, {Grunhut}, \&
  {Landstreet}}]{alecian08b}
{Alecian}, E., {Wade}, G.~A., {Catala}, C., {et~al.} 2008b, \aap, 481, L99

\bibitem[{{Alecian} {et~al.}(2009){Alecian}, {Wade}, {Catala}, {Bagnulo},
  {B{\"o}hm}, {Bouret}, {Donati}, {Folsom}, {Grunhut}, \&
  {Landstreet}}]{alecian09}
{Alecian}, E., {Wade}, G.~A., {Catala}, C., {et~al.} 2009, \mnras, 400, 354

\bibitem[{{Alecian} {et~al.}(2013){Alecian}, {Wade}, {Catala}, {Grunhut},
  {Landstreet}, {Bagnulo}, {B{\"o}hm}, {Folsom}, {Marsden}, \&
  {Waite}}]{alecian13}
{Alecian}, E., {Wade}, G.~A., {Catala}, C., {et~al.} 2013, \mnras, 429, 1001

\bibitem[{{Auri{\`e}re} {et~al.}(2008){Auri{\`e}re}, {Konstantinova-Antova},
  {Petit}, {Charbonnel}, {Dintrans}, {Ligni{\`e}res}, {Roudier}, {Alecian},
  {Donati}, {Landstreet}, \& {Wade}}]{auriere08}
{Auri{\`e}re}, M., {Konstantinova-Antova}, R., {Petit}, P., {et~al.} 2008,
  \aap, 491, 499

\bibitem[{{Auri{\`e}re} {et~al.}(2011){Auri{\`e}re}, {Konstantinova-Antova},
  {Petit}, {Roudier}, {Donati}, {Charbonnel}, {Dintrans}, {Ligni{\`e}res},
  {Wade}, {Morgenthaler}, \& {Tsvetkova}}]{auriere11}
{Auri{\`e}re}, M., {Konstantinova-Antova}, R., {Petit}, P., {et~al.} 2011,
  \aap, 534, A139

\bibitem[{{Auri{\`e}re} {et~al.}(2010){Auri{\`e}re}, {Wade}, {Ligni{\`e}res},
  {Hui-Bon-Hoa}, {Landstreet}, {Iliev}, {Donati}, {Petit}, {Roudier}, \&
  {Th{\'e}ado}}]{auriere10}
{Auri{\`e}re}, M., {Wade}, G.~A., {Ligni{\`e}res}, F., {et~al.} 2010, \aap,
  523, A40

\bibitem[{{Babel} \& {Montmerle}(1997)}]{babel97}
{Babel}, J. \& {Montmerle}, T. 1997, \aap, 323, 121

\bibitem[{{Bagnulo} {et~al.}(2009){Bagnulo}, {Landolfi}, {Landstreet}, {Landi
  Degl'Innocenti}, {Fossati}, \& {Sterzik}}]{bagnulo09}
{Bagnulo}, S., {Landolfi}, M., {Landstreet}, J.~D., {et~al.} 2009, \pasp, 121,
  993

\bibitem[{{Bagnulo} {et~al.}(2006){Bagnulo}, {Landstreet}, {Mason}, {Andretta},
  {Silaj}, \& {Wade}}]{bagnulo06}
{Bagnulo}, S., {Landstreet}, J.~D., {Mason}, E., {et~al.} 2006, \aap, 450, 777

\bibitem[{{Bailey} {et~al.}(2012){Bailey}, {Grunhut}, {Shultz}, {Wade},
  {Landstreet}, {Bohlender}, {Lim}, {Wong}, {Drake}, \& {Linsky}}]{bailey12}
{Bailey}, J.~D., {Grunhut}, J., {Shultz}, M., {et~al.} 2012, \mnras, 423, 328

\bibitem[{{Balona}(1975)}]{balona75}
{Balona}, L.~A. 1975, \memras, 78, 51

\bibitem[{{Bertiau}(1958)}]{bertiau58}
{Bertiau}, F.~C. 1958, \apj, 128, 533

\bibitem[{{Bohlender} {et~al.}(1987){Bohlender}, {Landstreet}, {Brown}, \&
  {Thompson}}]{bohlender87}
{Bohlender}, D.~A., {Landstreet}, J.~D., {Brown}, D.~N., \& {Thompson}, I.~B.
  1987, \apj, 323, 325

\bibitem[{{Bohlender} {et~al.}(1993){Bohlender}, {Landstreet}, \&
  {Thompson}}]{bohlender93}
{Bohlender}, D.~A., {Landstreet}, J.~D., \& {Thompson}, I.~B. 1993, \aap, 269,
  355

\bibitem[{{Borra} \& {Landstreet}(1979)}]{borra79}
{Borra}, E.~F. \& {Landstreet}, J.~D. 1979, \apj, 228, 809

\bibitem[{{Borra} {et~al.}(1982){Borra}, {Landstreet}, \& {Mestel}}]{borra82}
{Borra}, E.~F., {Landstreet}, J.~D., \& {Mestel}, L. 1982, \araa, 20, 191

\bibitem[{{Borra} {et~al.}(1983){Borra}, {Landstreet}, \& {Thompson}}]{borra83}
{Borra}, E.~F., {Landstreet}, J.~D., \& {Thompson}, I. 1983, \apjs, 53, 151

\bibitem[{{Braithwaite} \& {Nordlund}(2006)}]{braithwaite06}
{Braithwaite}, J. \& {Nordlund}, {\AA}. 2006, \aap, 450, 1077

\bibitem[{{Briquet} {et~al.}(2007){Briquet}, {Hubrig}, {Sch{\"o}ller}, \& {De
  Cat}}]{briquet07}
{Briquet}, M., {Hubrig}, S., {Sch{\"o}ller}, M., \& {De Cat}, P. 2007,
  Astronomische Nachrichten, 328, 41

\bibitem[{{Briquet} {et~al.}(2012){Briquet}, {Neiner}, {Aerts}, {Morel},
  {Mathis}, {Reese}, {Lehmann}, {Costero}, {Echevarria}, {Handler}, {Kambe},
  {Hirata}, {Masuda}, {Wright}, {Yang}, {Pintado}, {Mkrtichian}, {Lee}, {Han},
  {Bruch}, {De Cat}, {Uytterhoeven}, {Lefever}, {Vanautgaerden}, {de Batz},
  {Fr{\'e}mat}, {Henrichs}, {Geers}, {Martayan}, {Hubert}, {Thizy}, \&
  {Tijani}}]{briquet12}
{Briquet}, M., {Neiner}, C., {Aerts}, C., {et~al.} 2012, \mnras, 427, 483

\bibitem[{{Briquet} {et~al.}(2013){Briquet}, {Neiner}, {Leroy}, \&
  {P{\'a}pics}}]{briquet13}
{Briquet}, M., {Neiner}, C., {Leroy}, B., \& {P{\'a}pics}, P.~I. 2013, \aap,
  557, L16

\bibitem[{{Brown} \& {Verschueren}(1997)}]{brown97}
{Brown}, A.~G.~A. \& {Verschueren}, W. 1997, \aap, 319, 811

\bibitem[{{Buscombe} \& {Morris}(1960)}]{buscombe60}
{Buscombe}, W. \& {Morris}, P.~M. 1960, \mnras, 121, 263

\bibitem[{{Castelli}(1991)}]{castelli91}
{Castelli}, F. 1991, \aap, 251, 106

\bibitem[{{Castor} {et~al.}(1975){Castor}, {Abbott}, \& {Klein}}]{castor75}
{Castor}, J.~I., {Abbott}, D.~C., \& {Klein}, R.~I. 1975, \apj, 195, 157

\bibitem[{{Cowling}(1945)}]{cowling45}
{Cowling}, T.~G. 1945, \mnras, 105, 166

\bibitem[{{Cowling}(1953)}]{cowling53}
{Cowling}, T.~G. 1953, {Solar Electrodynamics}, ed. G.~P. {Kuiper}, 532

\bibitem[{{de Geus} {et~al.}(1989){de Geus}, {de Zeeuw}, \& {Lub}}]{degeus89}
{de Geus}, E.~J., {de Zeeuw}, P.~T., \& {Lub}, J. 1989, \aap, 216, 44

\bibitem[{{de Zeeuw} {et~al.}(1999){de Zeeuw}, {Hoogerwerf}, {de Bruijne},
  {Brown}, \& {Blaauw}}]{dezeeuw99}
{de Zeeuw}, P.~T., {Hoogerwerf}, R., {de Bruijne}, J.~H.~J., {Brown}, A.~G.~A.,
  \& {Blaauw}, A. 1999, \aj, 117, 354

\bibitem[{{Donati} \& {Landstreet}(2009)}]{donati09}
{Donati}, J.-F. \& {Landstreet}, J.~D. 2009, \araa, 47, 333

\bibitem[{{Donati} {et~al.}(1997){Donati}, {Semel}, {Carter}, {Rees}, \&
  {Collier Cameron}}]{donati97}
{Donati}, J.-F., {Semel}, M., {Carter}, B.~D., {Rees}, D.~E., \& {Collier
  Cameron}, A. 1997, \mnras, 291, 658

\bibitem[{{Duez} {et~al.}(2010){Duez}, {Braithwaite}, \& {Mathis}}]{duez10b}
{Duez}, V., {Braithwaite}, J., \& {Mathis}, S. 2010, \apjl, 724, L34

\bibitem[{{Duez} \& {Mathis}(2010)}]{duez10a}
{Duez}, V. \& {Mathis}, S. 2010, \aap, 517, A58

\bibitem[{{Elkin} {et~al.}(2010){Elkin}, {Mathys}, {Kurtz}, {Hubrig}, \&
  {Freyhammer}}]{elkin10}
{Elkin}, V.~G., {Mathys}, G., {Kurtz}, D.~W., {Hubrig}, S., \& {Freyhammer},
  L.~M. 2010, \mnras, 402, 1883

\bibitem[{{ESA}(1997)}]{esa97}
{ESA}. 1997, VizieR Online Data Catalog, 1239, 0

\bibitem[{{Folsom} {et~al.}(2008){Folsom}, {Wade}, {Kochukhov}, {Alecian},
  {Catala}, {Bagnulo}, {B{\"o}hm}, {Bouret}, {Donati}, {Grunhut}, {Hanes}, \&
  {Landstreet}}]{folsom08}
{Folsom}, C.~P., {Wade}, G.~A., {Kochukhov}, O., {et~al.} 2008, \mnras, 391,
  901

\bibitem[{{Garrison} {et~al.}(1977){Garrison}, {Hiltner}, \&
  {Schild}}]{garrison77}
{Garrison}, R.~F., {Hiltner}, W.~A., \& {Schild}, R.~E. 1977, \apjs, 35, 111

\bibitem[{{Gray}(1992)}]{gray92}
{Gray}, D.~F. 1992, {The observation and analysis of stellar photospheres}
  (Cambridge Astrophysics Series, Cambridge: Cambridge University Press, 1992,
  2nd ed., ISBN 0521403200.)

\bibitem[{{Grevesse} \& {Noels}(1993)}]{grevesse93}
{Grevesse}, N. \& {Noels}, A. 1993, {Origin and Evolution of the Elements}, ed.
  N.~{Prantzos}, E.~{Langioni-flam}, \& M.~{Classe} (Cambridge Univ. Press), p.
  14

\bibitem[{{Groote} {et~al.}(1980){Groote}, {Hunger}, \& {Schultz}}]{groote80}
{Groote}, D., {Hunger}, K., \& {Schultz}, G.~V. 1980, \aap, 83, L5

\bibitem[{{Groote} \& {Kaufmann}(1981)}]{groote81}
{Groote}, D. \& {Kaufmann}, J.~P. 1981, \aap, 94, L23

\bibitem[{{Grunhut} {et~al.}(2013){Grunhut}, {Wade}, {Leutenegger}, {Petit},
  {Rauw}, {Neiner}, {Martins}, {Cohen}, {Gagn{\'e}}, {Ignace}, {Mathis}, {de
  Mink}, {Moffat}, {Owocki}, {Shultz}, {Sundqvist}, \& {MiMeS
  Collaboration}}]{grunhut13}
{Grunhut}, J.~H., {Wade}, G.~A., {Leutenegger}, M., {et~al.} 2013, \mnras, 428,
  1686

\bibitem[{{Grunhut} {et~al.}(2009){Grunhut}, {Wade}, {Marcolino}, {Petit},
  {Henrichs}, {Cohen}, {Alecian}, {Bohlender}, {Bouret}, {Kochukhov}, {Neiner},
  {St-Louis}, \& {Townsend}}]{grunhut09}
{Grunhut}, J.~H., {Wade}, G.~A., {Marcolino}, W.~L.~F., {et~al.} 2009, \mnras,
  400, L94

\bibitem[{{Grunhut} {et~al.}(2012b){Grunhut}, {Wade}, {Sundqvist}, {ud-Doula},
  {Neiner}, {Ignace}, {Marcolino}, {Rivinius}, {Fullerton}, {Kaper},
  {Mauclaire}, {Buil}, {Garrel}, {Ribeiro}, \& {Ubaud}}]{grunhut12b}
{Grunhut}, J.~H., {Wade}, G.~A., {Sundqvist}, J.~O., {et~al.} 2012b, \mnras,
  426, 2208

\bibitem[{{Guti{\'e}rrez-Soto} {et~al.}(2009){Guti{\'e}rrez-Soto}, {Floquet},
  {Samadi}, {Neiner}, {Garrido}, {Fabregat}, {Fr{\'e}mat}, {Diago}, {Huat},
  {Leroy}, {Emilio}, {Hubert}, {Andrade}, {de Batz}, {Janot-Pacheco}, {Espinosa
  Lara}, {Martayan}, {Semaan}, {Suso}, {Auvergne}, {Chaintreuil}, {Michel}, \&
  {Catala}}]{gutierrez09}
{Guti{\'e}rrez-Soto}, J., {Floquet}, M., {Samadi}, R., {et~al.} 2009, \aap,
  506, 133

\bibitem[{{Hartkopf} {et~al.}(1993){Hartkopf}, {Mason}, {Barry}, {McAlister},
  {Bagnuolo}, \& {Prieto}}]{hartkopf93}
{Hartkopf}, W.~I., {Mason}, B.~D., {Barry}, D.~J., {et~al.} 1993, \aj, 106, 352

\bibitem[{{Hartkopf} {et~al.}(1996){Hartkopf}, {Mason}, {McAlister}, {Turner},
  {Barry}, {Franz}, \& {Prieto}}]{hartkopf96}
{Hartkopf}, W.~I., {Mason}, B.~D., {McAlister}, H.~A., {et~al.} 1996, \aj, 111,
  936

\bibitem[{{Hern{\'a}ndez} {et~al.}(2005){Hern{\'a}ndez}, {Calvet}, {Hartmann},
  {Brice{\~n}o}, {Sicilia-Aguilar}, \& {Berlind}}]{hernandez05}
{Hern{\'a}ndez}, J., {Calvet}, N., {Hartmann}, L., {et~al.} 2005, \aj, 129, 856

\bibitem[{{H{\"o}gbom}(1974)}]{hogbom74}
{H{\"o}gbom}, J.~A. 1974, \aaps, 15, 417

\bibitem[{{Houk}(1978)}]{houk78}
{Houk}, N. 1978, {Michigan catalogue of two-dimensional spectral types for the
  HD stars}

\bibitem[{{Hubeny}(1988)}]{hubeny88}
{Hubeny}, I. 1988, Comput. Phys. Comm., 52, 103

\bibitem[{{Hubeny} \& {Lanz}(1992)}]{hubeny92}
{Hubeny}, I. \& {Lanz}, T. 1992, A\&A, 262, 501

\bibitem[{{Hubeny} \& {Lanz}(1995)}]{hubeny95}
{Hubeny}, I. \& {Lanz}, T. 1995, ApJ, 439, 875

\bibitem[{{Jakate}(1979)}]{jakate79}
{Jakate}, S.~M. 1979, \aj, 84, 552

\bibitem[{{Jilinski} {et~al.}(2006){Jilinski}, {Daflon}, {Cunha}, \& {de La
  Reza}}]{jilinski06}
{Jilinski}, E., {Daflon}, S., {Cunha}, K., \& {de La Reza}, R. 2006, \aap, 448,
  1001

\bibitem[{{Jones}(1971)}]{jones71}
{Jones}, D.~H.~P. 1971, \mnras, 152, 231

\bibitem[{{Kharchenko} {et~al.}(2004){Kharchenko}, {Piskunov}, {R{\"o}ser},
  {Schilbach}, \& {Scholz}}]{kharchenko04}
{Kharchenko}, N.~V., {Piskunov}, A.~E., {R{\"o}ser}, S., {Schilbach}, E., \&
  {Scholz}, R.-D. 2004, Astronomische Nachrichten, 325, 740

\bibitem[{{Kharchenko} {et~al.}(2005){Kharchenko}, {Piskunov}, {R{\"o}ser},
  {Schilbach}, \& {Scholz}}]{kharchenko05}
{Kharchenko}, N.~V., {Piskunov}, A.~E., {R{\"o}ser}, S., {Schilbach}, E., \&
  {Scholz}, R.-D. 2005, \aap, 438, 1163

\bibitem[{{Khokhlova} {et~al.}(1997){Khokhlova}, {Vasilchenko}, {Stepanov}, \&
  {Tsymbal}}]{khokhlova97}
{Khokhlova}, V.~L., {Vasilchenko}, D.~V., {Stepanov}, V.~V., \& {Tsymbal},
  V.~V. 1997, Astronomy Letters, 23, 465

\bibitem[{{Kochukhov}(2006)}]{kochukhov06b}
{Kochukhov}, O. 2006, \aap, 454, 321

\bibitem[{{Kochukhov} \& {Bagnulo}(2006)}]{kochukhov06a}
{Kochukhov}, O. \& {Bagnulo}, S. 2006, \aap, 450, 763

\bibitem[{{Kochukhov} {et~al.}(2013){Kochukhov}, {Makaganiuk}, {Piskunov},
  {Jeffers}, {Johns-Krull}, {Keller}, {Rodenhuis}, {Snik}, {Stempels}, \&
  {Valenti}}]{kochukhov13}
{Kochukhov}, O., {Makaganiuk}, V., {Piskunov}, N., {et~al.} 2013, \aap, 554,
  A61

\bibitem[{{Koen} \& {Eyer}(2002)}]{koen02}
{Koen}, C. \& {Eyer}, L. 2002, \mnras, 331, 45

\bibitem[{{Kunzli} {et~al.}(1997){Kunzli}, {North}, {Kurucz}, \&
  {Nicolet}}]{kunzli97}
{Kunzli}, M., {North}, P., {Kurucz}, R.~L., \& {Nicolet}, B. 1997, \aaps, 122,
  51

\bibitem[{{Kupka} {et~al.}(1999){Kupka}, {Piskunov}, {Ryabchikova}, {Stempels},
  \& {Weiss}}]{kupka99}
{Kupka}, F., {Piskunov}, N., {Ryabchikova}, T.~A., {Stempels}, H.~C., \&
  {Weiss}, W.~W. 1999, \aaps, 138, 119

\bibitem[{{Kurucz}(1993)}]{kurucz93}
{Kurucz}, R. 1993, Opacities for Stellar Atmospheres:
  [-3.5],[-4.0],[-4.5].~Kurucz CD-ROM No.~7.~Cambridge, Mass.: Smithsonian
  Astrophysical Observatory, 1993., 7

\bibitem[{{Landstreet}(1992)}]{landstreet92}
{Landstreet}, J.~D. 1992, \aapr, 4, 35

\bibitem[{{Landstreet} {et~al.}(1989){Landstreet}, {Barker}, {Bohlender}, \&
  {Jewison}}]{landstreet89}
{Landstreet}, J.~D., {Barker}, P.~K., {Bohlender}, D.~A., \& {Jewison}, M.~S.
  1989, \apj, 344, 876

\bibitem[{{Landstreet} \& {Borra}(1978)}]{landstreet78}
{Landstreet}, J.~D. \& {Borra}, E.~F. 1978, \apjl, 224, L5

\bibitem[{{Landstreet} {et~al.}(2008){Landstreet}, {Silaj}, {Andretta},
  {Bagnulo}, {Berdyugina}, {Donati}, {Fossati}, {Petit}, {Silvester}, \&
  {Wade}}]{landstreet08}
{Landstreet}, J.~D., {Silaj}, J., {Andretta}, V., {et~al.} 2008, \aap, 481, 465

\bibitem[{{Levato} {et~al.}(1987){Levato}, {Malaroda}, {Morrell}, \&
  {Solivella}}]{levato87}
{Levato}, H., {Malaroda}, S., {Morrell}, N., \& {Solivella}, G. 1987, \apjs,
  64, 487

\bibitem[{{MacConnell} {et~al.}(1970){MacConnell}, {Frye}, \&
  {Bidelman}}]{macconnell70}
{MacConnell}, D.~J., {Frye}, R.~L., \& {Bidelman}, W.~P. 1970, \pasp, 82, 730

\bibitem[{{Maeder} \& {Meynet}(2000)}]{maeder00}
{Maeder}, A. \& {Meynet}, G. 2000, \araa, 38, 143

\bibitem[{{Makaganiuk} {et~al.}(2011){Makaganiuk}, {Kochukhov}, {Piskunov},
  {Jeffers}, {Johns-Krull}, {Keller}, {Rodenhuis}, {Snik}, {Stempels}, \&
  {Valenti}}]{makaganiuk11}
{Makaganiuk}, V., {Kochukhov}, O., {Piskunov}, N., {et~al.} 2011, \aap, 525,
  A97

\bibitem[{{Mathys} \& {Lanz}(1997)}]{mathys97}
{Mathys}, G. \& {Lanz}, T. 1997, \aap, 323, 881

\bibitem[{{Mayor} {et~al.}(2003){Mayor}, {Pepe}, {Queloz}, {Bouchy},
  {Rupprecht}, {Lo Curto}, {Avila}, {Benz}, {Bertaux}, {Bonfils}, {Dall},
  {Dekker}, {Delabre}, {Eckert}, {Fleury}, {Gilliotte}, {Gojak}, {Guzman},
  {Kohler}, {Lizon}, {Longinotti}, {Lovis}, {Megevand}, {Pasquini}, {Reyes},
  {Sivan}, {Sosnowska}, {Soto}, {Udry}, {van Kesteren}, {Weber}, \&
  {Weilenmann}}]{mayor03}
{Mayor}, M., {Pepe}, F., {Queloz}, D., {et~al.} 2003, The Messenger, 114, 20

\bibitem[{{Mel'Nik} \& {Dambis}(2009)}]{melnik09}
{Mel'Nik}, A.~M. \& {Dambis}, A.~K. 2009, \mnras, 400, 518

\bibitem[{{Mermilliod} {et~al.}(1997){Mermilliod}, {Mermilliod}, \&
  {Hauck}}]{mermilliod97}
{Mermilliod}, J.-C., {Mermilliod}, M., \& {Hauck}, B. 1997, \aaps, 124, 349

\bibitem[{{Mihalas} \& {Athay}(1973)}]{mihalas73}
{Mihalas}, D. \& {Athay}, R.~G. 1973, \araa, 11, 187

\bibitem[{{Molenda-Zakowicz} \& {Polubek}(2004)}]{molenda04}
{Molenda-Zakowicz}, J. \& {Polubek}, G. 2004, \actaa, 54, 281

\bibitem[{{Morel}(1997)}]{morel97}
{Morel}, P. 1997, A\&AS, 124, 597

\bibitem[{{Morossi} \& {Malagnini}(1985)}]{morossi85}
{Morossi}, C. \& {Malagnini}, M.~L. 1985, \aaps, 60, 365

\bibitem[{{Moss}(2001)}]{moss01}
{Moss}, D. 2001, in Astronomical Society of the Pacific Conference Series, Vol.
  248, Magnetic Fields Across the Hertzsprung-Russell Diagram, ed. G.~{Mathys},
  S.~K. {Solanki}, \& D.~T. {Wickramasinghe}, 305--+

\bibitem[{{Napiwotzki} {et~al.}(1993){Napiwotzki}, {Schoenberner}, \&
  {Wenske}}]{napiwotzki93}
{Napiwotzki}, R., {Schoenberner}, D., \& {Wenske}, V. 1993, \aap, 268, 653

\bibitem[{{Neiner} {et~al.}(2013){Neiner}, {Degroote}, {Coste}, {Briquet}, \&
  {Mathis}}]{neiner13}
{Neiner}, C., {Degroote}, P., {Coste}, B., {Briquet}, M., \& {Mathis}, S. 2013,
  ArXiv e-prints

\bibitem[{{Neiner} {et~al.}(2012){Neiner}, {Landstreet}, {Alecian}, {Owocki},
  {Kochukhov}, {Bohlender}, \& {MiMeS Collaboration}}]{neiner12}
{Neiner}, C., {Landstreet}, J.~D., {Alecian}, E., {et~al.} 2012, \aap, 546, A44

\bibitem[{{Niemczura}(2003)}]{niemczura03}
{Niemczura}, E. 2003, \aap, 404, 689

\bibitem[{{Nissen}(1974)}]{nissen74}
{Nissen}, P.~E. 1974, \aap, 36, 57

\bibitem[{{Oksala} {et~al.}(2010){Oksala}, {Wade}, {Marcolino}, {Grunhut},
  {Bohlender}, {Manset}, {Townsend}, \& {Mimes Collaboration}}]{oksala10}
{Oksala}, M.~E., {Wade}, G.~A., {Marcolino}, W.~L.~F., {et~al.} 2010, \mnras,
  405, L51

\bibitem[{{Oksala} {et~al.}(2012){Oksala}, {Wade}, {Townsend}, {Owocki},
  {Kochukhov}, {Neiner}, {Alecian}, \& {Grunhut}}]{oksala12}
{Oksala}, M.~E., {Wade}, G.~A., {Townsend}, R.~H.~D., {et~al.} 2012, \mnras,
  419, 959

\bibitem[{{Perryman} {et~al.}(1997){Perryman}, {Lindegren}, {Kovalevsky},
  {Hoeg}, {Bastian}, {Bernacca}, {Cr{\'e}z{\'e}}, {Donati}, {Grenon},
  {Grewing}, {van Leeuwen}, {van der Marel}, {Mignard}, {Murray}, {Le Poole},
  {Schrijver}, {Turon}, {Arenou}, {Froeschl{\'e}}, \& {Petersen}}]{perryman97}
{Perryman}, M.~A.~C., {Lindegren}, L., {Kovalevsky}, J., {et~al.} 1997, \aap,
  323, L49

\bibitem[{{Petit} {et~al.}(2013){Petit}, {Owocki}, {Wade}, {Cohen},
  {Sundqvist}, {Gagn{\'e}}, {Ma{\'{\i}}z Apell{\'a}niz}, {Oksala}, {Bohlender},
  {Rivinius}, {Henrichs}, {Alecian}, {Townsend}, {ud-Doula}, \& {MiMeS
  Collaboration}}]{petit13}
{Petit}, V., {Owocki}, S.~P., {Wade}, G.~A., {et~al.} 2013, \mnras, 429, 398

\bibitem[{{Piskunov} {et~al.}(2011){Piskunov}, {Snik}, {Dolgopolov},
  {Kochukhov}, {Rodenhuis}, {Valenti}, {Jeffers}, {Makaganiuk}, {Johns-Krull},
  {Stempels}, \& {Keller}}]{piskunov11}
{Piskunov}, N., {Snik}, F., {Dolgopolov}, A., {et~al.} 2011, The Messenger,
  143, 7

\bibitem[{{Piskunov}(1992)}]{piskunov92}
{Piskunov}, N.~E. 1992, in Stellar Magnetism, 92--+

\bibitem[{{Piskunov} {et~al.}(1995){Piskunov}, {Kupka}, {Ryabchikova}, {Weiss},
  \& {Jeffery}}]{piskunov95}
{Piskunov}, N.~E., {Kupka}, F., {Ryabchikova}, T.~A., {Weiss}, W.~W., \&
  {Jeffery}, C.~S. 1995, \aaps, 112, 525

\bibitem[{{Piskunov} \& {Valenti}(2002)}]{piskunov02}
{Piskunov}, N.~E. \& {Valenti}, J.~A. 2002, \aap, 385, 1095

\bibitem[{{Przybilla} {et~al.}(2011){Przybilla}, {Nieva}, \&
  {Butler}}]{przybilla11}
{Przybilla}, N., {Nieva}, M.-F., \& {Butler}, K. 2011, Journal of Physics
  Conference Series, 328, 012015

\bibitem[{{Renson} \& {Manfroid}(2009)}]{renson09}
{Renson}, P. \& {Manfroid}, J. 2009, \aap, 498, 961

\bibitem[{{Roberts} {et~al.}(1987){Roberts}, {Lehar}, \& {Dreher}}]{roberts87}
{Roberts}, D.~H., {Lehar}, J., \& {Dreher}, J.~W. 1987, \aj, 93, 968

\bibitem[{{Roslund}(1969)}]{roslund68}
{Roslund}, C. 1969, Arkiv for Astronomi, 5, 209

\bibitem[{{Schnerr} {et~al.}(2006){Schnerr}, {Verdugo}, {Henrichs}, \&
  {Neiner}}]{schnerr06}
{Schnerr}, R.~S., {Verdugo}, E., {Henrichs}, H.~F., \& {Neiner}, C. 2006, \aap,
  452, 969

\bibitem[{{Shatsky} \& {Tokovinin}(2002)}]{shatsky02}
{Shatsky}, N. \& {Tokovinin}, A. 2002, \aap, 382, 92

\bibitem[{{Shorlin} {et~al.}(2002){Shorlin}, {Wade}, {Donati}, {Landstreet},
  {Petit}, {Sigut}, \& {Strasser}}]{shorlin02}
{Shorlin}, S.~L.~S., {Wade}, G.~A., {Donati}, J.-F., {et~al.} 2002, \aap, 392,
  637

\bibitem[{{Silvester} {et~al.}(2014){Silvester}, {Kochukhov}, \&
  {Wade}}]{silvester14}
{Silvester}, J., {Kochukhov}, O., \& {Wade}, G.~A. 2014, ArXiv e-prints

\bibitem[{{Silvester} {et~al.}(2012){Silvester}, {Wade}, {Kochukhov},
  {Bagnulo}, {Folsom}, \& {Hanes}}]{silvester12}
{Silvester}, J., {Wade}, G.~A., {Kochukhov}, O., {et~al.} 2012, \mnras, 426,
  1003

\bibitem[{{Smith} \& {Groote}(2001)}]{smith01}
{Smith}, M.~A. \& {Groote}, D. 2001, \aap, 372, 208

\bibitem[{{Sokolov}(1995)}]{sokolov95}
{Sokolov}, N.~A. 1995, \aaps, 110, 553

\bibitem[{{Stibbs}(1950)}]{stibbs50}
{Stibbs}, D.~W.~N. 1950, \mnras, 110, 395

\bibitem[{{Telting} {et~al.}(2006){Telting}, {Schrijvers}, {Ilyin},
  {Uytterhoeven}, {De Ridder}, {Aerts}, \& {Henrichs}}]{telting06}
{Telting}, J.~H., {Schrijvers}, C., {Ilyin}, I.~V., {et~al.} 2006, \aap, 452,
  945

\bibitem[{{Thompson} {et~al.}(1987){Thompson}, {Brown}, \&
  {Landstreet}}]{thompson87}
{Thompson}, I.~B., {Brown}, D.~N., \& {Landstreet}, J.~D. 1987, \apjs, 64, 219

\bibitem[{{Tokovinin} {et~al.}(2010){Tokovinin}, {Mason}, \&
  {Hartkopf}}]{tokovinin10}
{Tokovinin}, A., {Mason}, B.~D., \& {Hartkopf}, W.~I. 2010, \aj, 139, 743

\bibitem[{{Townsend}(2010)}]{townsend10}
{Townsend}, R.~H.~D. 2010, \apjs, 191, 247

\bibitem[{{Townsend} {et~al.}(2005){Townsend}, {Owocki}, \&
  {Groote}}]{townsend05}
{Townsend}, R.~H.~D., {Owocki}, S.~P., \& {Groote}, D. 2005, \apjl, 630, L81

\bibitem[{{ud-Doula} \& {Owocki}(2002)}]{uddoula02}
{ud-Doula}, A. \& {Owocki}, S.~P. 2002, \apj, 576, 413

\bibitem[{{ud-Doula} {et~al.}(2008){ud-Doula}, {Owocki}, \&
  {Townsend}}]{uddoula08}
{ud-Doula}, A., {Owocki}, S.~P., \& {Townsend}, R.~H.~D. 2008, \mnras, 385, 97

\bibitem[{{ud-Doula} {et~al.}(2009){ud-Doula}, {Owocki}, \&
  {Townsend}}]{uddoula09}
{ud-Doula}, A., {Owocki}, S.~P., \& {Townsend}, R.~H.~D. 2009, \mnras, 392,
  1022

\bibitem[{{van Hoof} {et~al.}(1963){van Hoof}, {Bertiau}, \&
  {Deurinck}}]{vanhoof63}
{van Hoof}, A., {Bertiau}, F.~C., \& {Deurinck}, R. 1963, \apj, 137, 824

\bibitem[{{van Leeuwen}(2007)}]{vanleeuwen07}
{van Leeuwen}, F. 2007, \aap, 474, 653

\bibitem[{{Wade} {et~al.}(2007){Wade}, {Bagnulo}, {Drouin}, {Landstreet}, \&
  {Monin}}]{wade07}
{Wade}, G.~A., {Bagnulo}, S., {Drouin}, D., {Landstreet}, J.~D., \& {Monin}, D.
  2007, \mnras, 376, 1145

\bibitem[{{Wade} {et~al.}(2000b){Wade}, {Donati}, {Landstreet}, \&
  {Shorlin}}]{wade00b}
{Wade}, G.~A., {Donati}, J.-F., {Landstreet}, J.~D., \& {Shorlin}, S.~L.~S.
  2000b, \mnras, 313, 823

\bibitem[{{Wade} {et~al.}(2000c){Wade}, {Donati}, {Landstreet}, \&
  {Shorlin}}]{wade00c}
{Wade}, G.~A., {Donati}, J.-F., {Landstreet}, J.~D., \& {Shorlin}, S.~L.~S.
  2000c, MNRAS, 313, 851

\bibitem[{{Wade} {et~al.}(2005){Wade}, {Drouin}, {Bagnulo}, {Landstreet},
  {Mason}, {Silvester}, {Alecian}, {B{\"o}hm}, {Bouret}, {Catala}, \&
  {Donati}}]{wade05}
{Wade}, G.~A., {Drouin}, D., {Bagnulo}, S., {et~al.} 2005, \aap, 442, L31

\bibitem[{{Wade} {et~al.}(2000a){Wade}, {Kudryavtsev}, {Romanyuk},
  {Landstreet}, \& {Mathys}}]{wade00a}
{Wade}, G.~A., {Kudryavtsev}, D., {Romanyuk}, I.~I., {Landstreet}, J.~D., \&
  {Mathys}, G. 2000a, \aap, 355, 1080

\bibitem[{{Wade} {et~al.}(2006){Wade}, {Smith}, {Bohlender}, {Ryabchikova},
  {Bolton}, {Lueftinger}, {Landstreet}, {Petit}, {Strasser}, {Blake}, \&
  {Hill}}]{wade06}
{Wade}, G.~A., {Smith}, M.~A., {Bohlender}, D.~A., {et~al.} 2006, \aap, 458,
  569

\bibitem[{{Waelkens} {et~al.}(1998){Waelkens}, {Aerts}, {Kestens}, {Grenon}, \&
  {Eyer}}]{waelkens98}
{Waelkens}, C., {Aerts}, C., {Kestens}, E., {Grenon}, M., \& {Eyer}, L. 1998,
  \aap, 330, 215

\bibitem[{{Walborn}(1983)}]{walborn83}
{Walborn}, N.~R. 1983, \apj, 268, 195

\bibitem[{{Wiegert} \& {Garrison}(1998)}]{wiegert98}
{Wiegert}, P. \& {Garrison}, R.~F. 1998, \jrasc, 92, 134

\bibitem[{{Wolff}(1968)}]{wolff68}
{Wolff}, S.~C. 1968, \pasp, 80, 281

\bibitem[{{Wolff}(1990)}]{wolff90}
{Wolff}, S.~C. 1990, \aj, 100, 1994

\bibitem[{{Wolff} \& {Heasley}(1985)}]{wolff85}
{Wolff}, S.~C. \& {Heasley}, J.~N. 1985, \apj, 292, 589

\bibitem[{{Worthey} \& {Lee}(2011)}]{worthey11}
{Worthey}, G. \& {Lee}, H.-c. 2011, \apjs, 193, 1

\bibitem[{{Zboril} \& {North}(2000)}]{zboril00}
{Zboril}, M. \& {North}, P. 2000, Contributions of the Astronomical Observatory
  Skalnate Pleso, 30, 12

\bibitem[{{Zorec} {et~al.}(2009){Zorec}, {Cidale}, {Arias}, {Fr{\'e}mat},
  {Muratore}, {Torres}, \& {Martayan}}]{zorec09}
{Zorec}, J., {Cidale}, L., {Arias}, M.~L., {et~al.} 2009, \aap, 501, 297

\end{thebibliography}

{
\appendix
\section{Synthetic spectra of the triple system HD 156324}

To interpret the complex spectrum of HD 156324, we have modified the code BINMAG developed by Oleg Kochukhov to compute the combined spectrum of a triple-lined spectroscopic binary. This code uses individual ATLAS9/SYNTH spectra (Sec. 3.1) and the stellar fluxes computed by SYNTH to calculate the normalised spectrum of the system. Instrumental, macroturbulent, and rotational broadening, as well as radial velocity shifts have been applied to each individual spectrum using a Gaussian function (for the instrumental and macroturbulent broadening), and Gray (1992)'s formalism (for the rotation). The code also requires as input the radii ratios of the primary to the secondary ($R_{\rm P}/R_{\rm S}$), and of the secondary to the tertiary ($R_{\rm S}/R_{\rm T}$). The macrotubulent and radial velocities, the \vsini\ , and the radii ratios are all adjustable parameters. The effective temperatures and gravities of the components can be adjusted by loading different synthetic spectra.

Little is known about the system HD 156324. While some parameters can be unambiguously estimated using our observations (\vsini and \vrad), some others are much more difficult to constrain because of the degenerated effects they have on a spectrum ($T_{\rm eff}$, abundances, radii ratios). An additional difficulty can appear when the spectra are variables and distorted, which seem to be the case for the primary component of HD 156324. Such degeneracy and difficulties can be overcome when a large dataset, in which variability can be studied and understood is obtained. With our small dataset we can therefore only perform an approximate fit to the data, and derive very approximate values of the parameters. We have not attempted to fit the surface gravities and the macro turbulent velocities of the stars, and we fixed their values to 4.0 (cgs) and 0~\kms, respectively, which are typical for main-sequence stars at these temperatures. While approximate, this eye-fitting procedure allowed us to attribute the three components identified in our spectra of HD 156324 to three different stellar photospheres, and to identify strong abundance peculiarities in the primary and the tertiary.

Figures \ref{fig:sb3} and \ref{fig:sb3si} show the results of some of the best fits that we could achieve for the three observations of HD~156324 around 4470~\AA\ and 6370~\AA. They illustrate the over- and under-abundances of the respective \ion{He}{i} and \ion{Si}{ii} lines from the primary, while the \ion{Ne}{I}~6402~\AA\ line is relatively well fit for the primary and secondary. The \ion{Si}{ii} lines are also relatively well fit for the secondary ; however the depth of the lines of the third component are in general not well reproduced. In Table \ref{tab:rr} are reported the effective temperatures and radii ratio that were used in these plots. We note that we can also achieve a reasonable good fit of our data using different values of the secondary and tertiary temperatures (between 14000 and 17000 K), and different values of radii ratio (between 1 and 3 for $R_{\rm P}/R_{\rm S}$, and between 0.5 and 1.5 for $R_{\rm S}/R_{\rm P}$). Table \ref{tab:rr} also illustrates our inability to fit all of our observations at the same time using a unique set of values. The complexity and the variability of the spectra are most likely the reason for this difficulty. We note that a unique set of parameters could be determined for both observations obtained when the primary and secondary components are not blended. Additional observations at various orbital phases are required for a better estimation of the stellar parameters.

\begin{table}
\caption{Effective temperatures and radii ratio of our synthetic spectra}             
\label{tab:rr}      
\centering          
\begin{tabular}{cccccc}
\hline\hline
HJD & $T_{\rm eff}$ & $T_{\rm eff}$ & $T_{\rm eff}$ & $R_{\rm P}/R_{\rm S}$ & $R_{\rm S}/R_{\rm T}$ \\
(2450000+) & P & S & T & \\
\hline
6\,127.62863 & 22000 & 15000 & 14000 & 1.7 & 0.5 \\
6\,128.78131 & 22000 & 15000 & 14000 & 2.0 & 1.0\\
6\,462.92963 & 22000 & 15000 & 14000 & 2.0 & 1.0\\
\hline                  
\end{tabular}
\end{table}


\begin{figure*}
\centering
\includegraphics[width=6.5cm,angle=90]{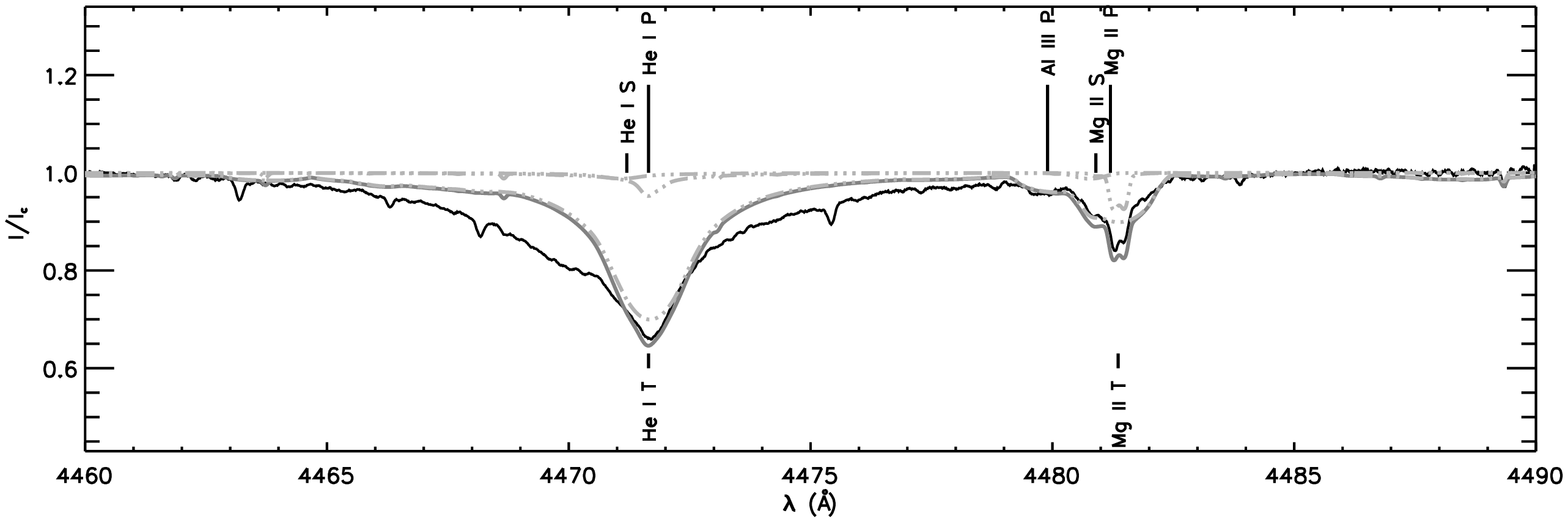}
\includegraphics[width=6.5cm,angle=90]{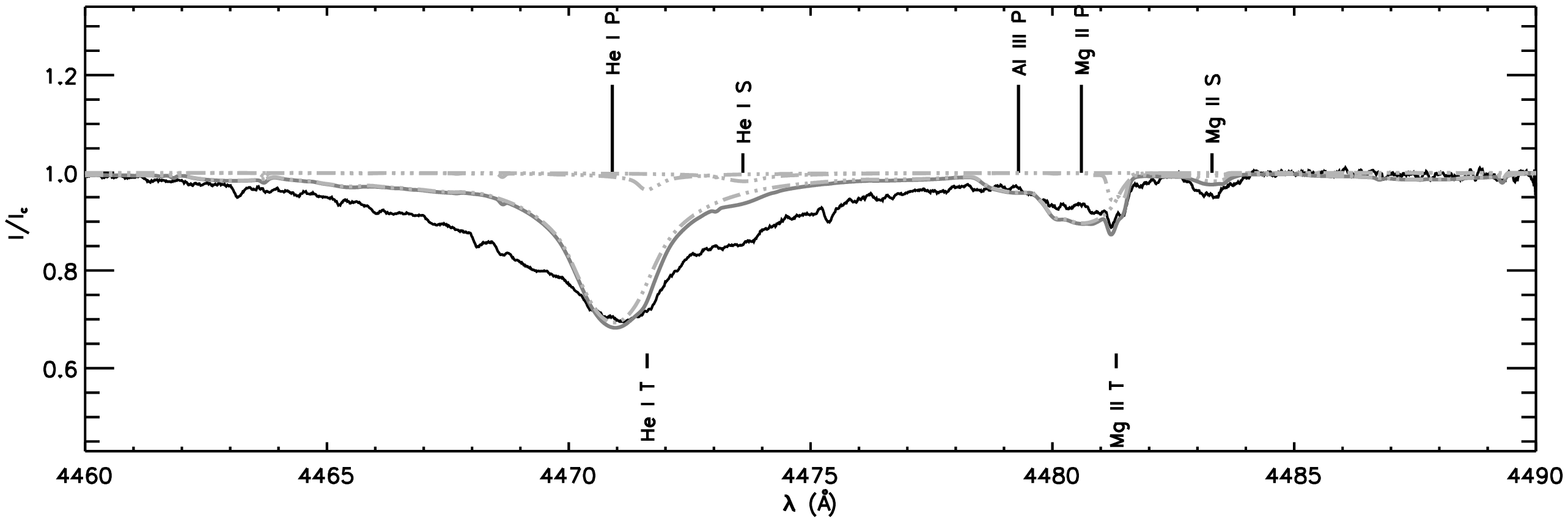}
\includegraphics[width=6.5cm,angle=90]{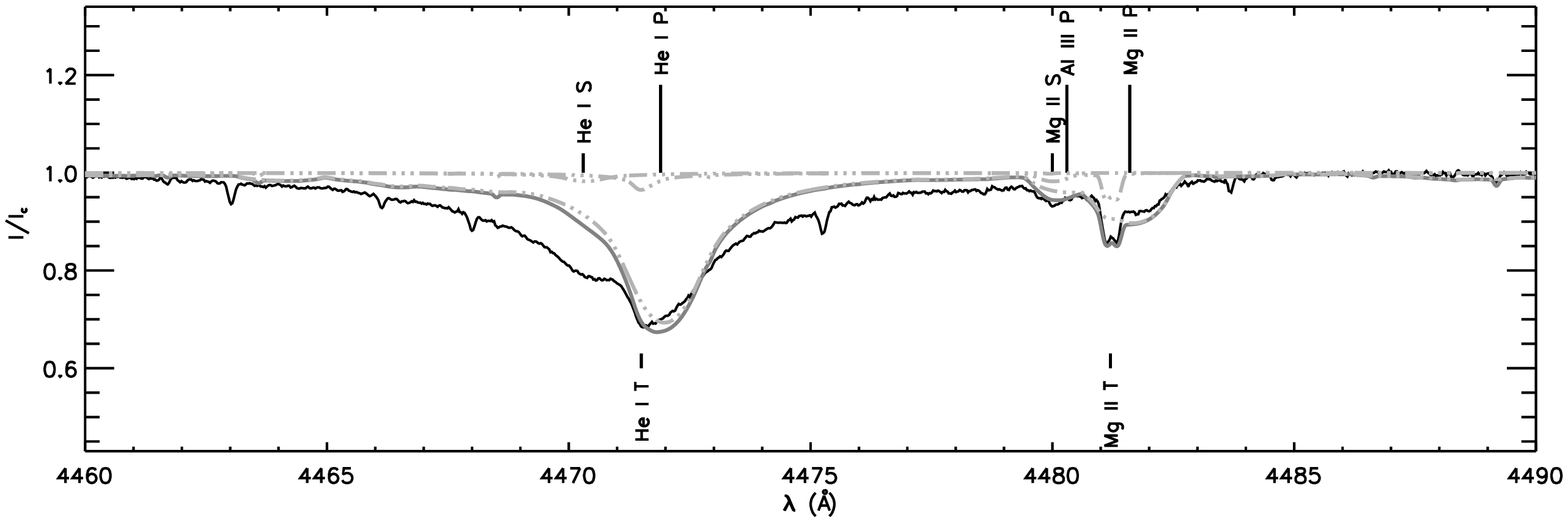}
\caption{Intensity spectra of HD 156324 around \ion{He}{i}~4471~\AA\ and \ion{Mg}{ii}~4481~\AA\ of the first, second, and third (top to bottom) observations. The combined synthetic spectra of the triple system are over plotted in each panel with full dark grey lines. The  contributions of the three components relative to the combined stellar flux of the system are over plotted in dot-dashed light-grey lines.The position of the strongest lines of the primary (P), secondary (S), and tertiary (T) components are indicated with vertical bars.}
\label{fig:sb3}
\end{figure*}

\begin{figure*}
\centering
\includegraphics[width=6.5cm,angle=90]{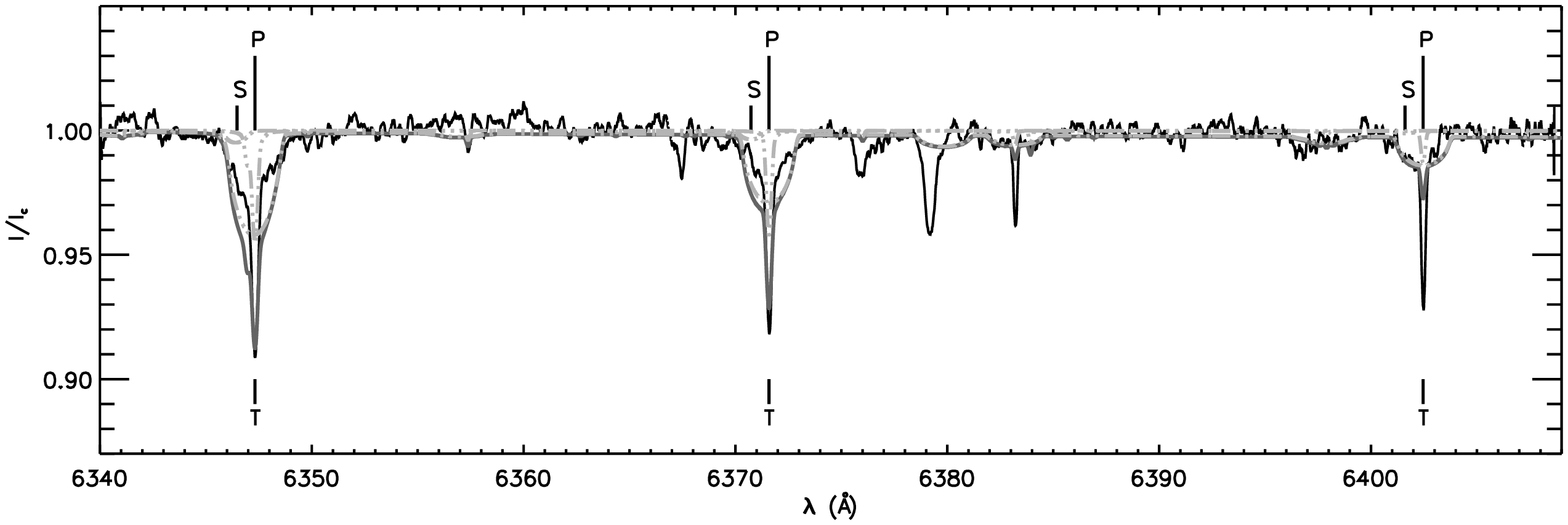}
\includegraphics[width=6.5cm,angle=90]{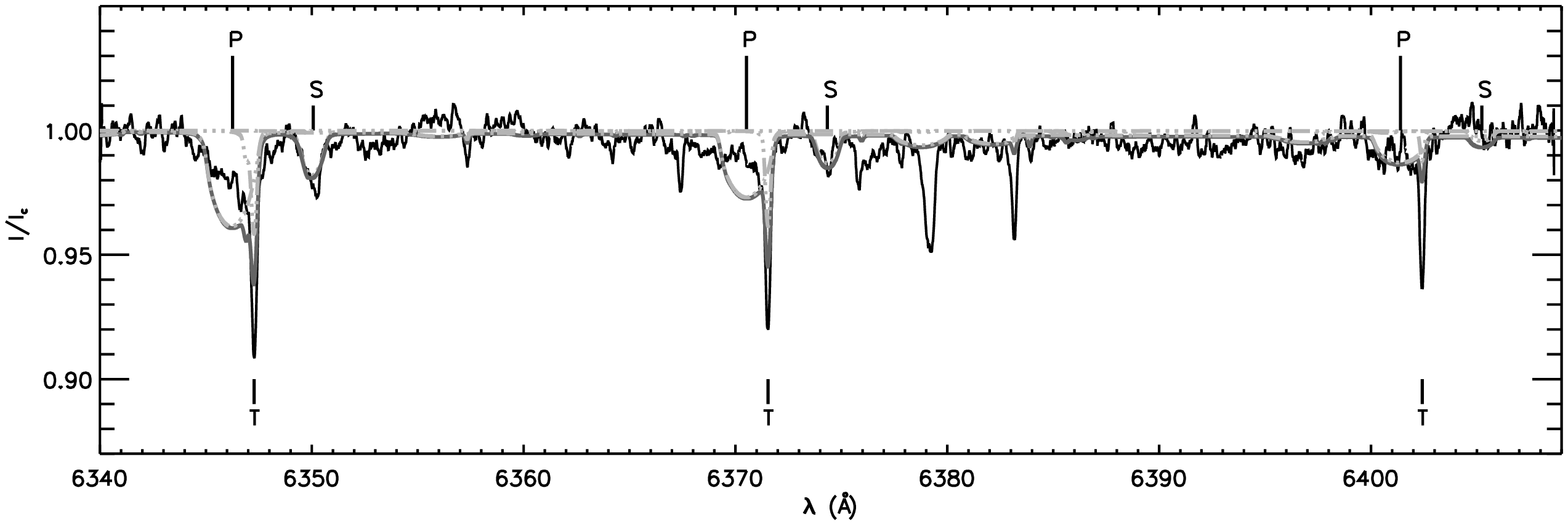}
\includegraphics[width=6.5cm,angle=90]{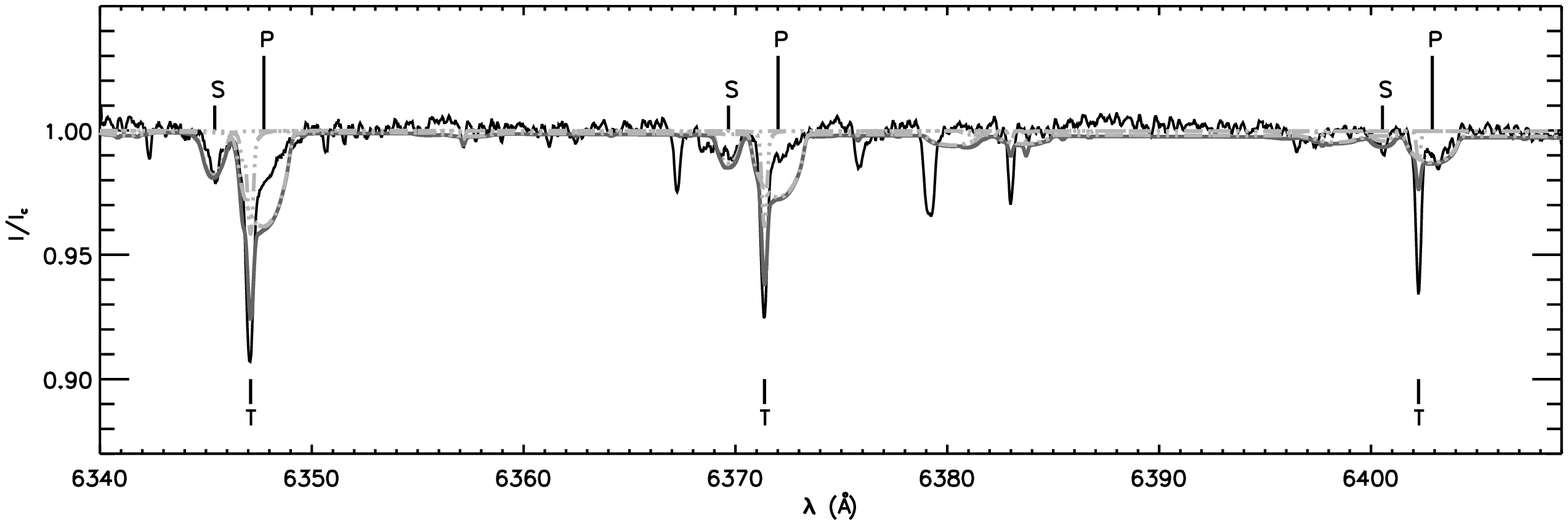}
\caption{Same as Fig. \ref{fig:sb3} but in a region of the spectrum containing the lines \ion{Si}{ii}~6347~\AA, \ion{Si}{ii}~6371~\AA\ and \ion{Ne}{ii}~6402~\AA.}
\label{fig:sb3si}
\end{figure*}

}

\end{document}